\documentstyle[epsfig,aps]{revtex}
\newcommand{\be}{\begin{equation}}
\newcommand{\ee}{\end{equation}}
\newcommand{\ber}{\begin{eqnarray}}
\newcommand{\eer}{\end{eqnarray}}
\newcommand{\ba}{\begin{array}}
\newcommand{\ea}{\end{array}}
\newcommand{\bacc}{\begin{array}{cc}}
\newcommand{\baccc}{\begin{array}{ccc}}
\newcommand{\bacccc}{\begin{array}{cccc}}
\newcommand{\lp}{\left(}
\newcommand{\rp}{\right)}
\newcommand{\ls}{\left[}
\newcommand{\rs}{\right]}\newcommand{\lb}{\left\{}
\newcommand{\rb}{\right\}}

\newcommand{\pivec}{\vec{\pi}}

\newcommand{\Psivec}{\vec{\Psi}}
\newcommand{\pihat}{\hat{\pi}}
\newcommand{\sighat}{\hat{\sigma}}
\newcommand{\dd}{\partial}
\newcommand{\dmpi}{\partial_\mu \pi}
\newcommand{\dnpi}{\partial_\nu \pi}
\newcommand{\dapi}{\partial_\alpha \pi}

\newcommand{\dMpi}{\partial^\mu \pi}

\newcommand{\dnsig}{\partial_\nu \sigma}
\newcommand{\dasig}{\partial_\alpha \sigma}

\newcommand{\dmph}{\partial_\mu \pihat}

\newcommand{\daph}{\partial_\alpha \pihat}
\newcommand{\dbph}{\partial_\beta \pihat}

\newcommand{\dMph}{\partial^\mu \pihat}

\newcommand{\dmsh}{\partial_\mu \sighat}

\newcommand{\dash}{\partial_\alpha \sighat}
\newcommand{\dbsh}{\partial_\beta \sighat}

\newcommand{\dMsh}{\partial^\mu \sighat}

\newcommand{\fpi}{f_\pi}
\newcommand{\tauvec}{\vec{\tau}}
\newcommand{\pik}{\pi_k}
\newcommand{\tauk}{\tau_k}
\newcommand{\ppi}{\underline{\pi}}
\newcommand{\tr}{{\mathrm tr}}
\newcommand{\cU}{{\cal U}}
\newcommand{\cUd}{{\cal U}^\dagger}

\newcommand{\dpi}{\dot{\pi}}
\newcommand{\dph}{\dot{\pihat}}

\newcommand{\ddph}{\ddot{\pihat}}

\newcommand{\eMNAB}{\epsilon^{\mu \nu \alpha \beta}}

\begin{document}
\title{Two Skyrmion Dynamics with $\omega$ Mesons}

\author{R.D. Amado, M. \'{A}. Hal\'{a}sz, P. Protopapas}

\address{Department of Physics and Astronomy, University of Pennsylvania,
         Philadelphia, PA 19104}

\date{\today}

\draft

\maketitle

\begin{abstract}
We present our first results of numerical simulations of two
skyrmion dynamics
using an $\omega$-meson stabilized effective Lagrangian. 
We consider skyrmion-skyrmion scattering with a fixed initial velocity
of $\beta=0.5$, for various impact parameters and groomings. 
The physical picture that emerges is surprisingly rich,
while consistent with previous results and general conservation laws.
We find meson radiation, skyrmion scattering out of the 
scattering plane, orbiting and capture to bound states. 
\end{abstract}

\section{Introduction}
In the large $N_C$ or classical limit of QCD, nucleons 
may be identified as classical
solitons of the scalar-isovector $SU(N_f)$ pion field. The simplest theory 
that manifests these solitons is the non-linear sigma model. However the
solitons of this theory are not stable against collapse. The first 
attempt to provide a stable theory was by Skyrme (long before QCD) \cite{Skyrme}
who introduced a fourth order term. This term does indeed lead to stabilized
solitons that are called skyrmions, and there is a vast body of work on
their properties and on how to quantize them \cite{Review}.
Unfortunately the fourth order term introduces numerical instabilities
that make complex dynamical calculation nearly impossible \cite{Livermore}.
It is possible to stabilize the non-linear sigma model without
the fourth order, or Skyrme term as it is called, by coupling the baryon
current to the $\omega$ meson field \cite{Review}.
This not only 
provides stability, but does so
in the context of more reasonable physics. 
It is also possible to stabilize the solitons by introducing
a $\rho$ meson field, with a gauge coupling, 
or with both the $\omega$ and $\rho$, but 
the $\rho$ adds 
%little to the physics and 
a great deal to the numerical
complications. Hence in this first, exploratory work, we stabilize
with the $\omega$ only. We continue to refer to the solitons as skyrmions.

The Skyrme approach, either in its fourth order or $\omega$ stabilized
form, has much to recommend it as a model of low energy strong interaction
physics. This low energy or long wave length domain is notoriously difficult
to describe in the context of standard QCD. The Skyrme approach includes
chiral symmetry, baryons (and therefore also antibaryons), pions, the 
one pion exchange potential, and, with quantization, nucleons and 
deltas. The idea of having nucleons arise naturally from an effective
theory of pions and vector mesons is especially attractive as a 
path from QCD to such effective Lagrangians is better understood.

It has been shown that the Skyrme model can give a good account,
with very few parameters, of the low 
energy properties of nucleons \cite{Review}.
In the baryon number
two system, the Skyrme approach can describe the principal features of
the nucleon-nucleon static potential \cite{W&A}. Most of these problems
have also been successfully studied in more traditional nuclear physics
forms.

One problem that has not yielded easily to traditional physics
approaches and is out of the reach of standard lattice QCD is low energy 
annihilation. This is a problem ideally suited to the Skyrme approach. We have 
already shown that a general picture of post-annihilation dynamics including
branching ratios can be obtained from the Skyrme approach \cite{Lu&RDA}.
The nucleon-antinucleon potential can be as well \cite{LPA}. 
What remains is a full modeling
of annihilation from start to finish. The only attempts we know of to do that
in the Skyrme model had numerical problems associated with the 
Skyrme term \cite{CalTech:ann} \cite{Livermore}. We propose to study skyrmion
dynamics using $\omega$ stabilization, thus avoiding the usual 
numerical problems. As a prelude to studying annihilation, we have
studied scattering in the baryon number two system. It is that work we
report here.

In the Skyrme model the three components of the pion field, 
{$\pi_1,\pi_2,\pi_3$}, can be aligned with the spatial directions,
{$x,y,z$}, providing a correspondence between the  two spaces. This
simple alignment is called a hedgehog. A rotation of the pion field with
respect to space is called a grooming. The energy of a single free skyrmion
is independent of grooming, but the interaction between two skyrmions
depends critically, as we shall see, on their relative grooming. 
In this paper we present the results of calculations of skyrmion-skyrmion
scattering (with $\omega$ meson stabilization) at medium energy for a 
variety of impact parameters and groomings. We find rich structure. Some
of the channels have simple scattering, but some display 
radiation, scattering out of the scattering plane, orbiting
and ultimately capture to a bound $B=2$ state. These calculations are
numerically complex and require considerable computational resources and
computing time, but they are numerically stable. We know of no other 
calculations of $B=2$ skyrmion scattering that show the phenomena we 
find. Furthermore our success here bodes well for extending the method
to annihilation. 

In the next section we briefly review the model we use, presenting
both the Lagrangian for the $\omega$ stabilized non-linear sigma model,
and our equations of motion. In Section 3 we discuss our numerical 
strategies and methods. Section 4 presents our results, mostly in graphical
form, and Section 5 deals with conclusions and outlook. The reader 
interested only in results can go directly to Sections 4 and 5.

\section{Formalism}

\subsection{Model}

Our starting point is the non-linear $\sigma$ model Lagrangian,
\be
\label{Ulagr1}
{\cal L}_\sigma~=~ 
\frac14 \fpi^2 \tr \lp \partial_\mu \cU \partial^\mu \cUd \rp +
\frac12 m_\pi^2 \fpi^2 \tr \lp \cU - 1 \rp~~,
\ee
where the $SU(2)$ field ${\cU}$ is parameterized by the three 
real pion fields\footnote{
To avoid confusion, throughout this paper we reserve the plain letter $\pi$ for 
the pion field, and use the symbol $\ppi = 3.1415\dots$ for the mathematical constant.}
$\{ \pik \}_{k=1,3} = \pivec$,
\be
\label{pidef}
\cU~=\exp \lp i~ (\tauvec \cdot \pivec) \rp~=~\cos \pi + i \sin \pi ~(\tauvec \cdot \pihat)~~.
\ee
Here, $\{ \tauk \}_{k=1,3} = \tauvec$ are the Pauli matrices (in flavor space).
We identify the baryon current with
\be 
B^\mu ~=~ \frac{1}{24 \ppi^2 }
\epsilon^{\mu \nu \alpha \beta}
\tr \ls 
    \lp \cUd \, \dd_\nu \cU \rp
    \lp \cUd \, \dd_\alpha \cU \rp
    \lp \cUd \, \dd_\beta \cU \rp
\rs~~.
\ee
The full Lagrangian also contains the free $\omega$ and the interaction part,
\be
\label{Ulagr2}
{\cal L} = {\cal L}_\sigma + \frac32 g \omega_\mu B^\mu
-~\frac12 \dd_\mu \omega_\nu \lp \dd^\mu \omega^\nu - \dd^\nu \omega^\mu \rp
+ \frac12 m^2 \omega_\mu \omega^\mu ~~.
\ee
We take $\fpi = 62~{\rm MeV}$ following \cite{fpi} 
%close to that of a nucleon
 and $g=\frac{m}{\sqrt{2} \fpi}$ from VMD where $m=770~{\rm MeV}$ is the
vector meson mass.  This gives a skyrmion mass of $791~{\rm MeV}$. 

\subsection{Choice of Dynamical Variables}

The traditional approach to numerical simulations involving skyrmions 
uses the cartesian decomposition of the unitary matrix $\cU$,
\ber
\label{defpsi}
{\cU}~=~\Phi {\bf I}_2 + \Psivec \cdot \tauvec~~;~~
\Phi = \cos \pi ~,~\Psivec = \pihat \sin \pi
\eer
The quadruplet of real numbers $(\Phi,\Psivec)$ is constrained by the unitarity condition
$\cU \cUd~=~{\bf I}_2 \rightarrow  \Phi^2 + \Psivec \cdot \Psivec =1 $, also known as
the chiral condition.
The Lagrangian is usually written in terms of  $(\Phi,\Psivec)$ and the
chiral condition is imposed using a Lagrange multiplier field.
 The four coordinates of $\cU$ are similar to the cartesian 
coordinates of a point in ${\bf R}^4$ 
confined to the surface of a hypersphere of 
unit radius. 

An attractive idea is to use an approach which ensures naturally the unitarity
of $\cU$ at all times. Such a method is used successfully in lattice QCD in the 
context of the hybrid Monte-Carlo algorithm \cite{Go87}. There, the dynamical variables
are $\cU$ itself and the Hermitian matrix $\dot{\cU} \cUd$. 
In the present work, we employ the parameterization of $\cU$ that follows from (\ref{pidef}).
We will use $\pi$ and $\pihat$ and their time derivatives
$\dpi , \dph$ as  dynamical variables. This implements only in part the principle
mentioned above, since the unit vector $\pihat$ is subject to conditions similar
to the four-dimensional vector $(\Phi,\Psivec)$. 
However, one has better geometrical intuition
for the former. The connection to  $(\Phi,\Psivec)$ is straightforward via
(\ref{defpsi}).
The Lagrangian in terms of $\pi,\pihat$ is
\ber
{\cal L} ~&=&~{\cal L}_\sigma~+{\cal L}_{int}~+{\cal L}_{\omega} \nonumber \\
{\cal L}_\sigma~&=&~ \frac{1}{2} \fpi^2 
\lp \dmpi \dMpi + \sin^2 \pi \dmph \cdot \dMph \rp +
m_\pi^2 \fpi^2 ( \cos \pi - 1 ) \nonumber \\
{\cal L}_{int}~&=&~\frac{3 g }{ 8 \ppi^2 } \eMNAB
\sin^2 \pi \omega_\mu \dnpi \ls \pihat \cdot \lp \daph \times \dbph \rp \rs \nonumber \\
{\cal L}_{\omega} ~&=&~ - \frac12 
\partial_\mu \omega_\nu 
\lp \partial^\mu \omega^\nu - \partial^\nu \omega^\mu \rp
+ \frac12 m_{vec}^2 \omega_\mu \omega^\mu~~.
\eer

\subsection{Equations of Motion}

We wish to obtain the Euler-Lagrange 
equations for $\pi, \pihat$. The problem is that $\pihat$ is a
unit vector, $\pihat \cdot \pihat = 1$, so its components cannot be treated as true 
coordinates. One may still write down the Euler-Lagrange equations, 
by considering the action
\ber
S ~=~ \int d^3 x dt {\cal L}\lp \pi(x,t),\pihat(x,t),\omega_\alpha(x,t);
\dmpi(x,t),\dmph(x,t),\partial_\mu{\omega}_\alpha(x,t) \rp~~.
\eer
The Lagrange equation for $\pi$ is obtained by requiring that $S$ be stationary
with respect to a small variation $\delta \pi(x,t)$ and the corresponding variations
$\delta (\dmpi) = \partial_\mu (\delta \pi)$. The variation of $S$ which has to vanish
for any $\delta \pi$, is 
\ber
\label{preleqpi}
\delta S ~=~\int d^3 x dt 
\lb
\frac{\partial {\cal L}}{\partial \pi} \delta \pi +
\frac{\partial {\cal L}}{\partial (\dmpi)} \delta (\dmpi) \rb
~=~\int d^3 x dt 
\lb
\frac{\partial {\cal L}}{\partial \pi} -
\partial_\mu \ls \frac{\partial {\cal L}}{\partial (\dmpi)}\rs \rb \delta \pi~~.
\eer
Therefore the quantity in curly brackets in the last equality of (\ref{preleqpi}) has to
be identically zero, since $\delta \pi$ is completely arbitrary. Of course this leads 
exactly to the usual form of the Euler-Lagrange equations.
We may repeat formally the same steps for $\pihat$, leading to
\ber
\delta S ~=~\int d^3 x dt 
\lb
\frac{\partial {\cal L}}{\partial \pihat} -
\partial_\mu \ls \frac{\partial {\cal L}}{\partial (\dmph)}\rs \rb \delta \pihat~~.
\eer
Here, partial differentiation with respect to the vector $\pihat$ or its derivatives 
gives the vector obtained by differentiating with respect to the corresponding components of 
$\pihat$ or its derivatives. If we want $\delta S$ to vanish for any $\delta \pihat$, 
we do \em not \em need to require that the quantity in curly brackets vanish identically.
This is because $\delta \pihat$ is not completely arbitrary. Both $\pihat$ and 
$\pihat' = \pihat + \delta \pihat$ have to be unit vectors. The variation of
$\pihat \cdot \pihat$ must vanish,
\ber
0 ~=~ \delta \lp \pihat \cdot \pihat \rp ~=~ 2 ~\delta \pihat \cdot \pihat ~~.
\eer
In other words, $\delta \pihat(x,t) ~\perp~\pihat(x,t)$ for any $(x,t)$.
Therefore, the necessary condition for the stationarity of $S$ is simply
\ber
\lb
\frac{\partial {\cal L}}{\partial \pihat} -
\partial_\mu \ls \frac{\partial {\cal L}}{\partial 
(\dmph)}\rs \rb \parallel \pihat~~.
\eer
The above statement is easily turned into an equation by subtracting the 
projection of the left hand side onto $\pihat$,
\ber
\lb
\frac{\partial {\cal L}}{\partial \pihat} -
\partial_\mu \ls \frac{\partial {\cal L}}{\partial (\dmph)}\rs \rb
- \pihat \cdot 
\lb
\frac{\partial {\cal L}}{\partial \pihat} -
\partial_\mu \ls \frac{\partial {\cal L}}{\partial (\dmph)}\rs \rb~=~0~~.
\eer
The Euler-Lagrange equations we obtain finally are
\ber
\label{leqs1}
\partial_\nu \partial^\nu \pi ~&=&~
\frac12 \sin 2 \pi \dmph \cdot \dMph
- m_\pi^2 \sin \pi
+ \frac{ 3 g }{8 \ppi^2 \fpi^2 } \eMNAB \sin^2 \pi
\partial_\mu \omega_\nu \ls \pihat \cdot \lp \daph \times \dbph \rp \rs \nonumber \\
\partial_\mu \partial^\mu \pihat 
~&=&~ \pihat \lp \pihat \cdot \partial_\mu \partial^\mu \pihat \rp 
- \frac { 2 \cos \pi }{\sin \pi } \partial_\mu \pi \partial^\mu \pihat 
- \frac{3 g }{4 \ppi^2 \fpi^2 } \eMNAB 
\partial_\mu \omega_\nu \dapi \lp \dbph \times \pihat \rp \nonumber \\
\partial_\nu \partial^\nu \omega^\mu ~&=&~
- m_{vec}^2 \omega^\mu
- \frac{ 3 g }{ 8 \ppi^2 } \eMNAB 
\sin^2 \pi \dnpi \ls \pihat \cdot \lp \daph \times \dbph \rp \rs
~~.
\eer

\section{Computation}

Our calculation is based on the equations of motion (\ref{leqs1}), using 
$ ( \pi, \pihat, \omega_\mu ; \dpi, \dph, \dot{\omega}_\mu ) $ as variables.
This choice leads to two problems which have to be tackled by our discretization
scheme. 
First, we have a 
coordinate singularity at $\pi = 0$ and $\pi = \ppi$. Here, $\pihat$ is not defined.
The situation is similar to that of angles in polar coordinates when the radius
vanishes. While the equations of motion are correct for any non-zero $\pi$,
in the vicinity of the coordinate singularity small changes of $\pivec$ are translated
into very large variations of $\pihat$. A scheme based on the 
equations of motion (\ref{leqs1}) breaks down around the `poles' of the hypersphere
due to the large discretization errors involved.

One way out is to rotate the coordinate system in $SU(2)$ so as to avoid the problem
regions. We introduce a new field ${\cal V}$ obtained by rotation with a fixed 
$U_0 \in SU(2)$,
\ber
\label{Vdef}
{\cal U} ~=~ U_0 {\cal V}~~;~~{\cal V}~=~\exp \lp i \tauvec \cdot \vec{\sigma} \rp~~;
~~U_0~=~ \exp \lp  i \tauvec \cdot \vec{\Theta} \rp~~.
\eer
Substituting (\ref{Vdef}) into our Lagrangian (\ref{Ulagr1})(\ref{Ulagr2}),
we find that only the pion mass term is modified since everywhere else ${\cal U}$
is combined with $\cUd$ or its derivatives so $U_0$ drops out. 
The equations of motion can then be derived
in an identical fashion to (\ref{leqs1}). We cite them here for the sake of 
completeness,
\ber
\label{leqs1rot}
\partial_\nu \partial^\nu \sigma ~&=&~
\frac12 \sin 2 \pi \dmsh \cdot \dMsh
- m_\pi^2 \lp \cos \Theta ~\sin \sigma 
             + (\hat{\Theta} \cdot \sighat ) \sin \Theta ~ \cos \sigma \rp
+ \frac{ 3 g }{8 \ppi^2 \fpi^2 } \eMNAB \sin^2 \sigma
\partial_\mu \omega_\nu \ls \sighat \cdot \lp \dash \times \dbsh \rp \rs \nonumber \\
\partial_\mu \partial^\mu \sighat 
~&=&~ \sighat \lp \sighat \cdot \partial_\mu \partial^\mu \sighat \rp 
- \frac { 2 \cos \sigma }{\sin \sigma } \partial_\mu \sigma \partial^\mu \sighat 
- \frac{3 g }{4 \ppi^2 \fpi^2 } \eMNAB 
\partial_\mu \omega_\nu \dasig \lp \dbsh \times \sighat \rp 
- m_\pi^2 \lp \hat{\Theta} - \sighat(\hat{\Theta} \cdot \sighat) \rp
\sin \Theta \sin \sigma
\nonumber \\
\partial_\nu \partial^\nu \omega^\mu ~&=&~
- m_{vec}^2 \omega^\mu
- \frac{ 3 g }{ 8 \ppi^2 } \eMNAB 
\sin^2 \sigma \dnsig \ls \sighat \cdot \lp \dash \times \dbsh \rp \rs
~~.
\eer
In our calculation, we first rotate the field to be updated,
together with the surrounding fields, so that the 
corresponding $(\sigma,\sighat)$ is comfortably away from the 
coordinate singularities.
Then we apply the discretized equations of motion (see below) derived from
(\ref{leqs1rot}) and finally we rotate back the updated fields.
We stress that switching to rotated fields and (\ref{leqs1rot}) amounts to 
a mere changing of the coordinate system. The content of the equations is identical
irrespective of the coordinate system. The difference is in the 
discretization which in the vicinity of the `poles' leads to large errors which
are avoided in the rotated frame. The choice of $\vec{\Theta}$ is largely
arbitrary. For simplicity we choose it so that $\vec{\sigma}$ is always on the
`equator' of the $SU(2)$ hypersphere. 

The second important numerical issue follows from the unit vector nature\footnote{
We will continue to use $\pivec$ when referring to the 
pion field, even though we perform our updates using the rotated pion
fields $\vec{\sigma}$.} of 
$\pihat$,  $\pivec \cdot \pivec = 1$. 
This and the corresponding constraints on $\dph$ and $\ddph$, namely
$\pihat \cdot \dph = 0$, $\ddph \cdot \pihat + \dph \cdot \dph = 0$
are consistent with the equations of motion but are violated by the discretized
equations. Therefore they need to be imposed by projecting out the spurious
components of $\dph$ and $\ddph$ at every step in the update.
There are similar conditions for the spatial derivatives,
which also have to be taken into account. 
This is of course a rather technical point, but it is a reminder of the fact
that our approach does not completely conform to the idea of using a minimal 
set of coordinates. That would be achieved for instance by trading $\pihat$,
which describes a direction in three dimensions, for the two angles that define that
direction. That would have introduced another set of coordinate singularities to
be avoided. The opposite strategy would be to apply the same considerations
we followed for $\pihat$ in deriving the Euler-Lagrange equations and imposing
the unit length constraint, to the four-dimensional 
unit vector $(\Phi,\Psivec)$.
That also remains an open possibility, along with using unitary and Hermitian
matrices as dynamical variables and of course, the traditional path employing  
Lagrange multipliers. 
We have not found it necessary to employ any of these notions at this time. 
Besides the fact that it leads to a reasonably stable
calculation, our choice of variables has the advantage of fairly simple
equations of motion. The decomposition of $\pivec$ into its length and unit vector
follows quite naturally, and leads to great simplification in the much more
complicated case of including a vector-isovector $\rho$ meson field.

The main idea of our numerical scheme is the following.
The discretized time evolution for the fields themselves is quite straightforward,
given the knowledge of their time-derivatives or `velocities'. 
The evolution of the velocities is the core of our procedure. 
The velocities are assigned to half-integer timesteps. 
Thus, one timeslice contains the 
fields at a given time and the velocities half a timestep earlier. 
The time evolution of the velocities follows from solving the equations of motion for
the second time derivatives. The latter are written in terms of the retarded
velocity (at time $(t-\Delta t /2)$) and the one at $(t+\Delta t /2)$.
The rest of the equations of motion contain the local fields and their spatial derivatives,
(all defined at time $t$), but they also contain the velocities as it is evident
from (\ref{leqs1rot}). 
We approximate the velocities at $t$ by the average of their values at 
$(t \pm \Delta t /2)$. This leads to an implicit equation for the updated
velocities. 
We solve the implicit equation iteratively to cubic order in $\Delta t$ which exceeds
the order of accuracy in which it was derived.
Finally, the new velocities are used to update the fields.

Use of a flux-conservative form for the equations of motion
might have lead to improved accuracy. However, this way we can use the known conservation
laws, in particular, energy and baryon number conservation, as a check for the 
accuracy of our simulations. 
The long-term stability of the calculation is largely improved by the feedback
mechanism built into our implicit scheme. 
Calculations performed using an explicit scheme
involving a second timeslice shifted with one half timestep give virtually identical
results as far as the time evolution of the fields and energy and baryon number
conservation are concerned. 
Part of the calculations presented below -- those not involving complicated physical 
situations -- have been in fact performed with this earlier version of our code.
When pushed beyond $25 ~{\rm fm/c}$ in time (a few thousand timesteps), 
these calculations typically develop 
instabilities which build up fast to destructive levels.
By contrast, the feedback calculations ran without exception up to 5000 timesteps
or more without becoming unstable. 
This is achieved without adding an explicit numerical viscosity term to our equations.

The calculation based on the algorithm described above was implemented on a three-dimensional
grid of points. The physical size of the box used for the calculations shown in this
paper was $18 \times 10 \times 10$ fermi. We take advantage of the four-fold spatial symmetry
of the problem \cite{CalTech:sc,Ba96}, therefore our mesh covers only one quadrant of
the physical box. 
We used a fixed lattice spacing of $0.1~{\rm fm}$ therefore our lattice
has $91 \times 101 \times 51$ regular points. 
In addition to these, we have a layer of
unequally spaced points on each outside wall of the box simulating an absorbing
boundary.
Thus our full mesh has $101 \times 121 \times 61$ points. 
An indication of the better intrinsic stability of the model Lagrangian employed 
here is the fact that our calculations are fairly stable out to $50~{\rm fm/c}$ in time
with timesteps of $0.1$ to $0.4~{\rm fm/c}$, corresponding to a CFL ratio of 
$0.1$ to $0.4$.
This is larger than in the early works of  Verbaarschot \em et al \em 
\cite{centralsc} ($0.05$) and of the Caltech group \cite{CalTech:sc} ($0.075$ to $0.013$),
and comparable to the more recent calculations of \cite{Ba96}, which employ fourth-order 
spatial differences, in contrast with our second-order spatial difference scheme. 
In the absence of radiation, the total energy is conserved to within $3 \%$ for
typically $20~{\rm fm}$. Due to the emission of radiation which eventually leaves 
the box it is harder to assess the degree of energy conservation for those processes
which involve (quasi)bound states and have to be followed for a longer time. 
We can estimate the amount of energy being radiated
by calculating the energy contained in a sphere of given radius ($2~{\rm fm}$) 
around the origin, large enough to contain most of the field.
The radiated energy first leaves this sphere and then the box. Excluding loss through
radiation identified in this manner, the total energy is conserved to within $4 ~\%$
in all the calculations presented below, including
the long runs ($40\ldots 60~{\rm fm/c}$) involving bound states.
A check for consistency is to reverse the arrow of time at some
point in the calculation, which should lead back close to the initial state.
We performed this check successfully on one case with nontrivial dynamics but 
little radiation.
%After reversing the time at $15~{\rm fm/c}$ and running 
%for another $10~{\rm fm}$, the energy was still within $3 \%$ of the initial value
%and the trajectories were virtually mirrored.

We construct the initial state 
by first numerically solving the field equations in the 
spherically symmetric ansatz. 
This gives us the radial functions for the spherically
symmetric static skyrmion (hedgehog). 
We then place a boosted hedgehog configuration in the simulation box. 
The skyrmion is boosted towards its mirror image implemented via the boundary
conditions at the symmetry walls of the box, which corresponds only to one quadrant
of the physical region being simulated. 
One can dial all the relative groomings discussed below as well as the 
corresponding skyrmion-antiskyrmion configurations by simply choosing the appropriate 
signs in the boundary conditions for the various field components \cite{repel}.
The problem of the overlap of the tails of the two skyrmions 
in the initial state is not easy to solve. 
Instead, we chose the initial configuration with a large separation ($9~{\rm fm}$)
between the two centers, so that the overlap becomes truly negligible.

Our calculations have been performed on clusters of $6-12$ IBM SP-2 computers.
Depending on the number of processors and the timestep, one $20~{\rm fm/c}$ calculation
takes half a day to two days to complete. 

\section{Results}

For skyrmion-skyrmion scattering with finite impact parameter, there
are four relative groomings that are distinct. 
The first is no grooming
at all. This is the hedgehog-hedgehog channel. 
The second grooming
we consider is a relative rotation of $\ppi$ about an axis perpendicular
to the scattering plane (the plane formed by the incident direction and the
impact parameter). 
The third is a relative
grooming by $\ppi$ about the direction of motion. For zero impact parameter
this channel is known to be repulsive. \cite{repel} 
The fourth grooming consists of a relative rotation 
by $\ppi$ about the direction of the impact parameter. In the limit of
zero impact parameter the second and fourth groomings become equivalent,
they then correspond to a rotation of $\ppi$ about an axis normal to the
incident direction. In this zero impact parameter case, this channel
is known to be attractive. \cite{repel} 
All of our scattering studies are done at a relative velocity of
$\beta=v/c=0.5$, which corresponds to a center of mass kinetic
 energy of 230 MeV. In order to keep our study relatively small,
 we have not studied the effect of varying the incident energy. 
As the main means of presentation we chose energy density contour plots.
The baryon density plots are very similar to the energy density plots.

\subsection{Head-on collisions}

Let us begin, for simplicity, with the case of zero impact parameter.
For the hedgehog-hedgehog channel (HH) and the repulsive channel (rotation
by $\ppi$ about the incident direction) symmetry dictates that the
scattering can only be exactly forward (the skyrmions passing through
on another) or exactly backward (the skyrmions bouncing back off each
other). We find that in fact the scattering is backward in both
the HH and the repulsive channels. 

\begin{figure}
\centerline{
\vbox{
\vspace{-2.1cm}
\hbox{
\psfig{file=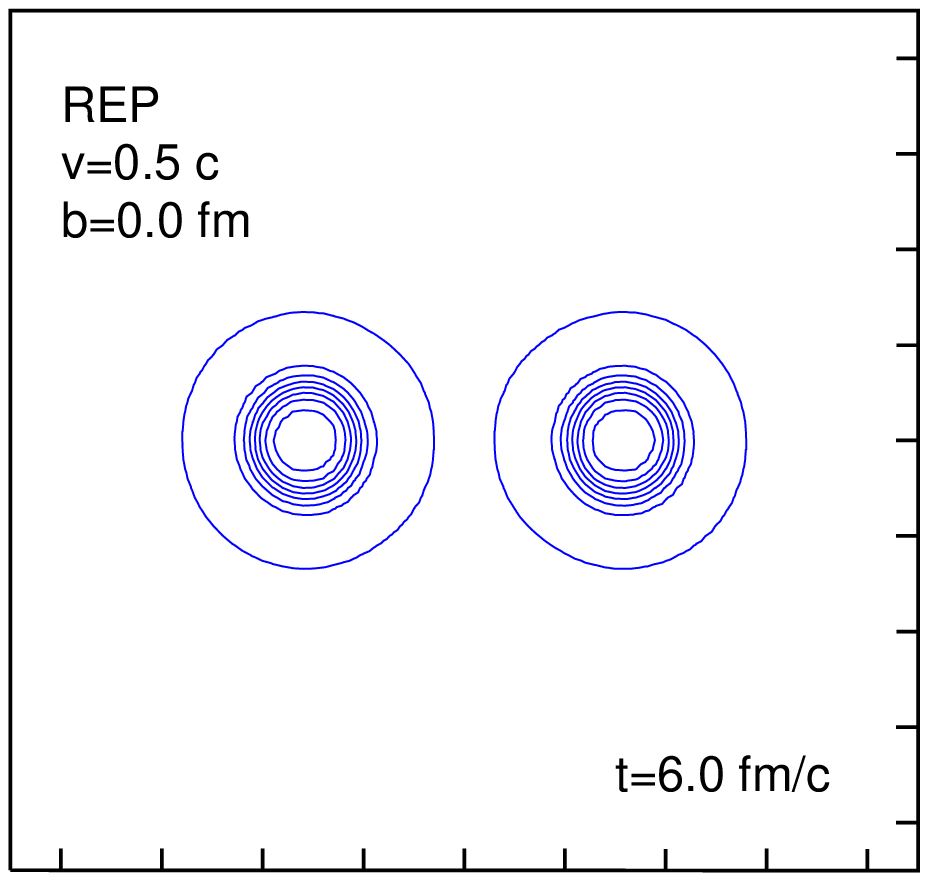,width=8.2cm,angle=0}
\hspace {-4.18cm}
\psfig{file=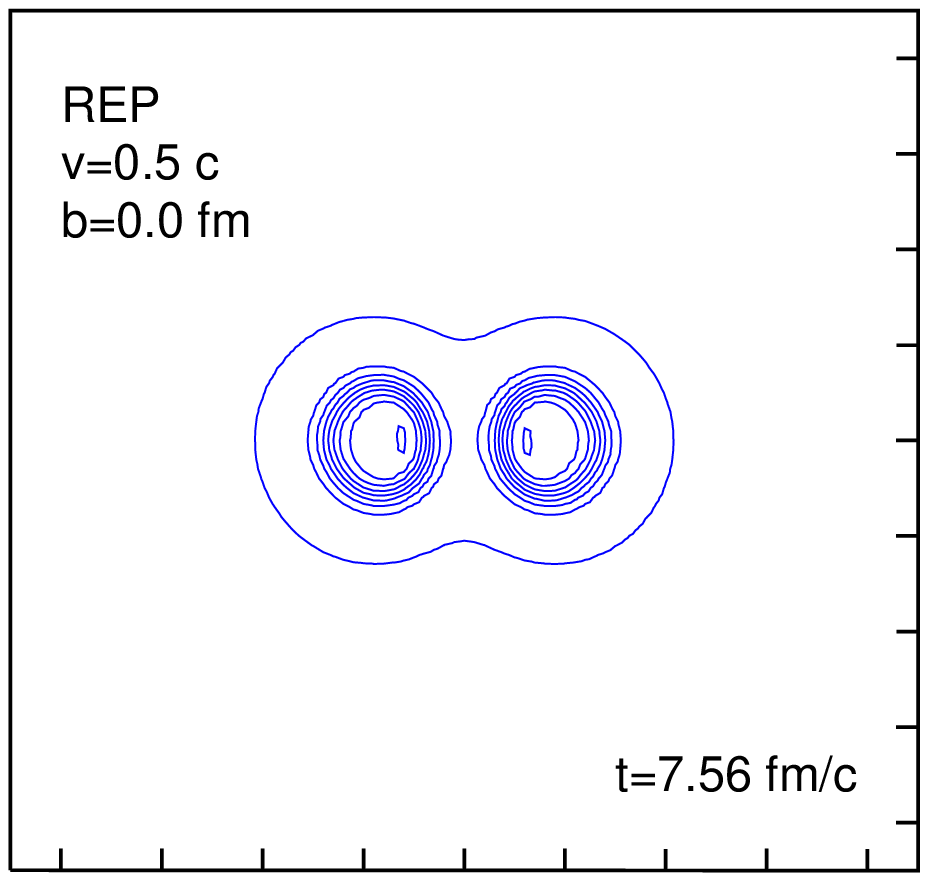,width=8.2cm,angle=0}
\hspace {-4.18cm}
\psfig{file=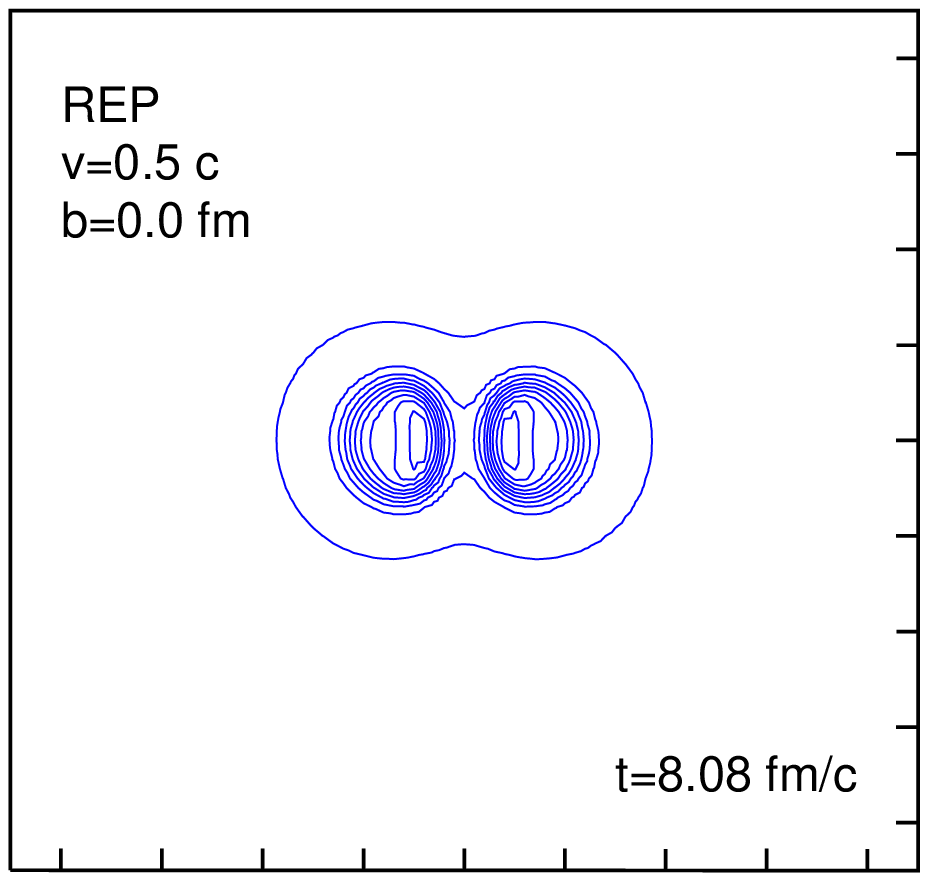,width=8.2cm,angle=0}
\hspace {-4.18cm}
\psfig{file=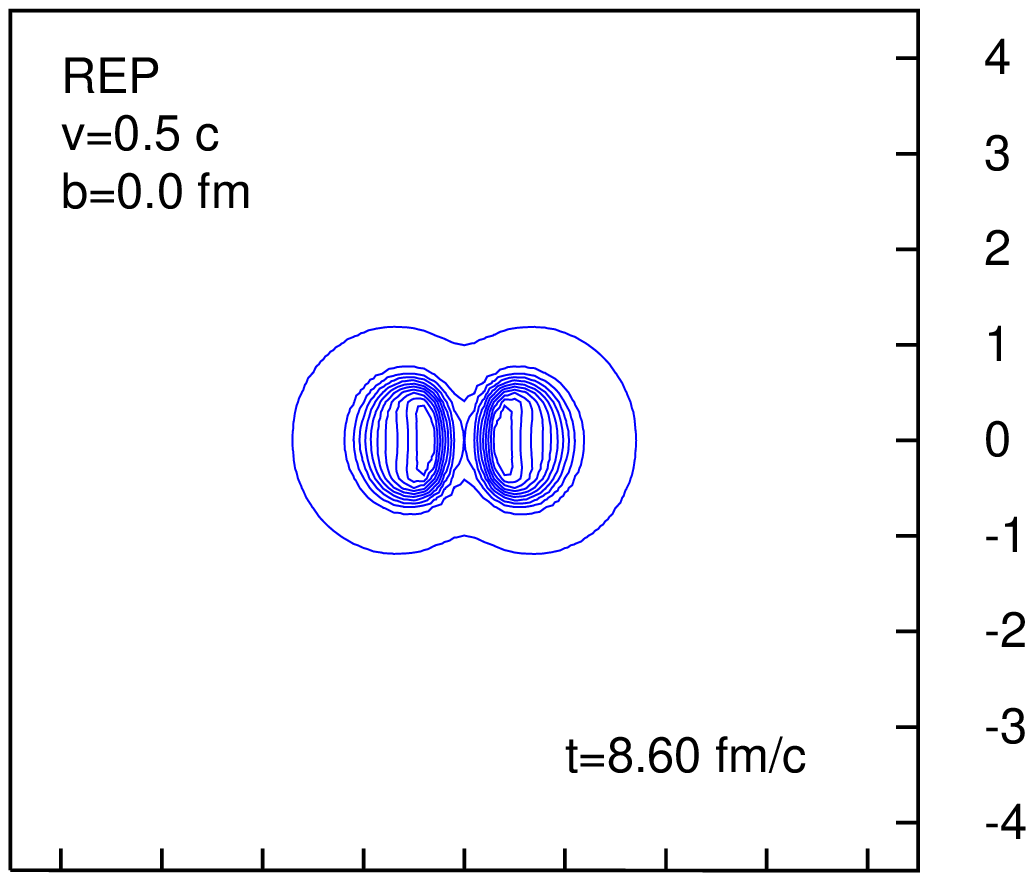,width=8.2cm,angle=0}
}
\vspace{-4.2cm}
\hbox{
\psfig{file=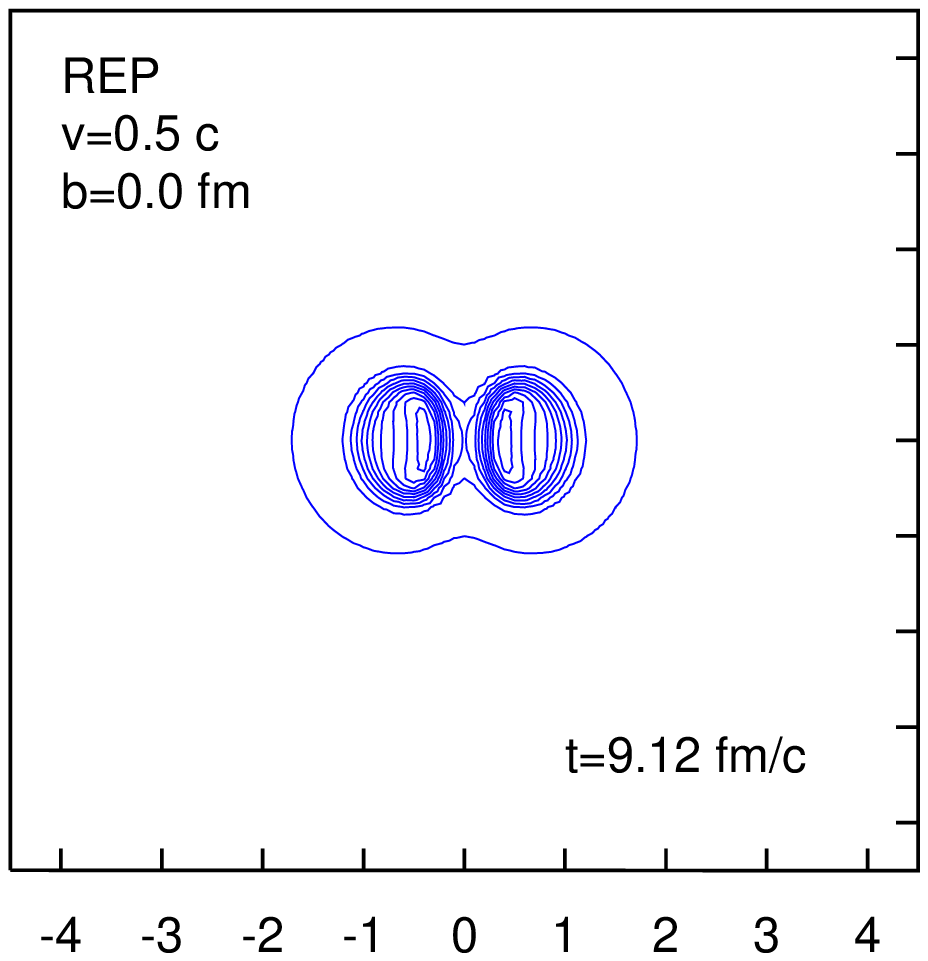,width=8.2cm,angle=0}
\hspace {-4.18cm}
\psfig{file=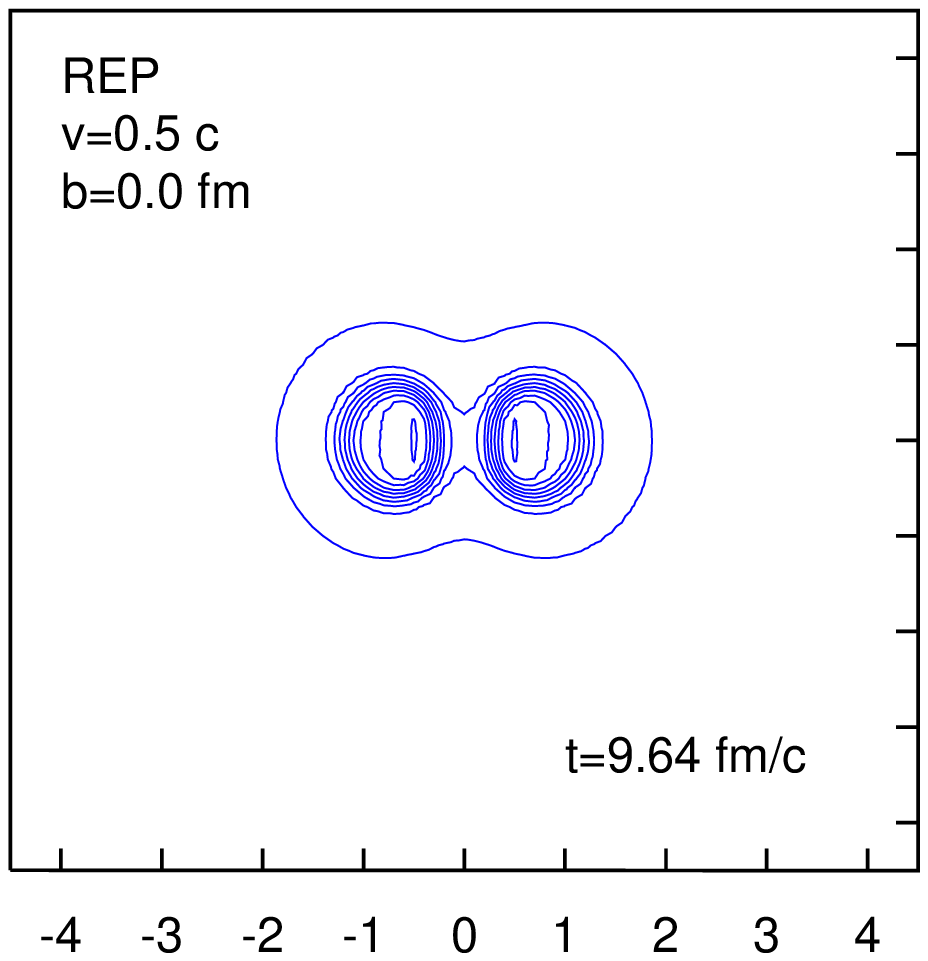,width=8.2cm,angle=0}
\hspace {-4.18cm}
\psfig{file=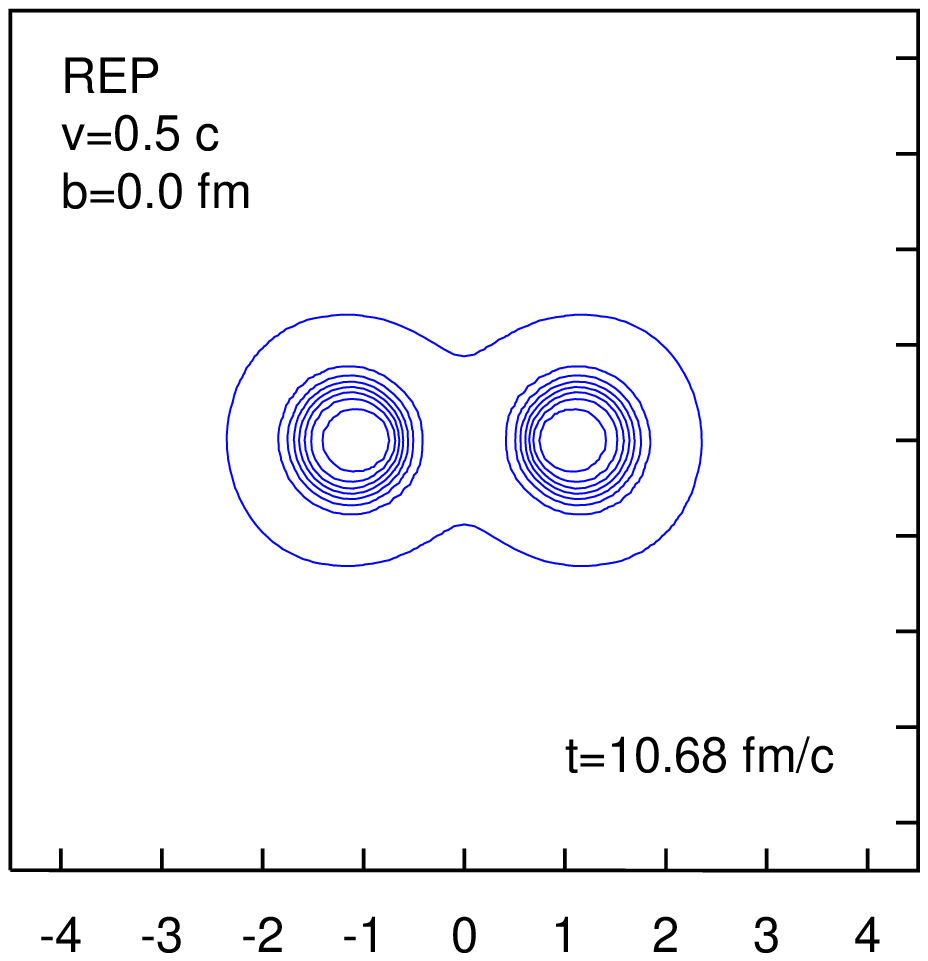,width=8.2cm,angle=0}
\hspace {-4.18cm}
\psfig{file=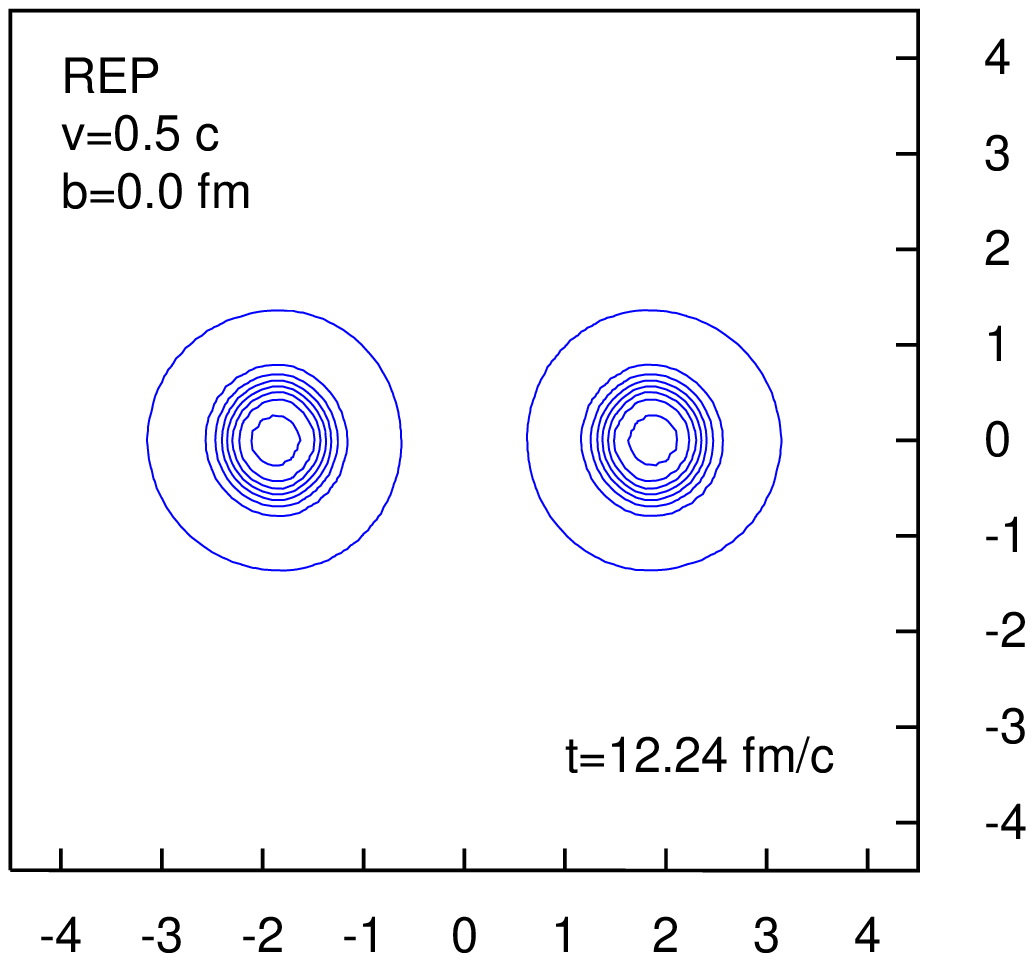,width=8.2cm,angle=0}
}
\vspace{-1.3cm}
}
}
\caption{Contour plots of the energy density in the $xy$ plane
for scattering in the 
repulsive channel with zero impact parameter.
The spacing between the contours is $100~{\rm MeV/fm^3}$.
The first contour is at the $5~{\rm MeV/fm^3}$ level.
Note that the frames are not evenly spaced in time.
Here and in all contour plots, the $x$ axis points to the right
and the $y$ axis points upwards, unless explicitly specified otherwise.
The length on both axes is measured in fermi.
}
\label{REP_b0.0_series}
\end{figure}

We illustrate this type of scattering in Figure \ref{REP_b0.0_series} 
with contour plots of the energy density in the $xy$ plane\footnote{
Throughout the paper we adopt the following convention. The direction of 
the initial motion is $x$, the impact parameter -- if nonzero -- 
points in the $y$ direction and $z$ is the direction perpendicular
to the [initial] plane of motion, $xy$. 
For zero impact parameter the choice of $y$ and $z$ is of course arbitrary.} 
for the repulsive channel.
Unless otherwise specified we keep the same choice for the energy contour
levels for all similar plots that will follow, namely, the first contour 
at $5 {\rm MeV/fm^3}$, and the others equally spaced at $100 {\rm MeV/fm^3}$.
The head-on scattering in the repulsive channel has nothing surprising.
The process is reminiscent of two tennis balls bouncing off each other.
The skyrmions start compressing as soon as they touch. 
They slow down, compress and stop, and then expand and move off in the 
direction opposite to the one they came in along. 
The collision is practically elastic. We were unable to detect any
energy loss through radiation and the velocities of the topological centers
 before and after collision are practically the same.

For the hedgehog-hedgehog channel we again find backward scattering
with an evolution very similar to that shown in Figure \ref{REP_b0.0_series},
hence we show no figure.

\bigskip

In the attractive
channel (rotation by $\ppi$ normal to the incident direction) the
skyrmions scatter at $90^0$ along an axis perpendicular to the plane 
formed by the incident direction and the grooming axis. This right
angle scattering is well known. \cite{rightangle} 
It proceeds through the axially symmetric $B=2$ configuration \cite{torus}.
The skyrmions lose their individual identity in this process.

\begin{figure}
\centerline{
\vbox{
\vspace{-2.1cm}
\hbox{
\psfig{file=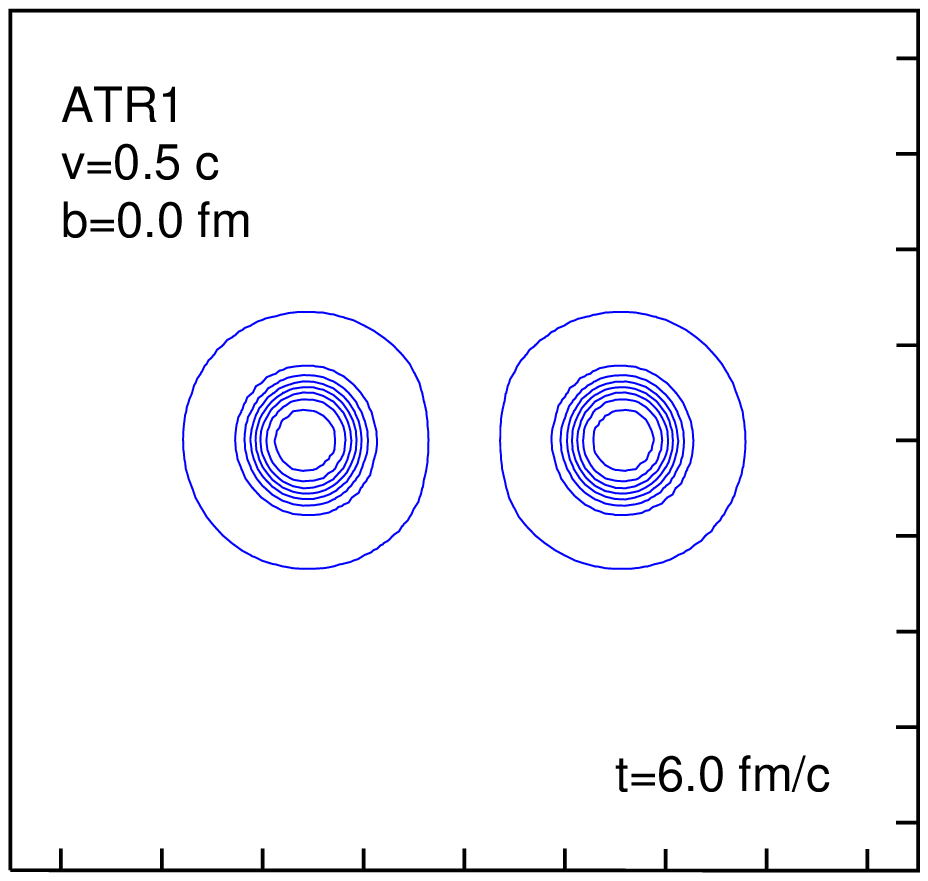,width=8.2cm,angle=0}
\hspace {-4.18cm}
\psfig{file=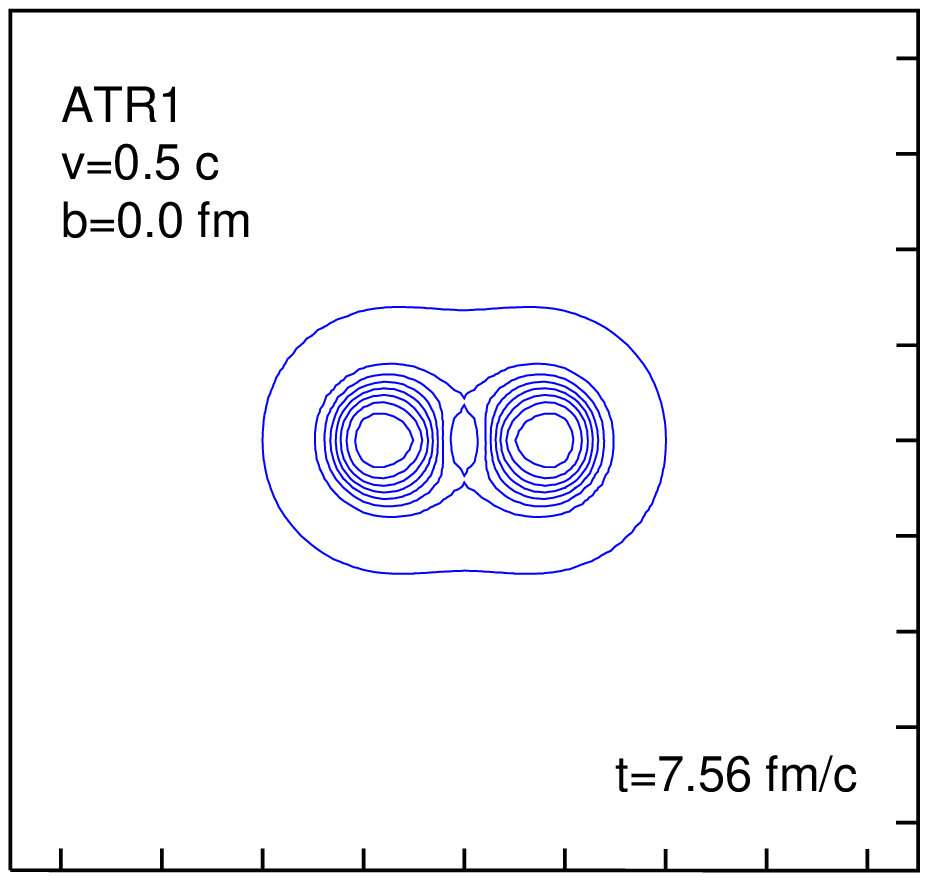,width=8.2cm,angle=0}
\hspace {-4.18cm}
\psfig{file=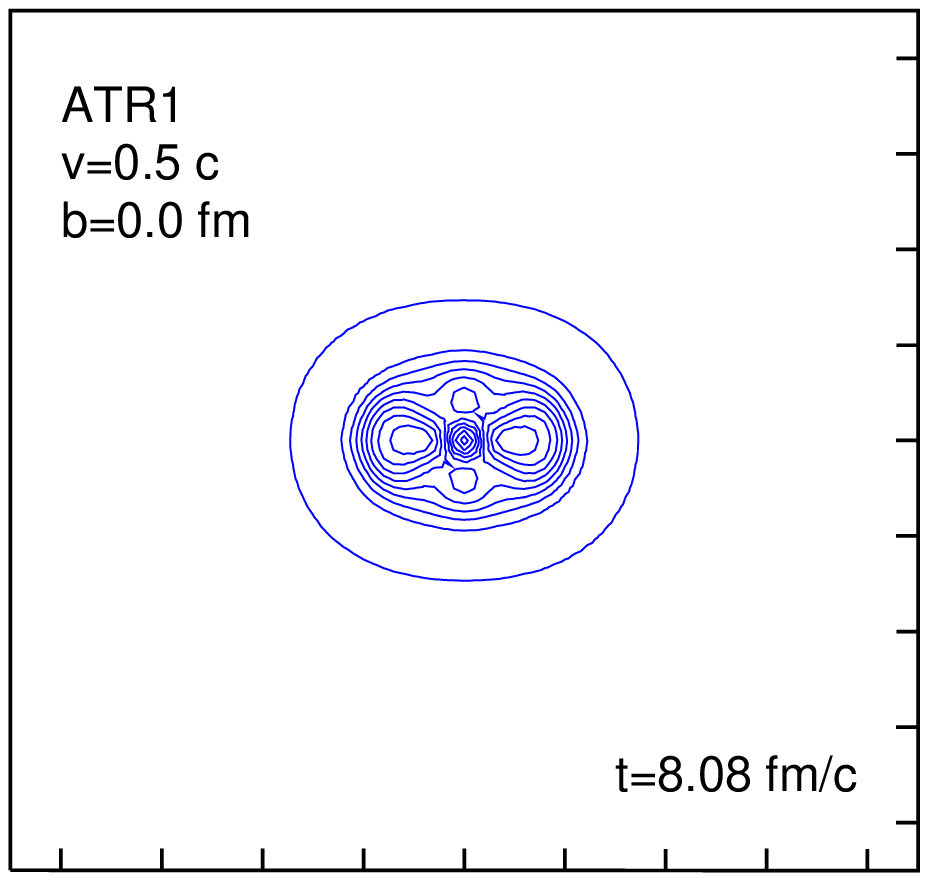,width=8.2cm,angle=0}
\hspace {-4.18cm}
\psfig{file=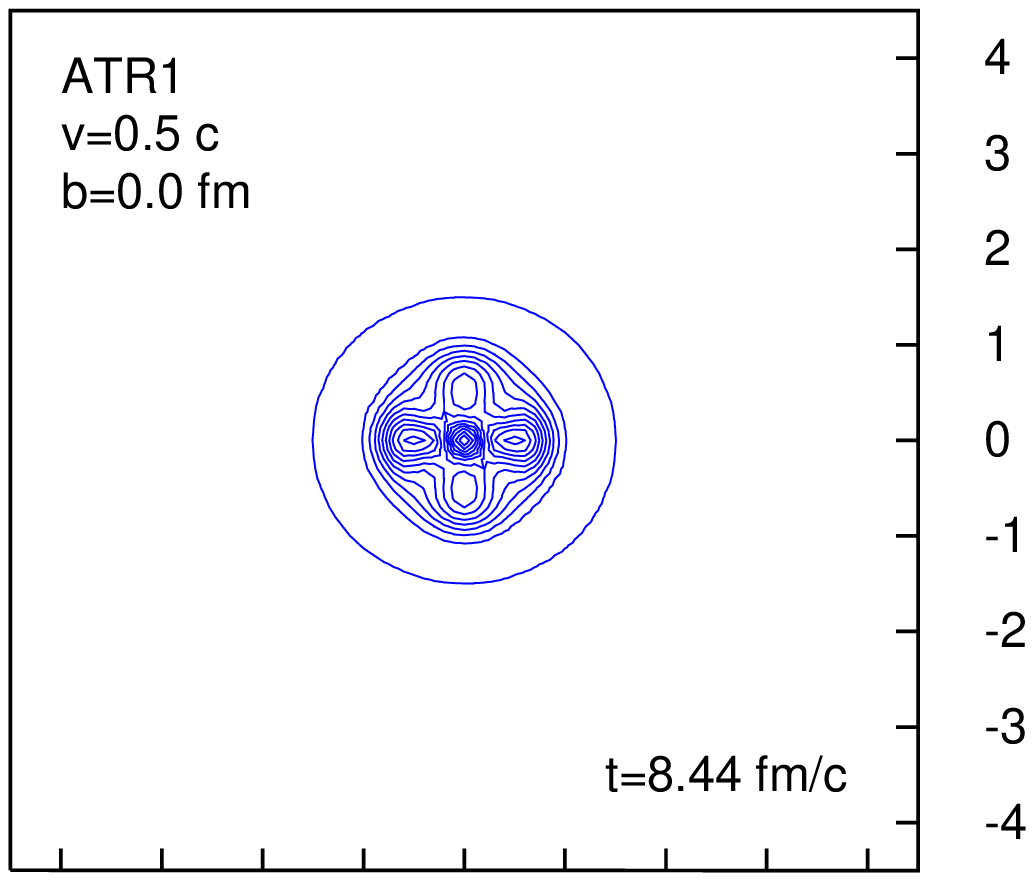,width=8.2cm,angle=0}
}
\vspace{-4.2cm}
\hbox{
\psfig{file=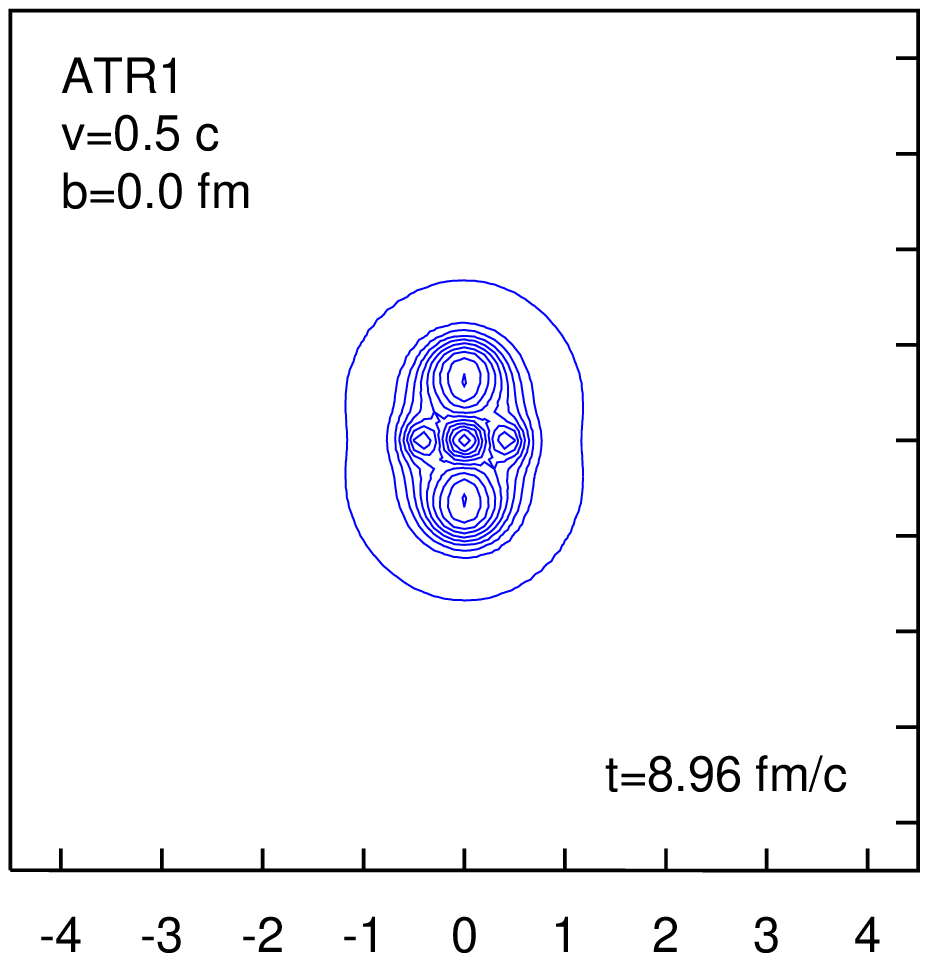,width=8.2cm,angle=0}
\hspace {-4.18cm}
\psfig{file=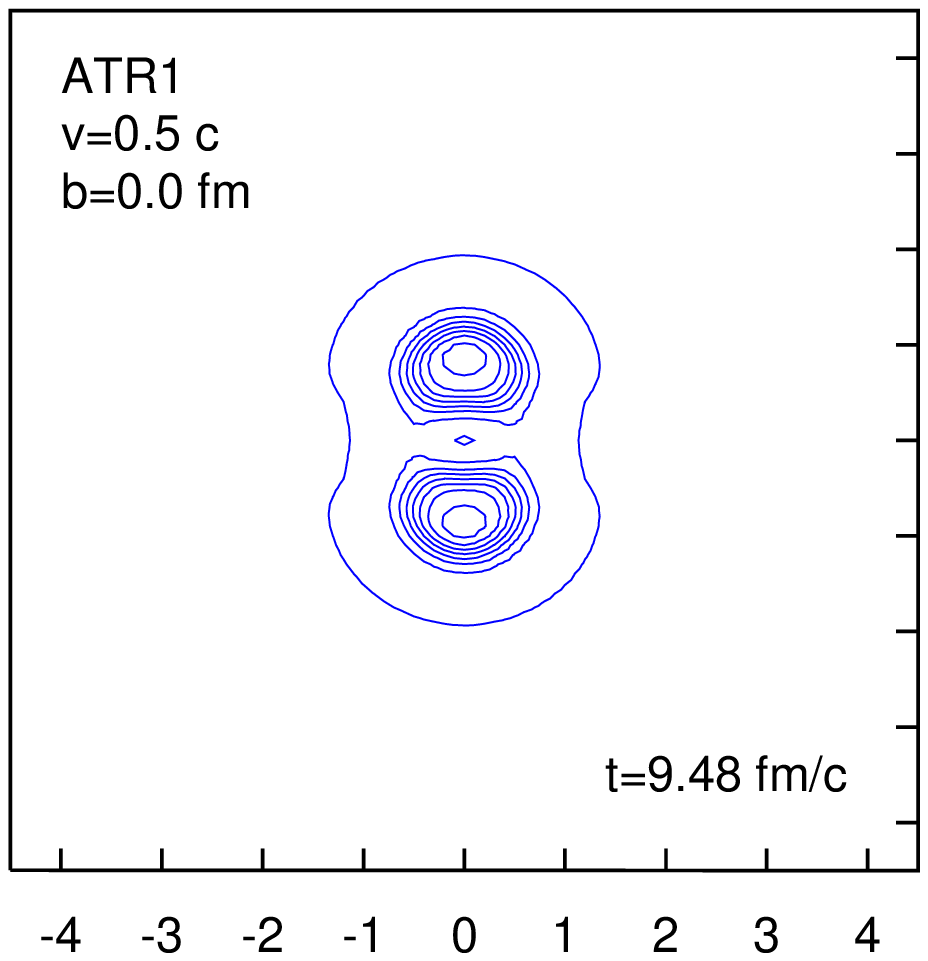,width=8.2cm,angle=0}
\hspace {-4.18cm}
\psfig{file=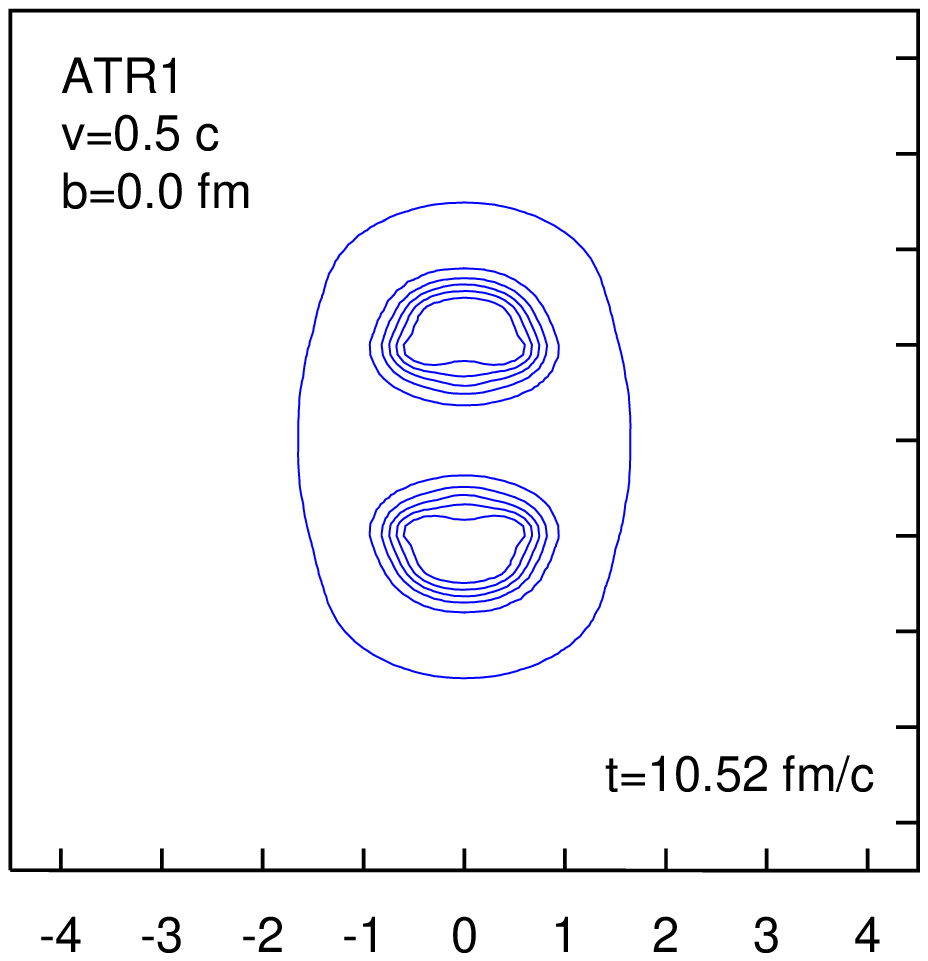,width=8.2cm,angle=0}
\hspace {-4.18cm}
\psfig{file=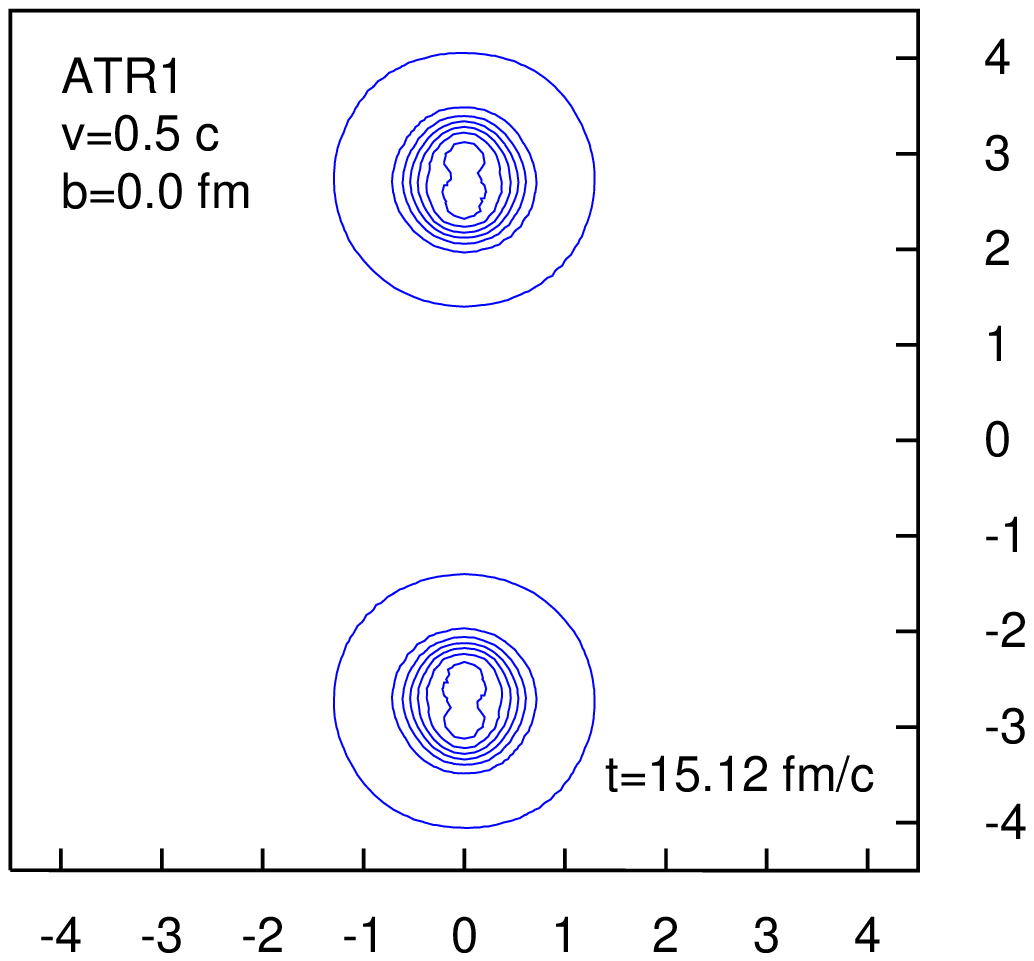,width=8.2cm,angle=0}
}
\vspace{-1.3cm}
}
}
\caption{Contour plots of the energy density in the $xy$ plane
for scattering in the 
attractive (1) channel with zero impact parameter.
The spacing between the contours is $100~{\rm MeV/fm^3}$.
The first contour is at the $5~{\rm MeV/fm^3}$ level.
Note that the frames are not evenly spaced in time.}
\label{ATR1_b0.0_series}
\end{figure}

\begin{figure}
\centerline{
\vbox{
\vspace{-1.1cm}
\hbox{
\hspace{0.0cm}
\psfig{file=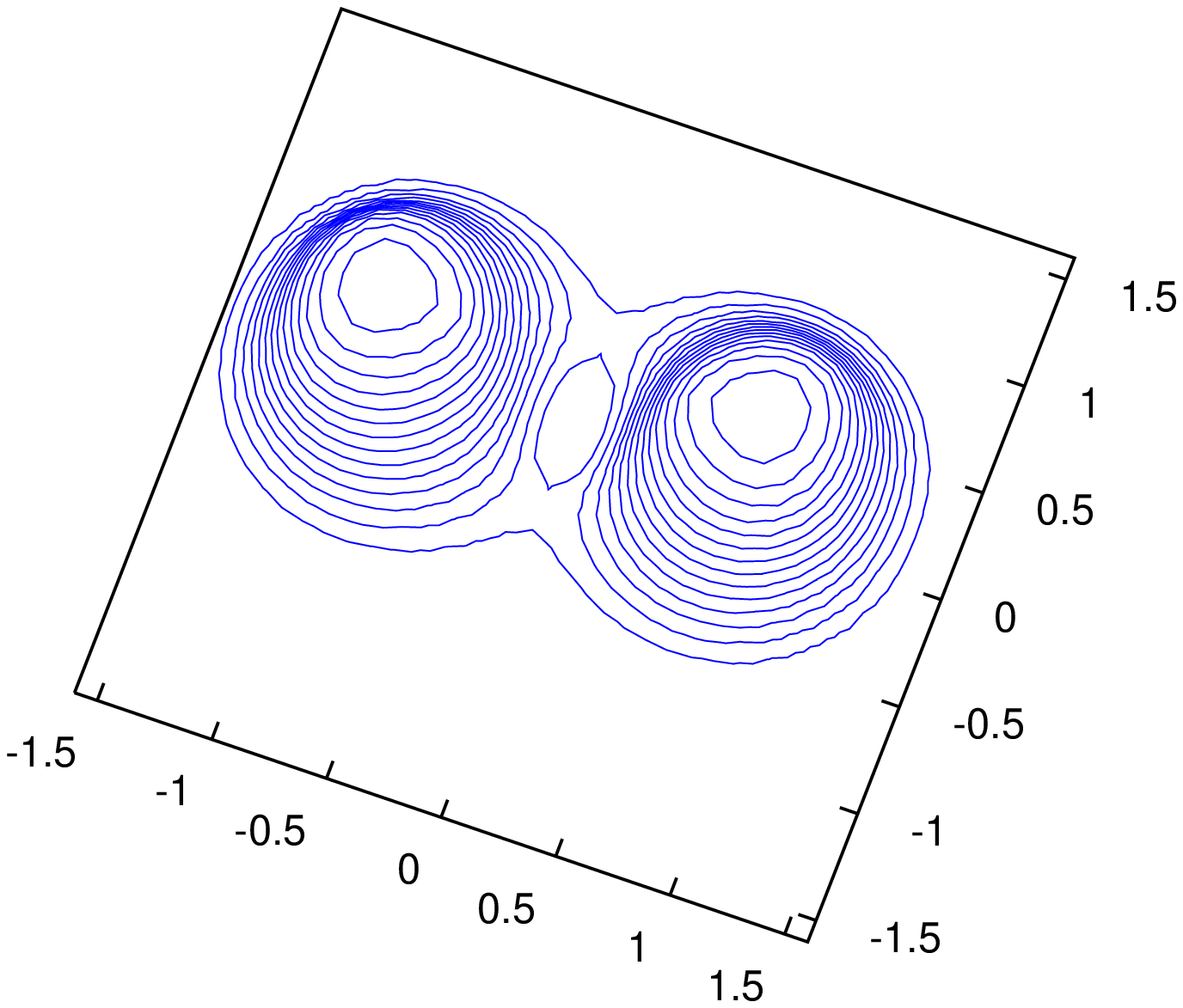,width=5cm,angle=0}
\hspace {-1.8cm}
\psfig{file=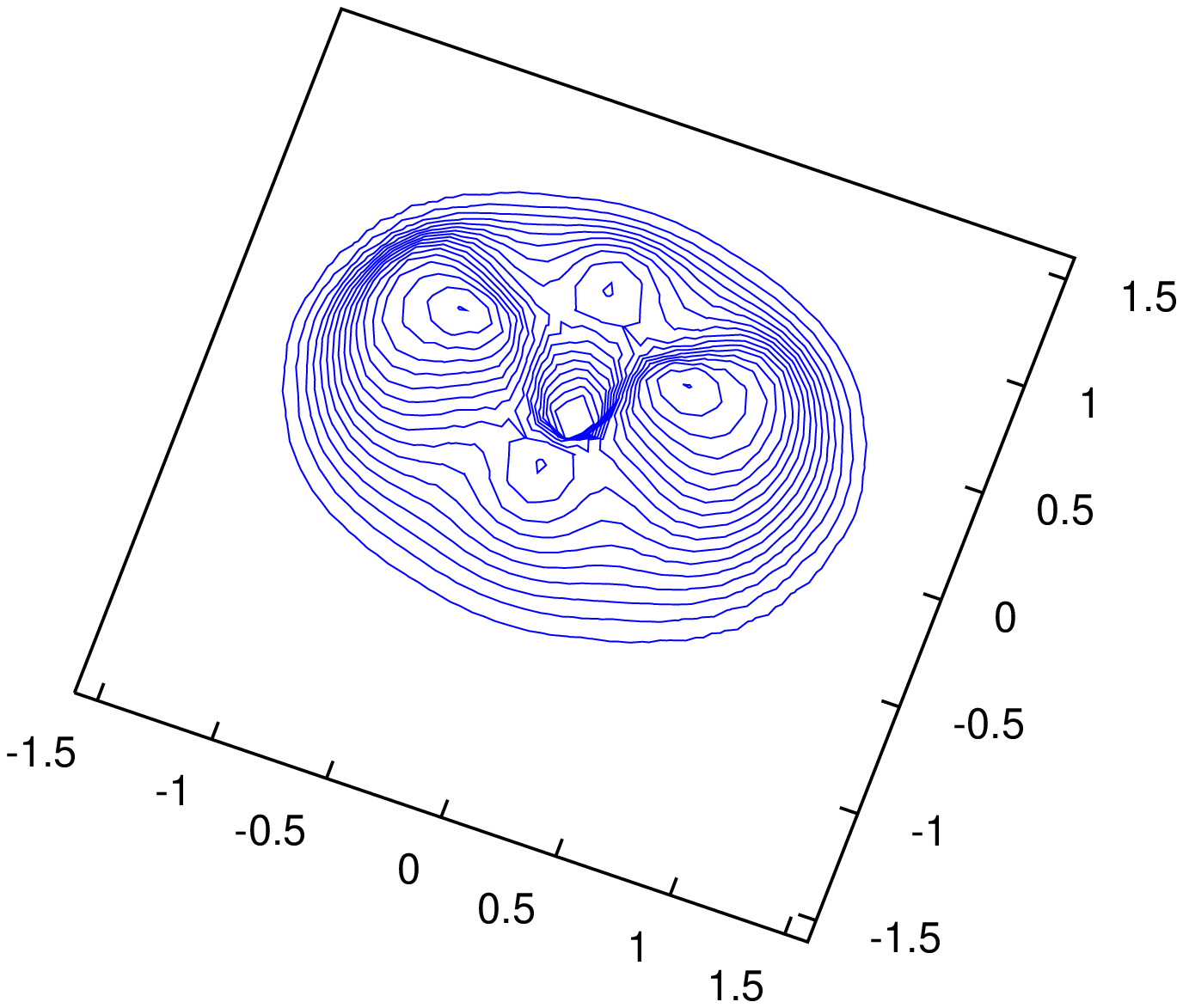,width=5cm,angle=0}
\hspace {-1.8cm}
\psfig{file=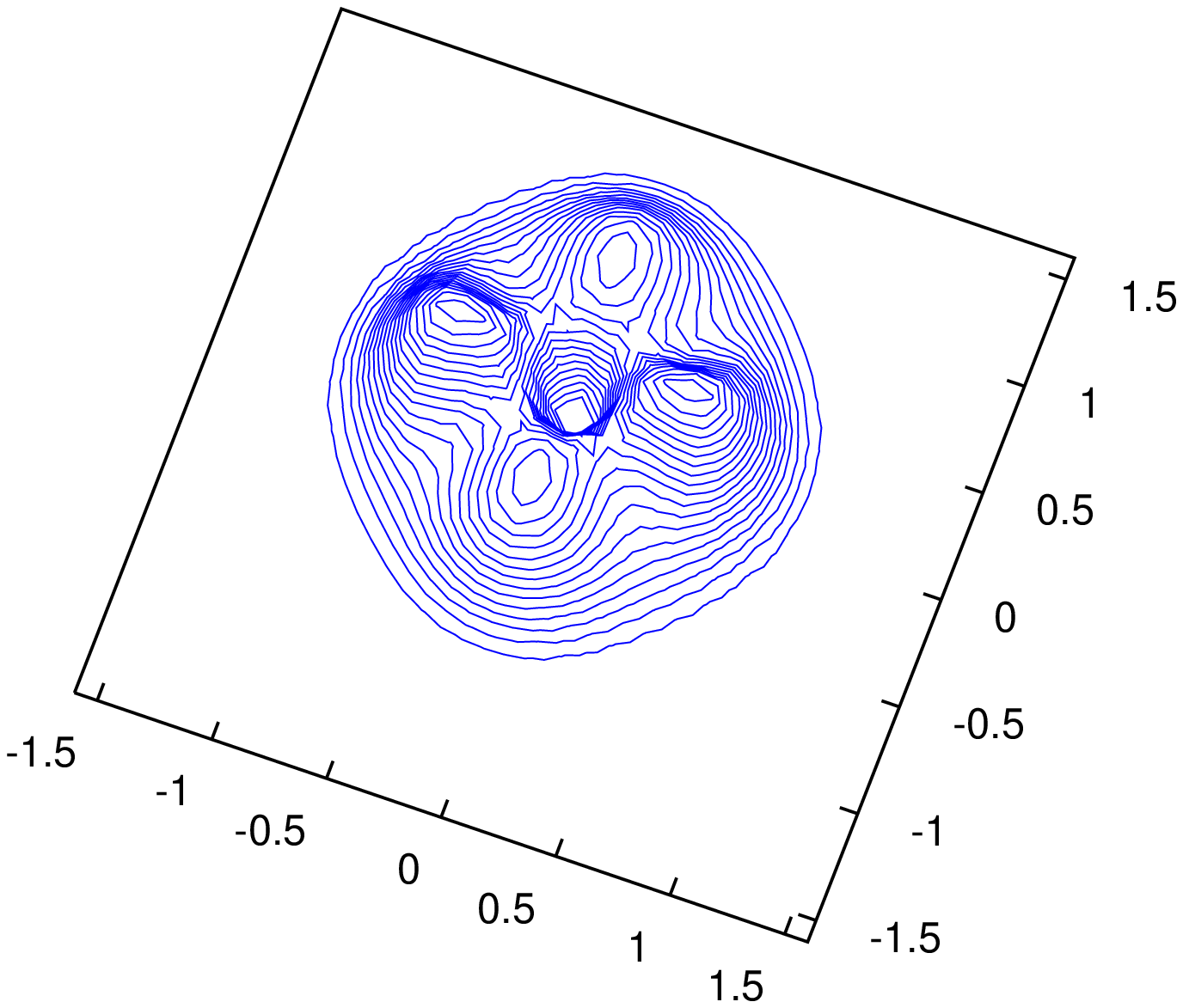,width=5cm,angle=0}
\hspace{-1.8cm}
\psfig{file=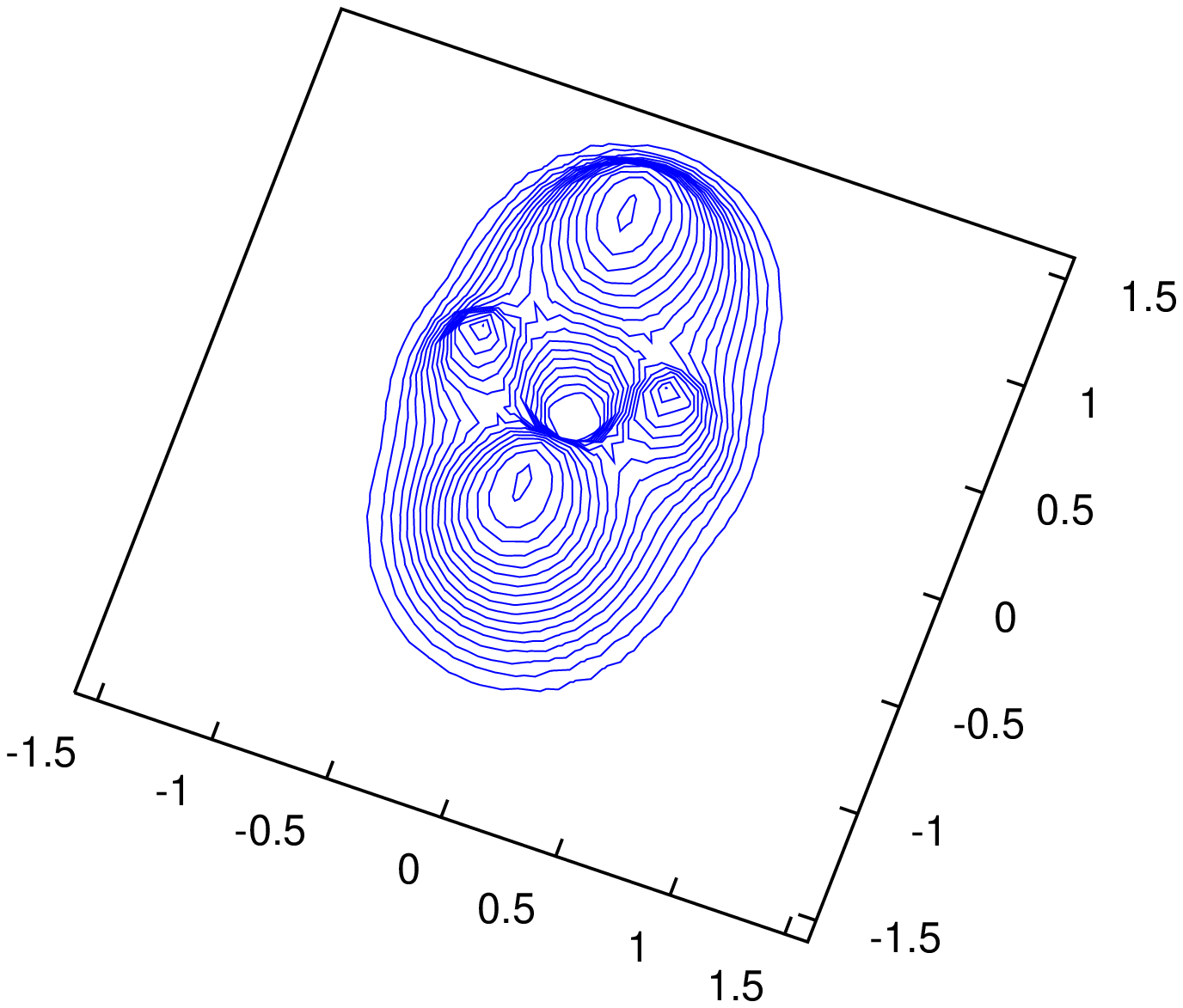,width=5cm,angle=0}
\hspace{-1.8cm}
\psfig{file=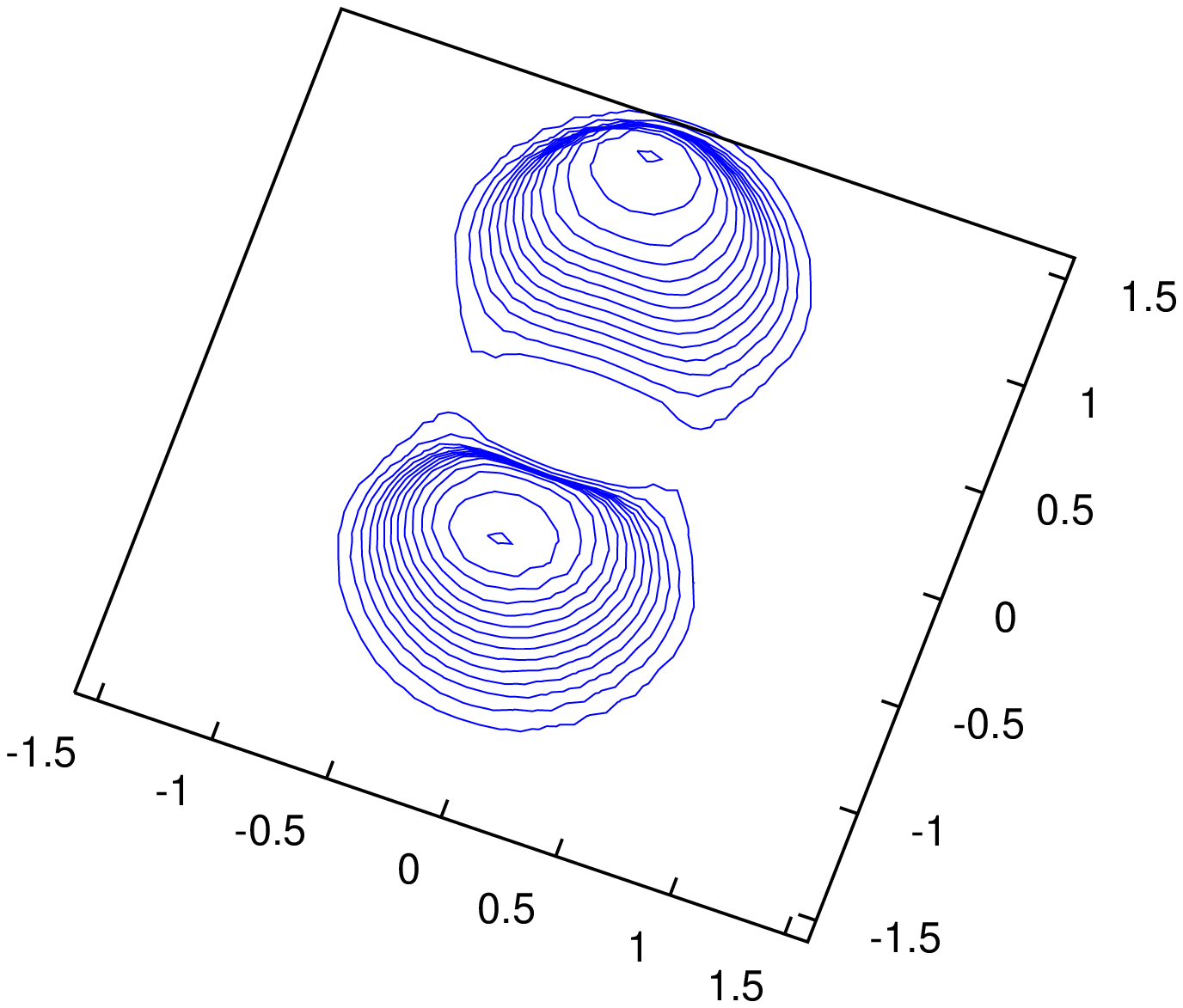,width=5cm,angle=0}
}
}
}
\caption{Contour plots of the energy density in the $xy$ plane
for scattering in the 
attractive (1) channel with zero impact parameter.
These are the same plots as frames 2 through 6 of Figure 
\ref{ATR1_b0.0_series}, at times $t=7.56,~8.08,~8.44,~8.96,~9.48~{\rm fm/c}$.
The spacing between the contours is $50~{\rm MeV/fm^3}$.
The first contour is at the $100~{\rm MeV/fm^3}$ level.
The distances are measured in $\rm fm$.
}
\label{ATR1_b0.0_seriesA}
\end{figure}

In Figure \ref{ATR1_b0.0_series} we show the energy contours for
head-on collision in the attractive channel. 
At the midpoint of this process, shown in
 in the fourth frame of Figure \ref{ATR1_b0.0_series},
one can clearly see the torus-shaped configuration.
It is situated in the plane defined by the incoming
and outgoing directions and perpendicular to the grooming axis.
The bulk of the energy density avoids the center of the doughnut as it 
shifts from the incoming direction to the one perpendicular to it.
To illustrate this point better we plot 
the middle frames of 
Figure \ref{ATR1_b0.0_series} 
in perspective in
Figure \ref{ATR1_b0.0_seriesA}.\footnote{ 
We remind the reader that we plot level contours of the total energy density
in the median plane of our three-dimensional system, as opposed to three-dimensional
surfaces of constant energy.}

\begin{figure}
\centerline{
\psfig{file=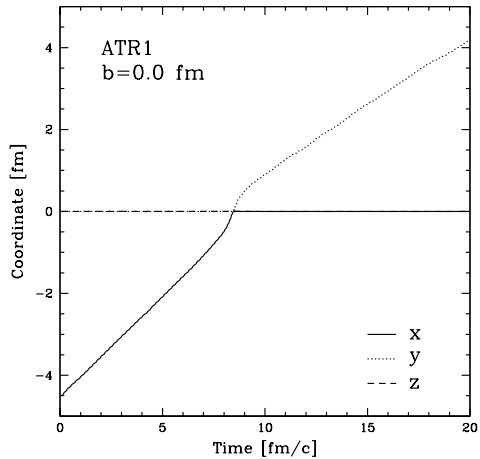,width=6.5cm,angle=0}
}
\caption{Time evolution of the three coordinates of the topological center
of one of the skyrmions, for the head-on collision in the attractive (1)
grooming.}
\label{ATR1_b0.0_timepath}
\end{figure}

In Figure \ref{ATR1_b0.0_timepath} we plot the coordinates of the 
center of one skyrmion (defined as the point where the pion field
amplitude is exactly $\ppi$) versus. time. 
The only nonzero coordinate is $x$, initially. Then, after the
right-angle scattering, $y$ is the only non-zero coordinate.
Straight lines indicate uniform motion. That is the case both before and
after the collision. The slopes of the $y$ line is noticeably smaller
than that of the $x$ line before collision, in other words, the outgoing
velocity is slightly smaller. 
This is due to a genuine physical process, radiation, 
rather than to a numerical artifact, because there
is no decrease in the total energy of the system.
In some of the processes we discuss
below, there are stronger examples of slowing down, accompanied by 
detectable radiation. 

As we have seen above, the scattering direction is determined by the incident
direction and the grooming direction. When the grooming direction is normal
to the incident direction, a torus is formed in the plane normal to the grooming 
direction. In the presence of a non-zero impact parameter, it matters whether
the grooming direction is parallel or normal to that impact parameter.
In both cases there is a tendency to the formation of a torus normal to the
grooming direction, but the ultimate evolution of the scattering is different
in the two cases. We will refer to the case where the grooming is normal to the
impact parameter as attractive (1) and attractive (2) when they are parallel.

\bigskip

Let us now look at scattering in each grooming as a function of
impact parameter. In each of the four cases we study impact parameters
of $0.4~{\rm fm}$, $0.8~{\rm fm}$, $1.6~{\rm fm}$, and $2.8~{\rm fm}$. 
In all cases we find that the scattering
at $2.8~{\rm fm}$ is ``routine''  and hence we do not go to larger impact parameter.

\subsection{Simple scattering: the hedgehog-hedgehog and attractive (1) channels}

\begin{figure}[t]
\centering
\epsfig{file=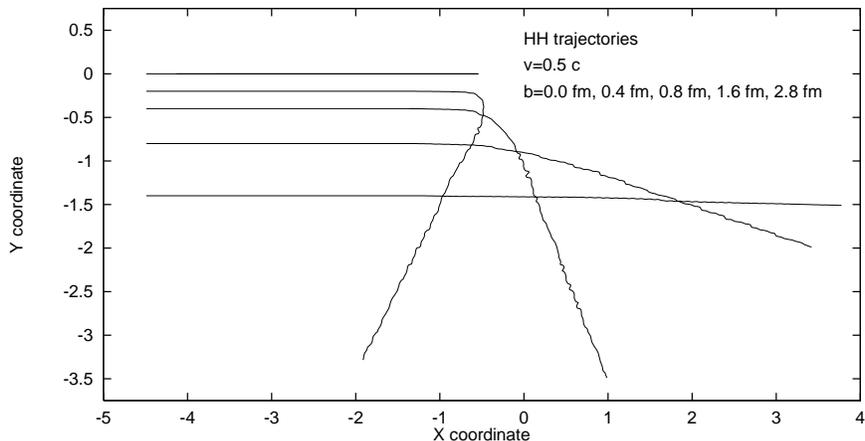,bbllx=80,bblly=50,bburx=750,bbury=410,width=11cm}
\caption{
Paths of the topological centers of one of the colliding skyrmions
in the $xy$ plane in the hedgehog-hedgehog (no grooming) case.
The $x$ axis points to the right and the $y$ axis points upwards.
The length on both axes is measured in fermi.
}
\label{HHpath}
\end{figure}

We begin with two groomings that lead to 
simple dynamics.
Consider first the hedgehog-hedgehog channel.
In Figure \ref{HHpath} we plot the trajectories of the topological  
center of one of the skyrmions for zero impact parameter 
and for each of the four non-zero impact parameters. 
Here and in the other similar plots, 
we define the center as the point where the norm of
the pion field reaches the value of $|\pivec|=\ppi$.
This is a good indicator of the global movement of a skyrmion, especially
when the two colliding objects are somewhat separated.\footnote{The 
topological center does not necessarily coincide with the center
of mass. While the latter is insensitive to internal oscillations
of the skyrmion, the topological center oscillates.}
We cut off the trajectories in Figure \ref{HHpath} 
after $t=17 ~{\rm fm/c}$.
We see normal scattering trajectories corresponding 
to a repulsive interaction. This channel is indeed known to be mildly 
repulsive. For the smallest impact parameter, $b=0.4~{\rm fm}$, 
the scattering is in the
backward direction (recall that for $b=0$ the scattering angle is
$180^o$) and gradually turns forward for increasing impact parameter.
This is exactly what one might expect, since this channel is the 
most similar to point-particle scattering. 
The interaction between the skyrmions is central in the HH channel, 
because it is independent of the direction of the relative position vectors
of the topological centers.

\begin{figure}
\centerline{
\vbox{
\vspace{-2.1cm}
\hbox{
\psfig{file=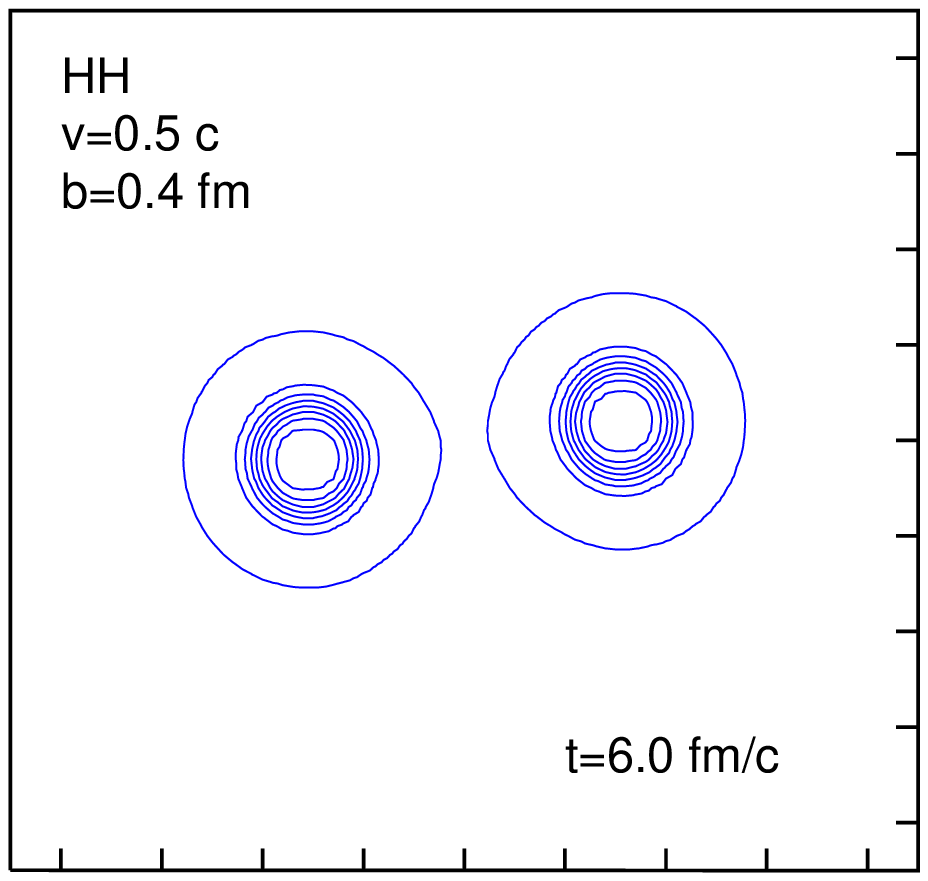,width=8.2cm,angle=0}
\hspace {-4.18cm}
\psfig{file=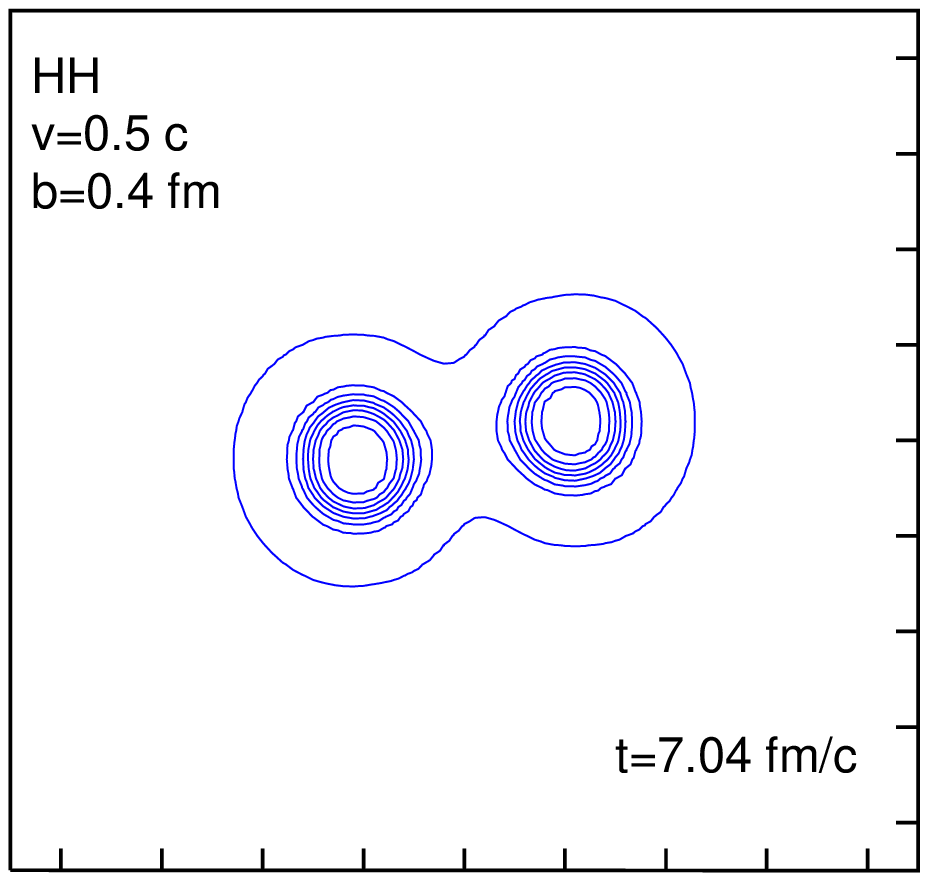,width=8.2cm,angle=0}
\hspace {-4.18cm}
\psfig{file=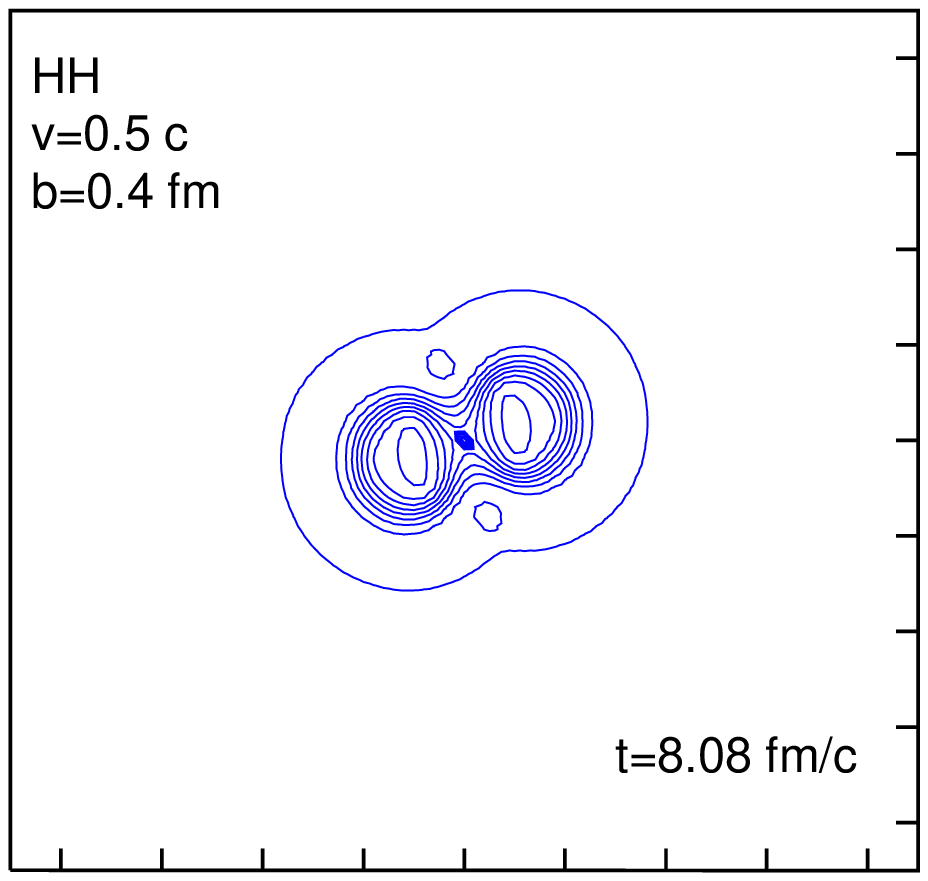,width=8.2cm,angle=0}
\hspace {-4.18cm}
\psfig{file=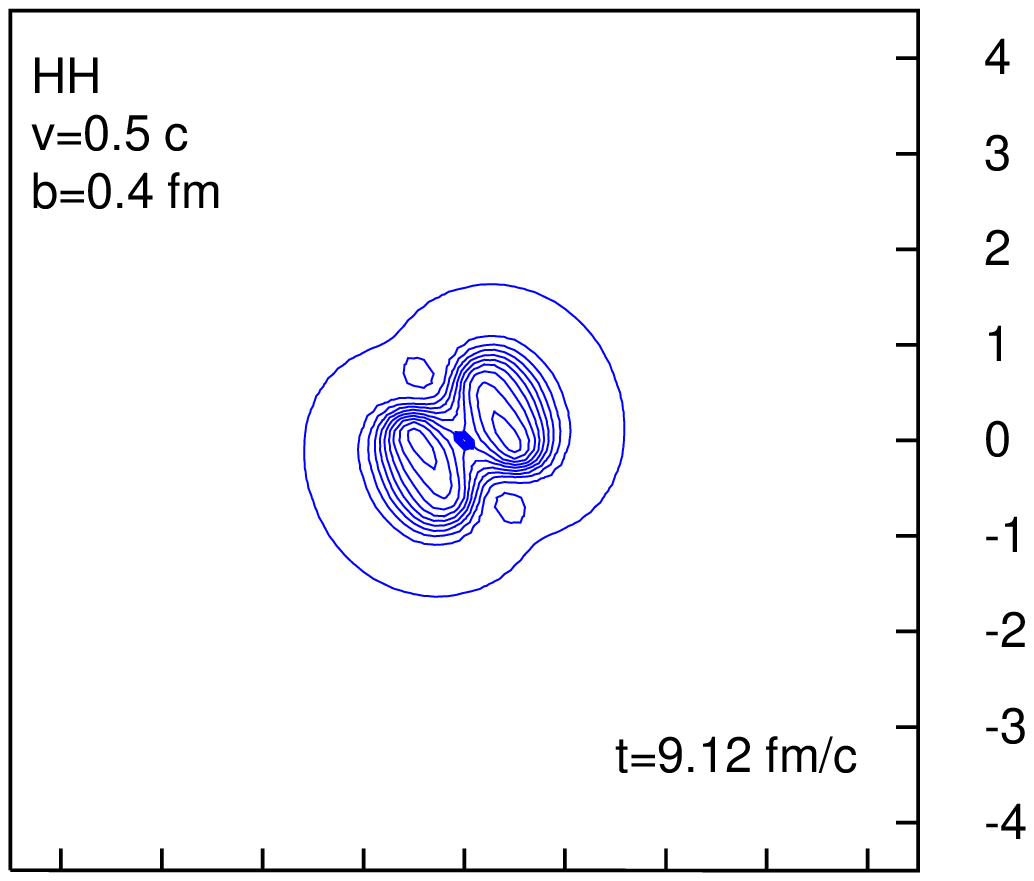,width=8.2cm,angle=0}
} 
\vspace{-4.2cm}
\hbox{
\psfig{file=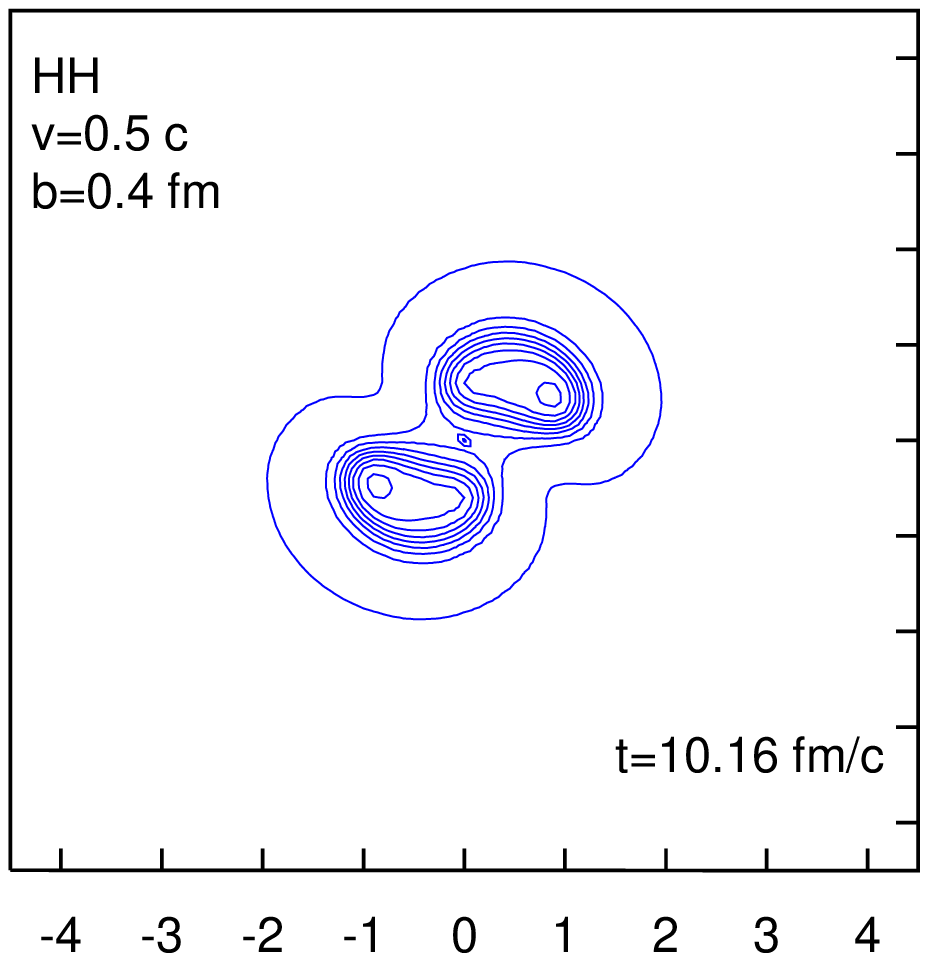,width=8.2cm,angle=0}
\hspace {-4.18cm}
\psfig{file=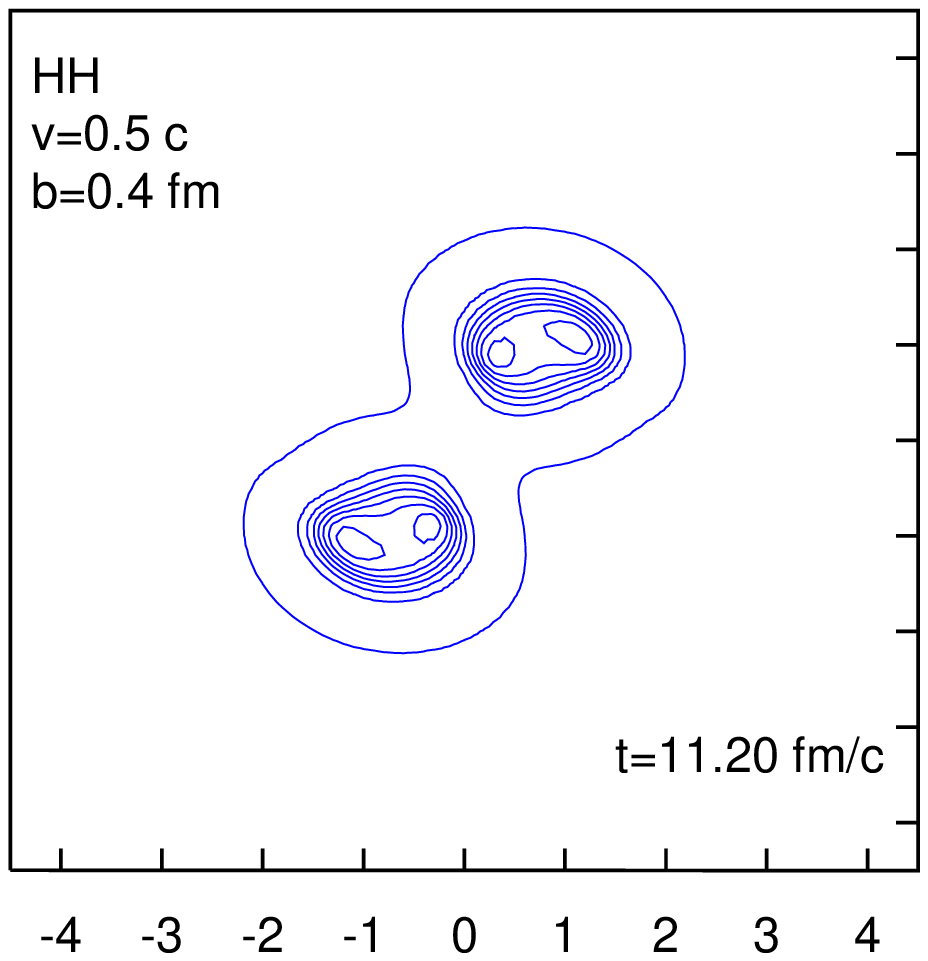,width=8.2cm,angle=0}
\hspace {-4.18cm}
\psfig{file=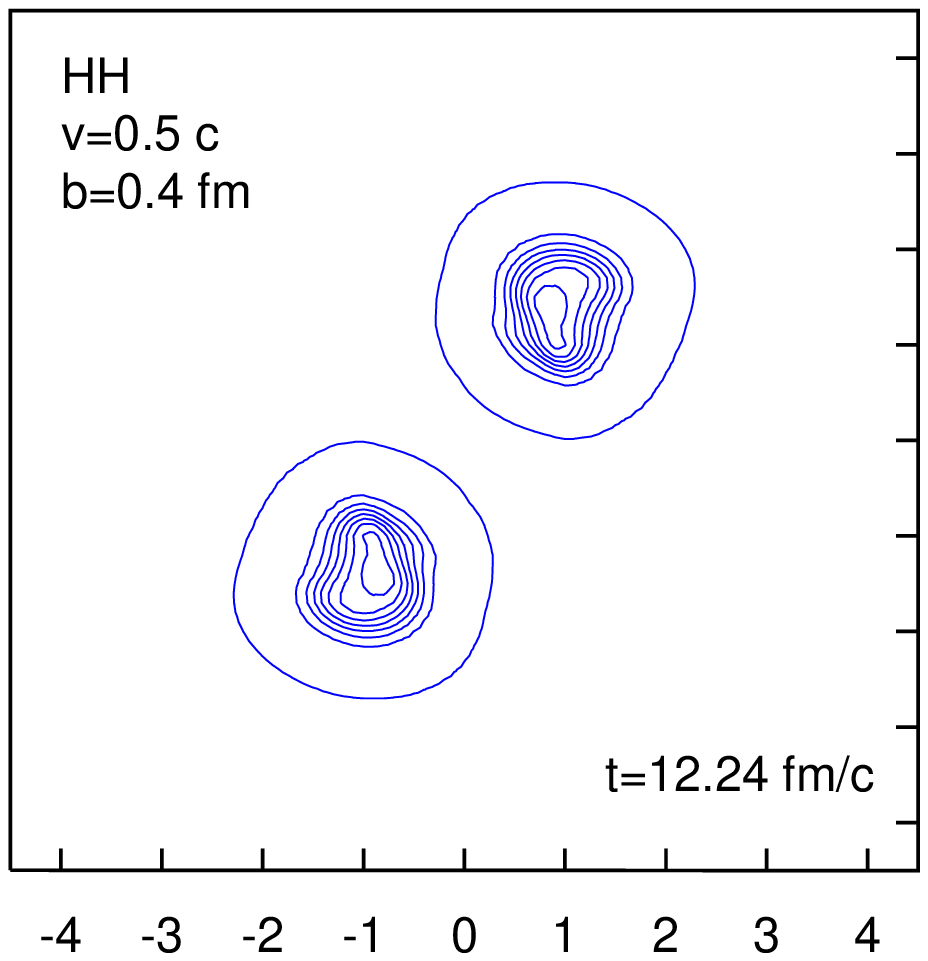,width=8.2cm,angle=0}
\hspace {-4.18cm}
\psfig{file=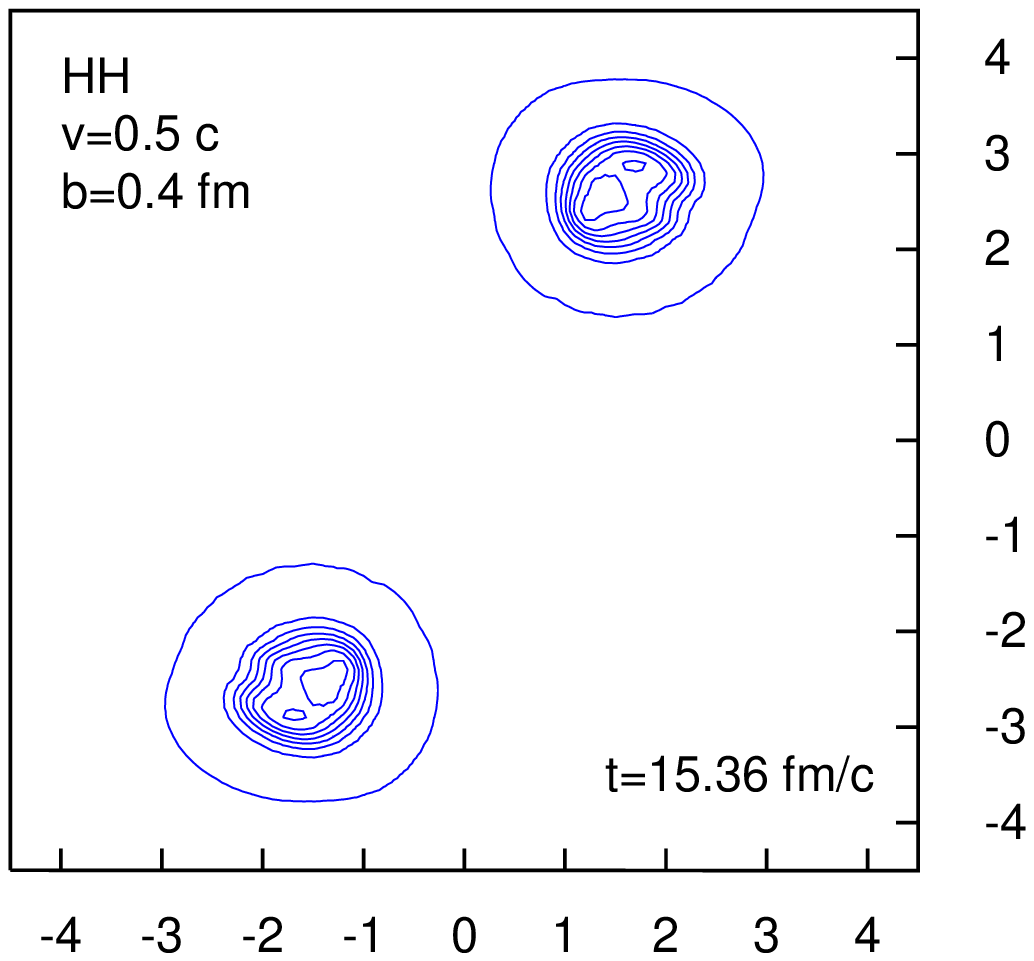,width=8.2cm,angle=0}
}
\vspace{-1.3cm}
}
}
\caption{Contour plots of the energy density in the $xy$ plane
for scattering in the hedgehog-hedgehog channel with impact parameter $b=0.4~{\rm fm}$.
The spacing between the contours is $100~{\rm MeV/fm^3}$.
The first contour is at the $5~{\rm MeV/fm^3}$ level.
The length on both axes is measured in fermi.
Note that the frames are not evenly spaced in time.}
\label{HH_b0.2_series}
\end{figure}

The large-angle scattering
for small impact parameters probes the interaction of the soft core
of the skyrmions. 
We illustrate this in Figure \ref{HH_b0.2_series} 
with energy contour plots from 
the $b=0.4~{\rm fm}$ case. Just like in the head-on case, the skyrmions
compress as they touch. They slow down and then accelerate and proceed in the
outgoing direction. 
Internal oscillations of the skyrmions \cite{centralsc} can be observed
after the collision, therefore this process is not entirely 
elastic. 

The other channel that exhibits only simple scattering is the
attractive (1) grooming -- where the grooming direction is perpendicular
to the plane of motion.

\begin{figure}[t]
\centering
\epsfig{file=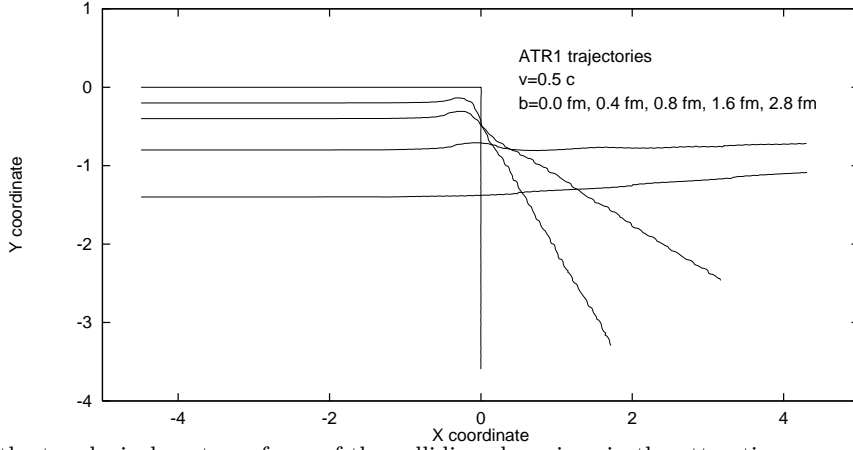,bbllx=80,bblly=50,bburx=750,bbury=410,width=11cm}
\caption{
Paths of the topological centers of one of the colliding skyrmions
in the attractive case with grooming around the direction perpendicular 
to the plane of initial movement.
The $x$ axis points to the right and the $y$ axis points upwards.
The length on both axes is measured in fermi.
}
\label{ATR1path}
\end{figure}

We have studied the same five impact parameters here. 
In Figure \ref{ATR1path} we show the trajectory of the skyrmion centers
for each of those impact parameters. 
For $b=0$ and this
grooming, the scattering angle is $90^o$. For large impact parameters
the scattering is nearly forward.  
Hence in between, we should
see something in between, which is precisely what the figure shows.
Note that in this channel, in contrast to the HH channel, the
trajectories begin their scattering by curving toward the
scattering center, as one would expect for an attractive channel. 
The attraction is seen to act practically unperturbed in 
the scattering with $b=2.8~{\rm fm}$.

\begin{figure}
\centerline{
\vbox{
\vspace{-2.1cm}
\hbox{
\psfig{file=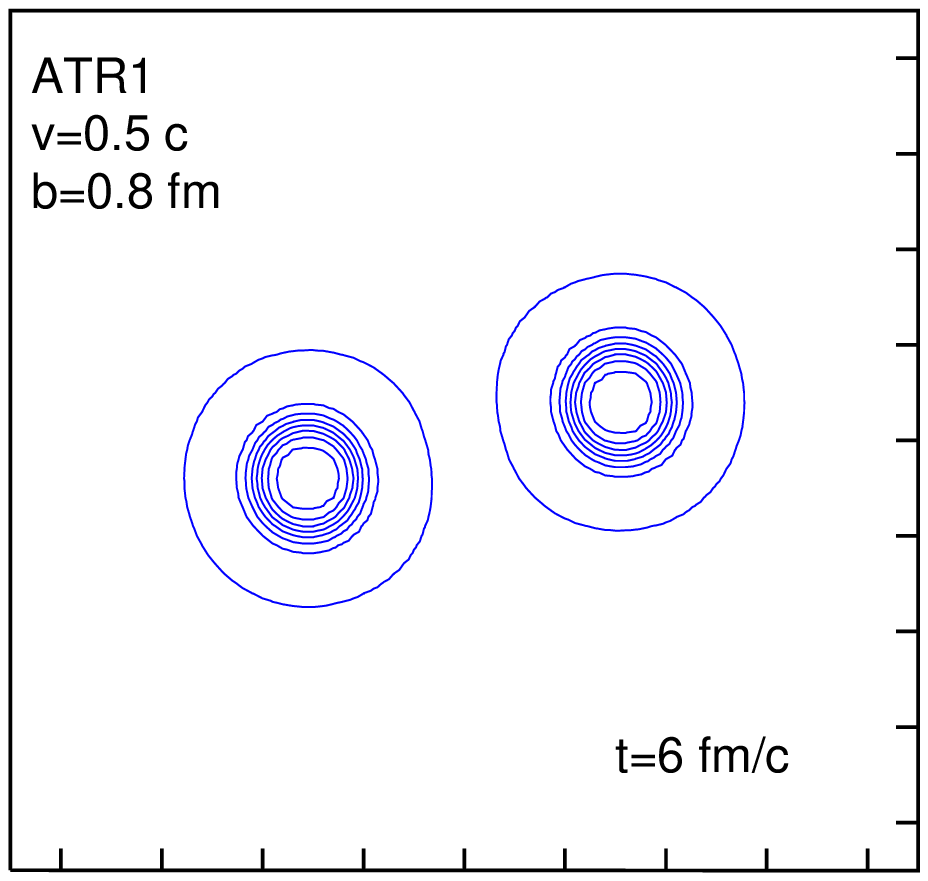,width=8.2cm,angle=0}
\hspace {-4.18cm}
\psfig{file=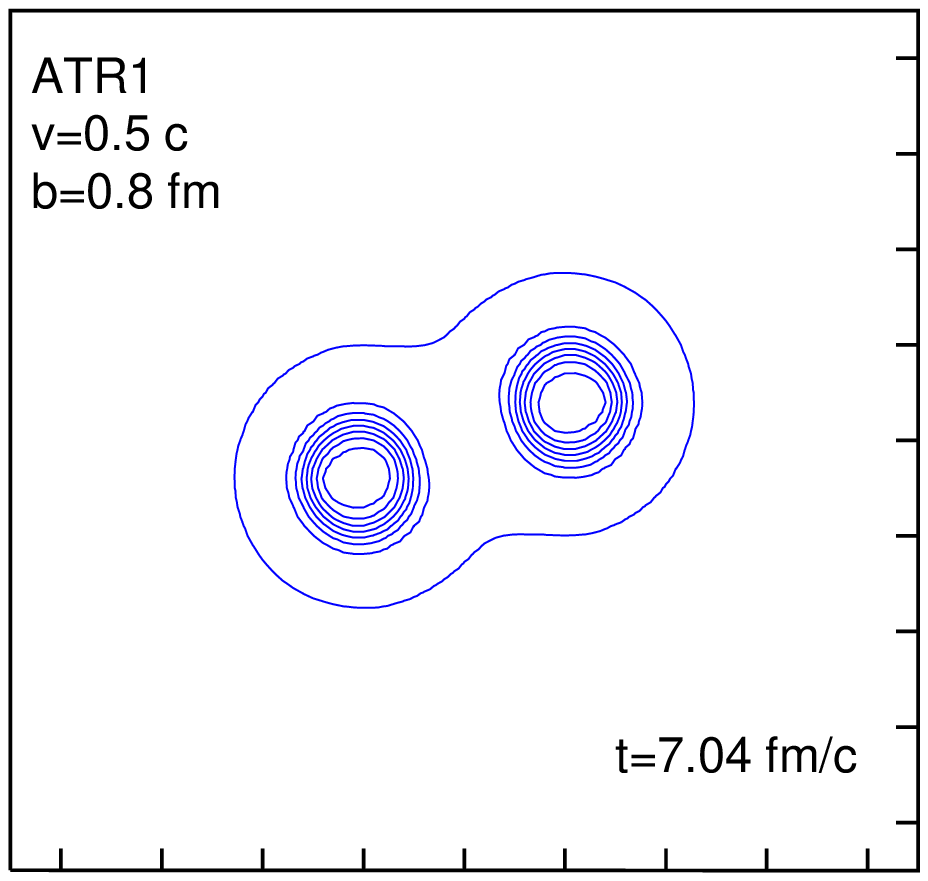,width=8.2cm,angle=0}
\hspace {-4.18cm}
\psfig{file=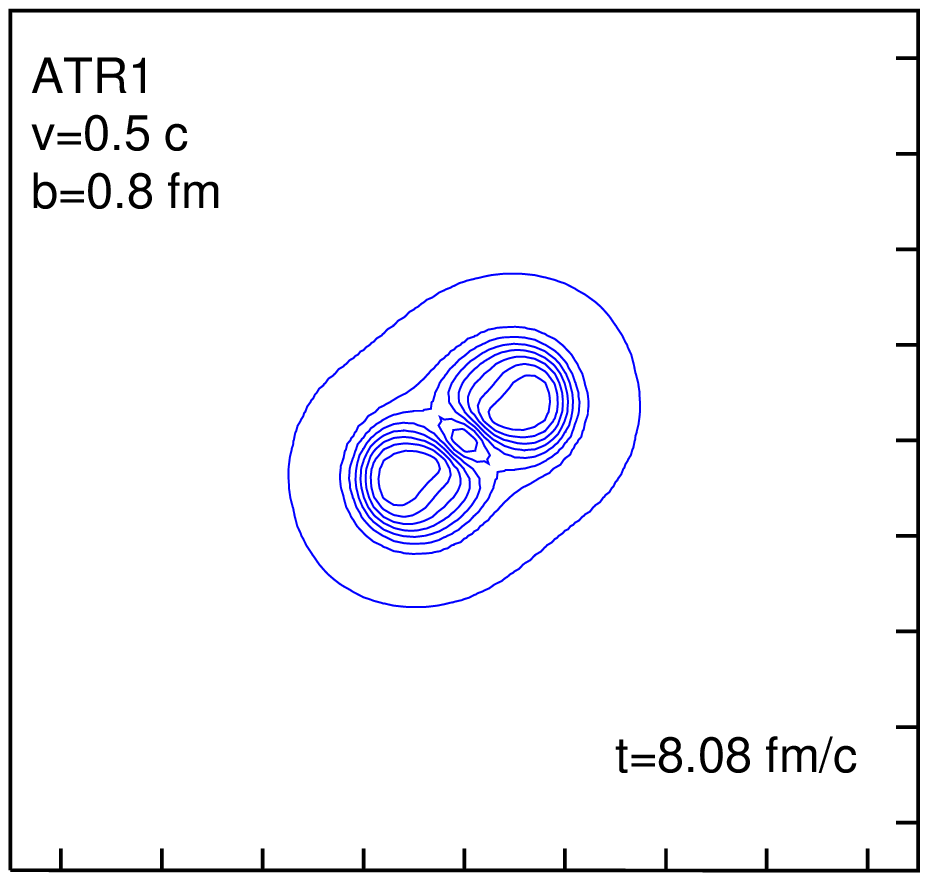,width=8.2cm,angle=0}
\hspace {-4.18cm}
\psfig{file=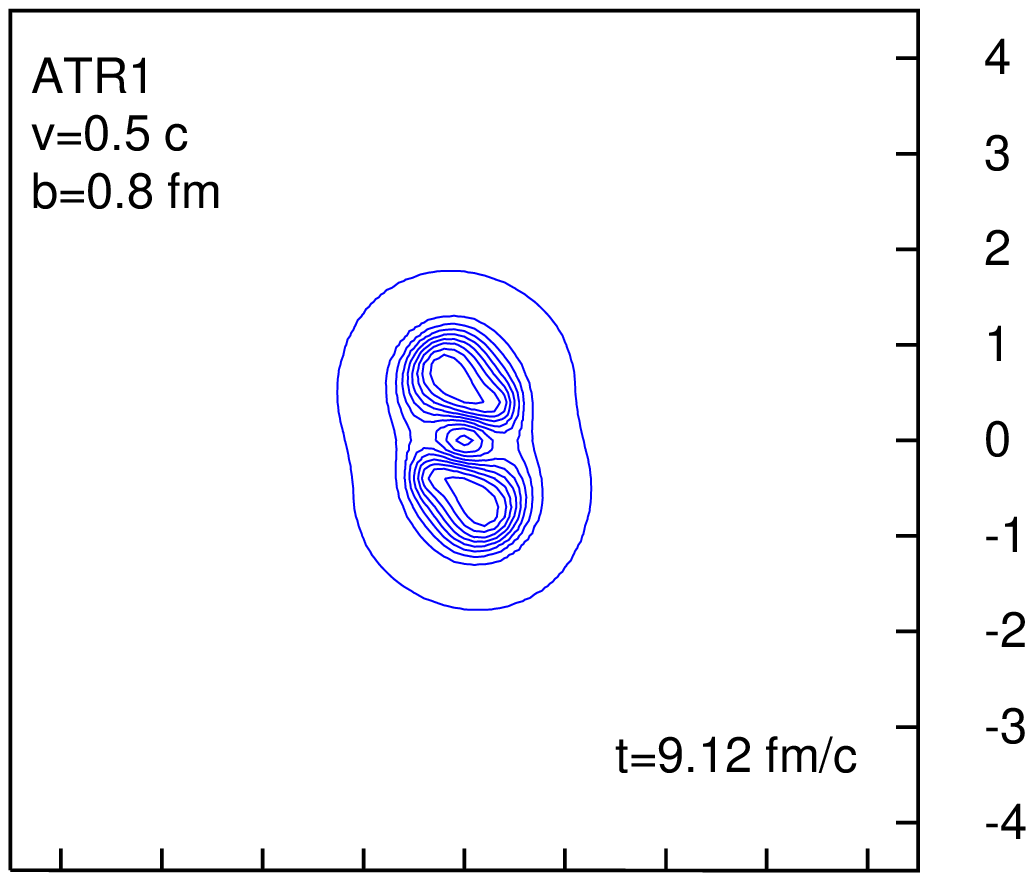,width=8.2cm,angle=0}
}
\vspace{-4.2cm}
\hbox{
\psfig{file=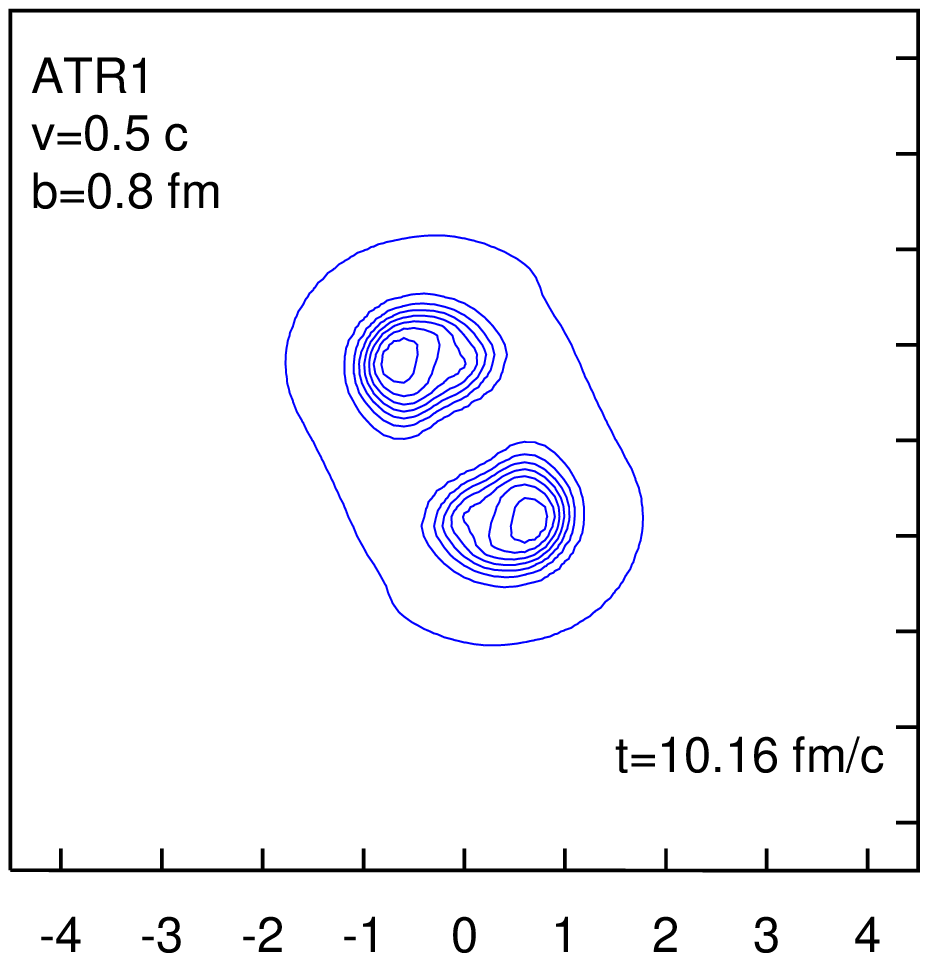,width=8.2cm,angle=0}
\hspace {-4.18cm}
\psfig{file=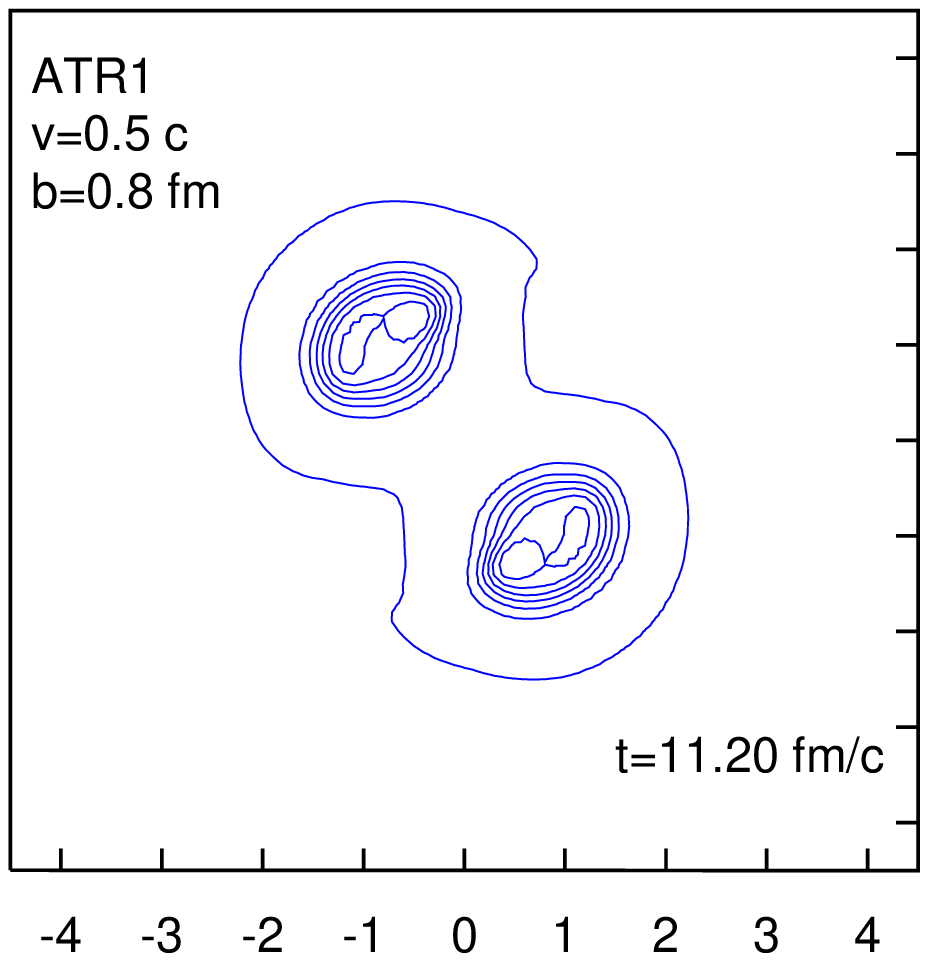,width=8.2cm,angle=0}
\hspace {-4.18cm}
\psfig{file=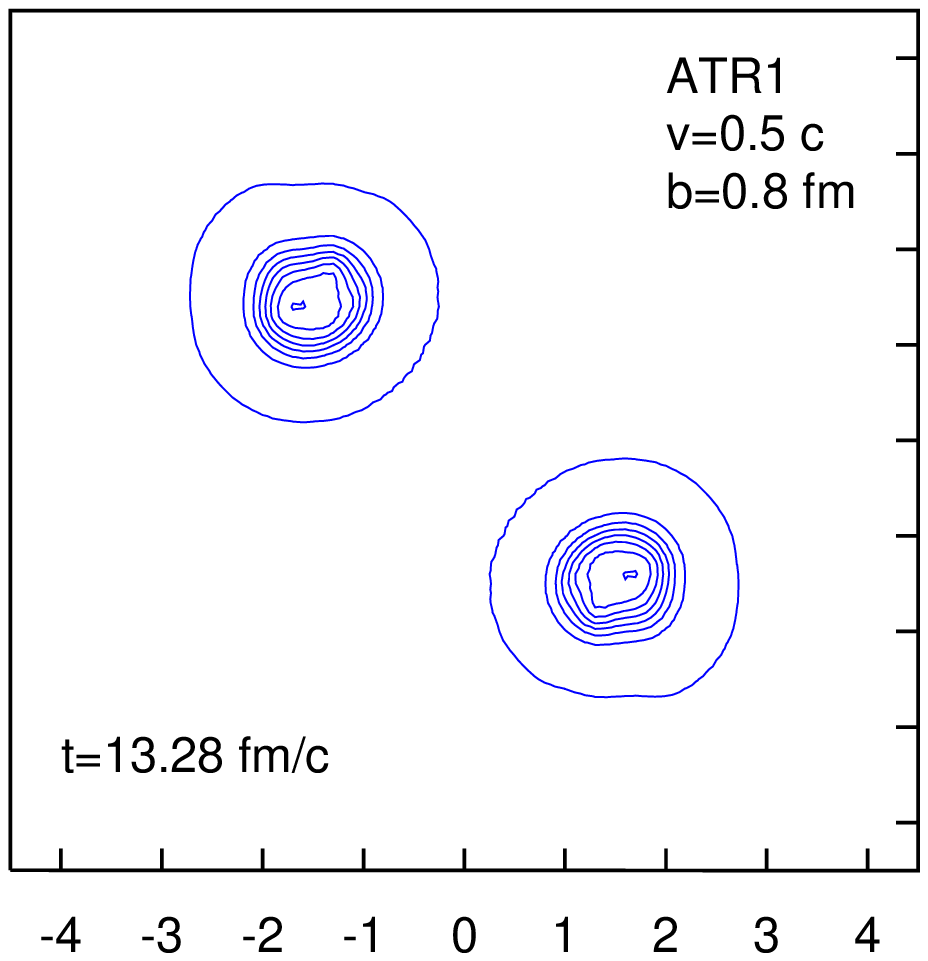,width=8.2cm,angle=0}
\hspace {-4.18cm}
\psfig{file=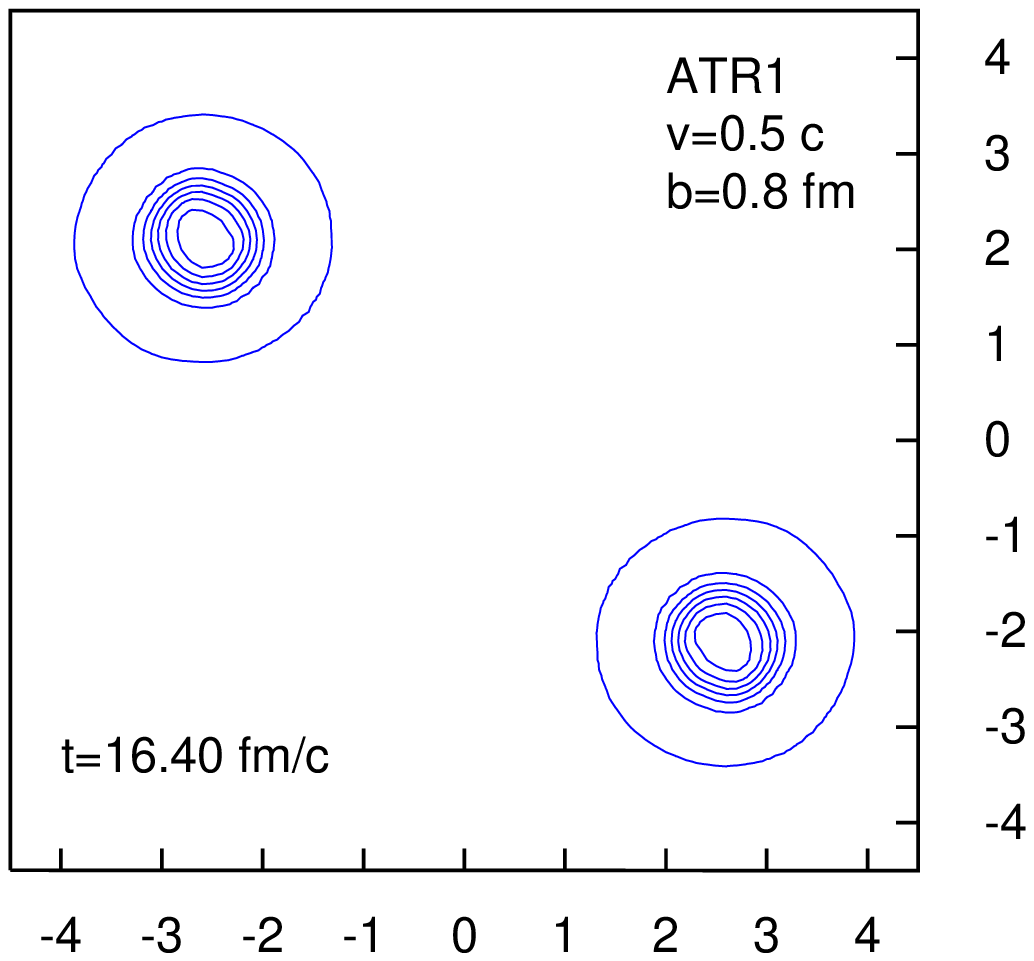,width=8.2cm,angle=0}
}
\vspace{-1.3cm}
}
}
\caption{Contour plots of the energy density in the $xy$ plane
for scattering in the attractive (1) channel with impact parameter $b=0.8~{\rm fm}$.
The spacing between the contours is $100~{\rm MeV/fm^3}$.
The first contour is at the $5~{\rm MeV/fm^3}$ level.
The length on both axes is measured in fermi.
Note that the frames are not evenly spaced in time.}
\label{ATR1_b0.4_series}
\end{figure}

Again we illustrate one scattering process in more detail.
In Figure \ref{ATR1_b0.4_series} we plot energy contours for
the $b=0.8~{\rm fm}$ scattering. 
The most important feature is the existence in the third and fourth
frames of a configuration 
reminiscent of the doughnut-shaped intermediate state identified in
the $b=0$ case for this grooming (Figures \ref{ATR1_b0.0_series} and
\ref{ATR1_b0.0_seriesA}). We are further away from 
axial symmetry than in Figure \ref{ATR1_b0.0_series}, but the relatively
large void in the middle is clearly identifiable. 
It is important to point out that even at this mutual distance of $0.8 ~{\rm fm}$,
the attractive interaction is strong enough to start forming the $B=2$
configuration. In the present case it is ripped apart by momentum.
As we shall see that is not always the case in all groomings.
Another feature to mention is the presence of internal oscillations.
They can be clearly seen in the last few frames of Figure \ref{ATR1_b0.4_series} 
deforming the outgoing skyrmions. 
In the case of $b=1.6~{\rm fm}$ trajectory, where the attraction is 
too weak to lead to a $B=2$ configuration, the topological centers 
are attracted toward each other but then they bounce back and 
oscillate transversally.

\subsection{Orbiting, capture and radiation: the repulsive channel}

\begin{figure}[t]
\centering
\epsfig{file=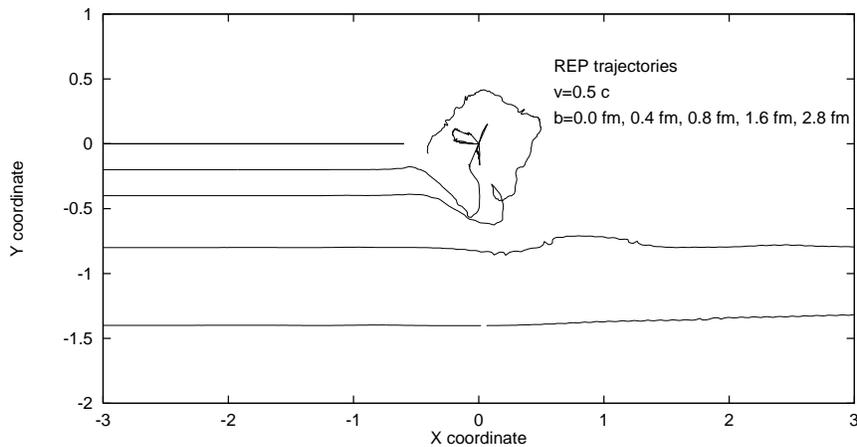,bbllx=75,bblly=50,bburx=750,bbury=410,width=11cm}
\caption{
Paths of the topological centers of one of the colliding skyrmions
in the repulsive case.
The $x$ axis points to the right and the $y$ axis points upwards.
The length on both axes is measured in fermi.
}
\label{REPpath}
\end{figure}

For the remaining two groomings, where the grooming direction is in the scattering plane,
the dynamics is more complex. 
Let us first consider grooming about the incident direction,
referred to as the repulsive channel.
For the smaller impact parameters $b=0.4~{\rm fm}$ and $b=0.8~{\rm fm}$,
we find a remarkable type of trajectory.
The scattering begins as repulsive
as we see in Figure \ref{REPpath}, but
as the skyrmions pass one another, they find themselves groomed
by a rotation of $\ppi$ about an axis normal to the line
joining them. This is the most attractive configuration. Hence as they
pass one another they begin to attract. 
The two skyrmions now, feeling this attraction, couple
and begin to orbit. 
In order to make the motion clearer in Figure \ref{REPpath}, 
we have decreased the spatial scale of our plot and increased the length of
time shown on the orbiting trajectories, which are cut off at  $28.0~{\rm fm/c}$.

\begin{figure}
\centerline{
\vbox{
\vspace{-2.1cm}
\hbox{
\psfig{file=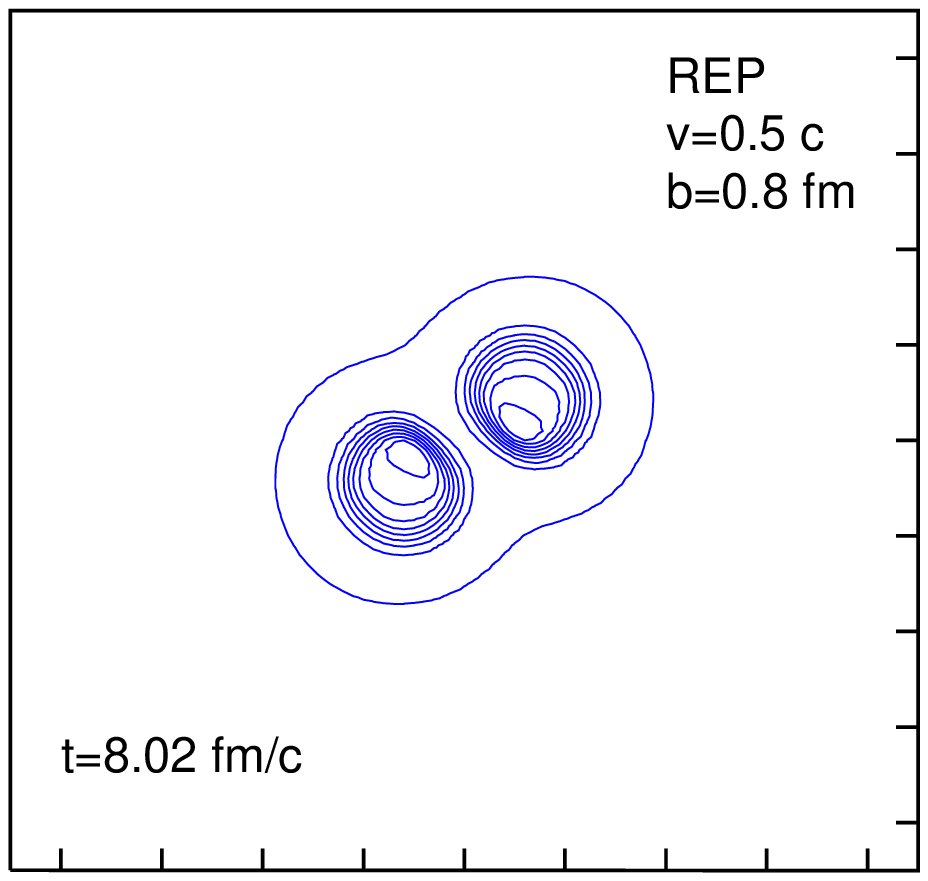,width=8.2cm,angle=0}
\hspace {-4.18cm}
\psfig{file=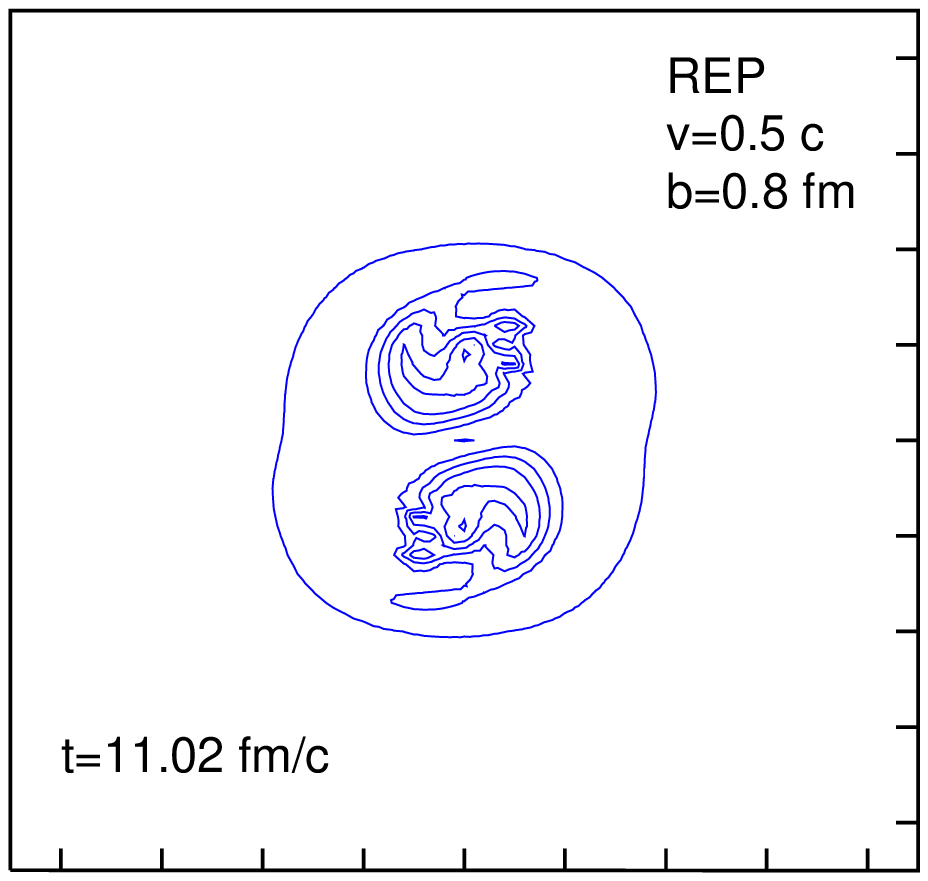,width=8.2cm,angle=0}
\hspace {-4.18cm}
\psfig{file=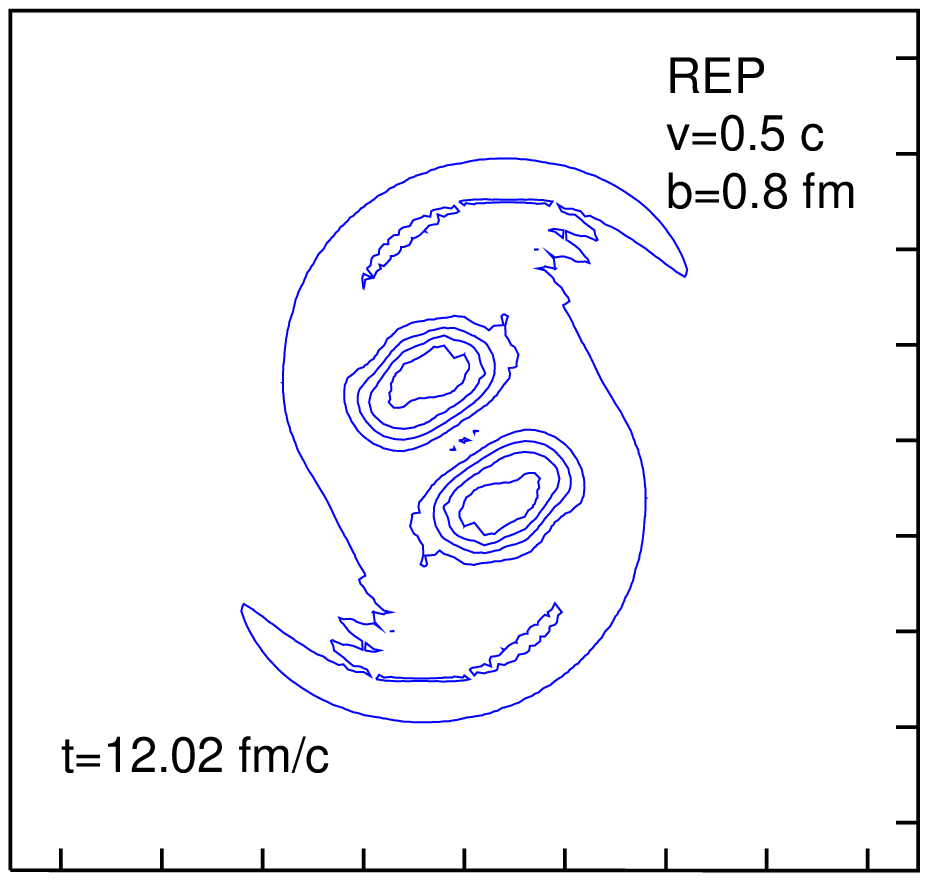,width=8.2cm,angle=0}
\hspace {-4.18cm}
\psfig{file=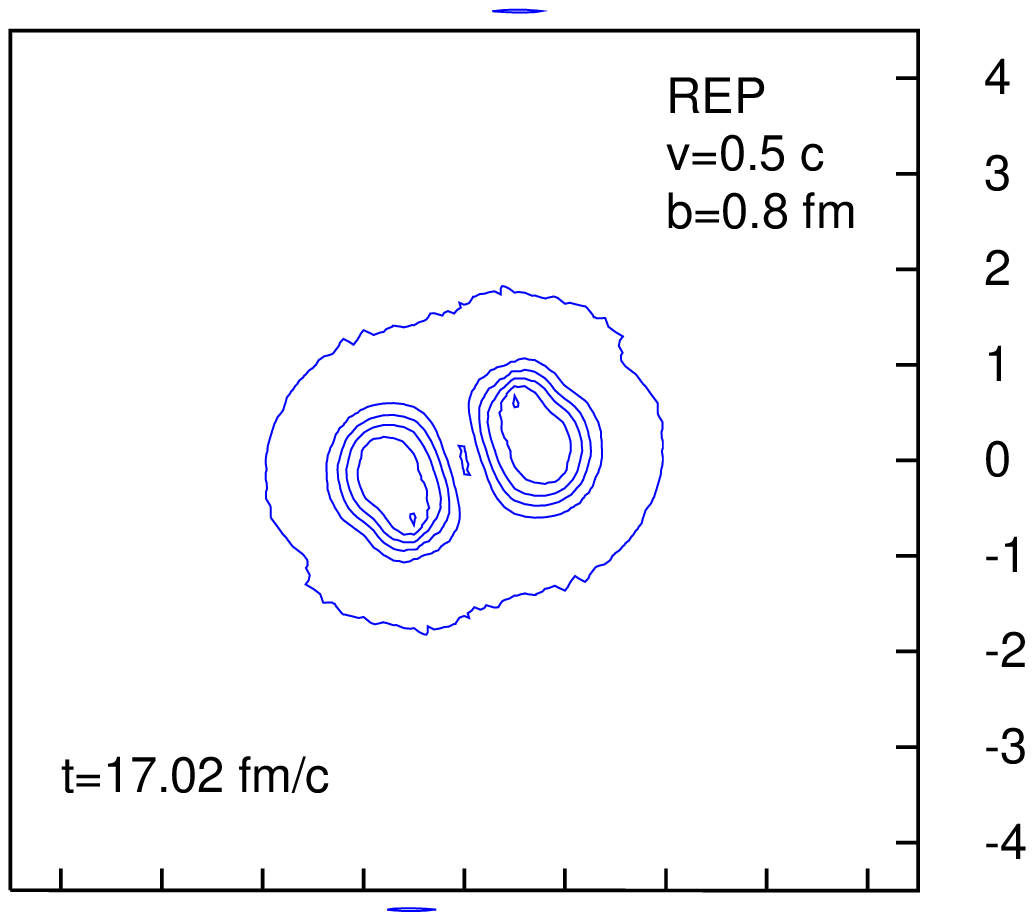,width=8.2cm,angle=0}
}
\vspace{-4.2cm}
\hbox{
\psfig{file=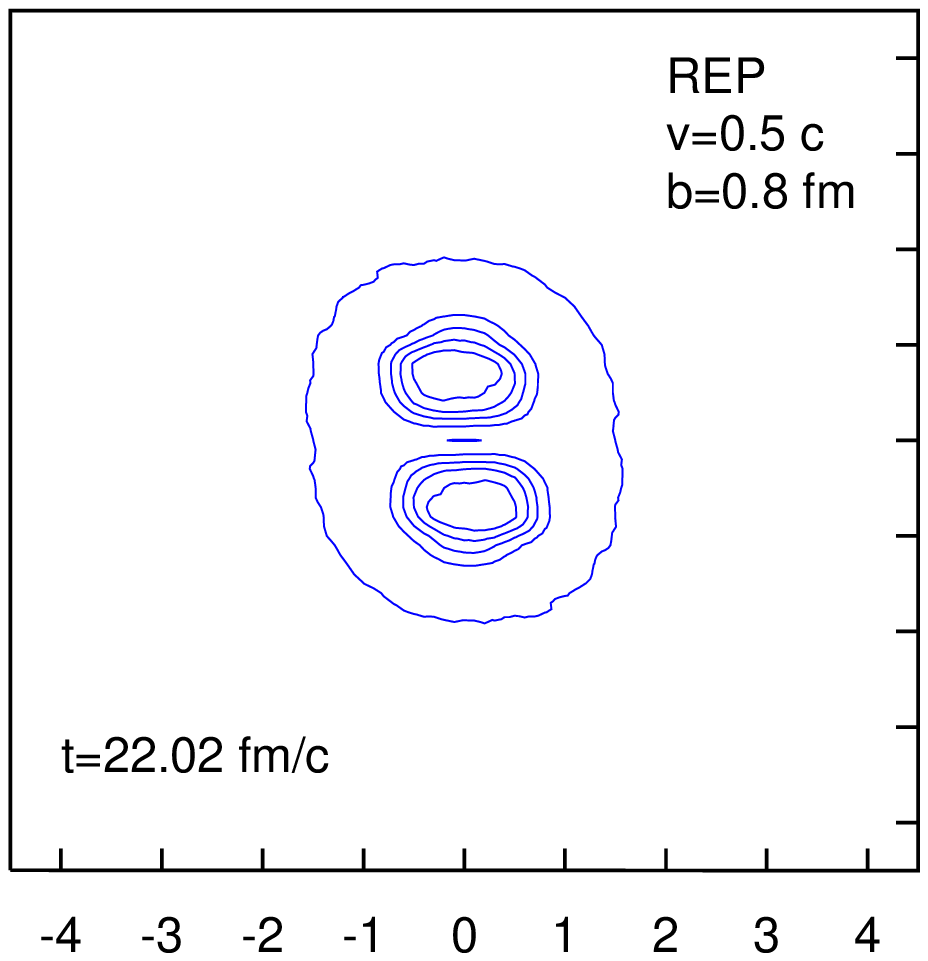,width=8.2cm,angle=0}
\hspace {-4.18cm}
\psfig{file=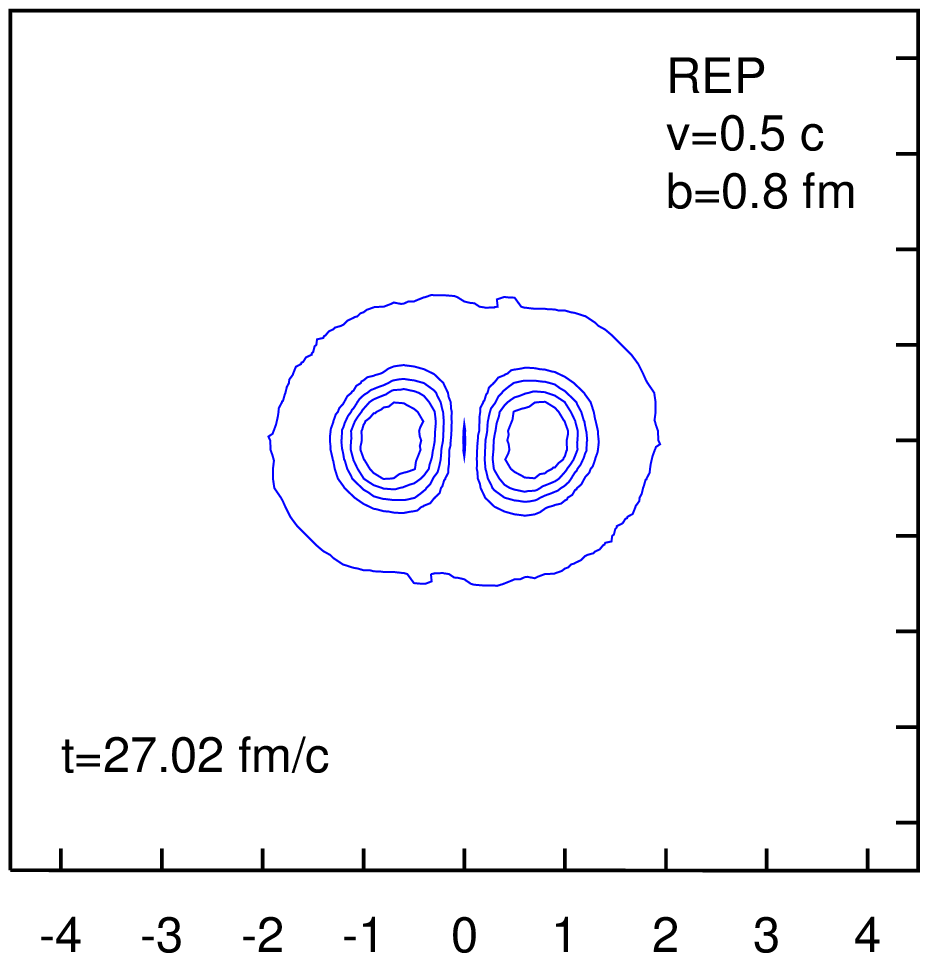,width=8.2cm,angle=0}
\hspace {-4.18cm}
\psfig{file=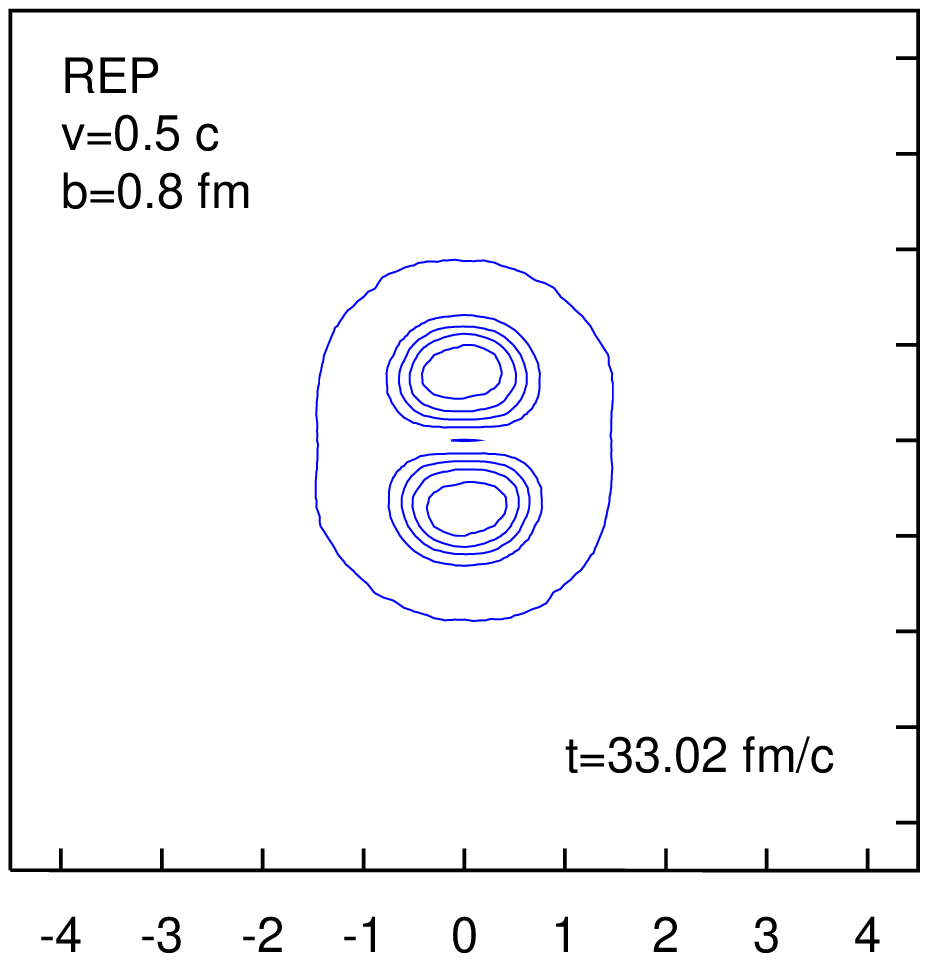,width=8.2cm,angle=0}
\hspace {-4.18cm}
\psfig{file=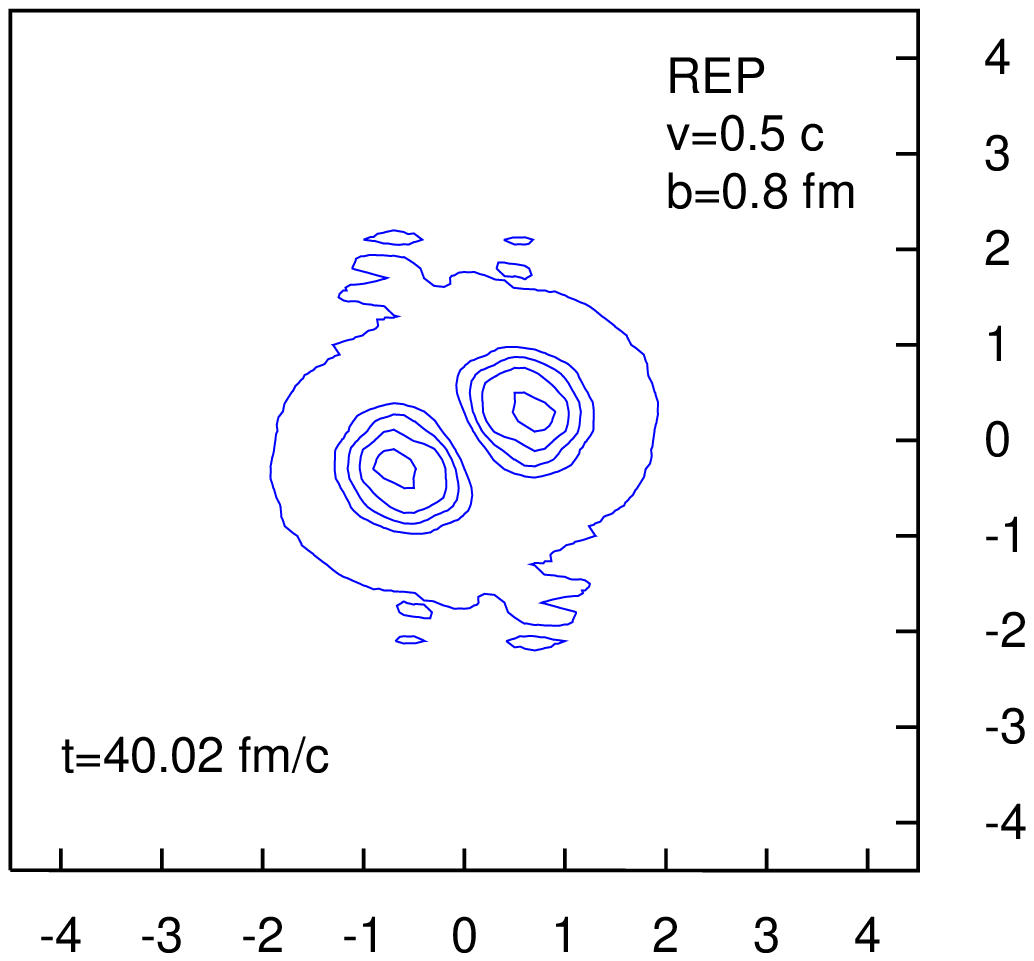,width=8.2cm,angle=0}
}
\vspace{-1.3cm}
}}
\caption{Contour plots of the energy density in the $xy$ plane
for scattering in the repulsive channel with impact parameter $b=0.8~{\rm fm}$.
The spacing between the contours is $100~{\rm MeV/fm^3}$.
The first contour is at the $5~{\rm MeV/fm^3}$ level so that
the low-amplitude waves corresponding to outgoing radiation
can be observed.
The length on both axes is measured in fermi.
Note that the frames are not evenly spaced in time.}
\label{REP_b0.4_series}
\end{figure}

In Figure \ref{REP_b0.4_series} we show energy density contour plots
for $b=0.8~{\rm fm}$ to illustrate our point. 
Notice that the plots cover a longer time
period than in the previous cases.
As the skyrmions orbit, they radiate. This radiation carries off
energy and the skyrmions couple more strongly. The radiation and 
coupling last for a long time. 
The first energy level we plot is $5~{\rm MeV/fm^3}$,
comparable to the amplitude of the radiation.
The first burst of radiation can be 
seen in the second and third frames of Figure \ref{REP_b0.4_series}.
It appears to consist of the spinning off of a region
of high local energy density.
A later burst is seen in the last frame.
These bursts accompany the orbiting movement of the now bound
skyrmions.

\begin{figure}
\centerline{
\vbox{
\hbox{
\psfig{file=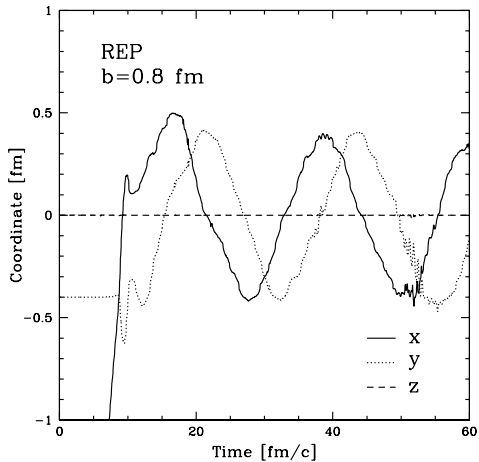,width=6.5cm,angle=0}
}
}}
\caption{
Time evolution of the three coordinates of the topological center of one of
the two skyrmions. They indicate orbiting motion in the $xy$ plane 
with slow damping. 
}
\label{REP_b0.4_timepath}
\end{figure}

The radiation takes away some of the angular momentum
and a significant part of the energy.
The total energy radiated is greater than the incident
kinetic energy, hence the two are now in a bound state. 
Thus for this choice of
parameters, ($b=0.8~{\rm fm}$, $v=0.5~c$, repulsive grooming) we 
have the phenomena of orbiting, radiation and capture. 
We assume that this $B=2$ system will eventually find its way to 
axial-symmetric torus shaped bound state at rest. 
To reach that state they will have
to radiate more energy and the remaining angular momentum. 
We have followed the orbiting for a time of more than $60$ fm/c.
We observed continued orbiting with slowly decreasing amplitude, 
as shown in Figure \ref{REP_b0.4_timepath}, and
continuous energy loss through bursts radiation.
We are not able, numerically, to follow the state to the very end.

\begin{figure}
\centerline{
\vbox{
\vspace{-2.1cm}
\hbox{
\psfig{file=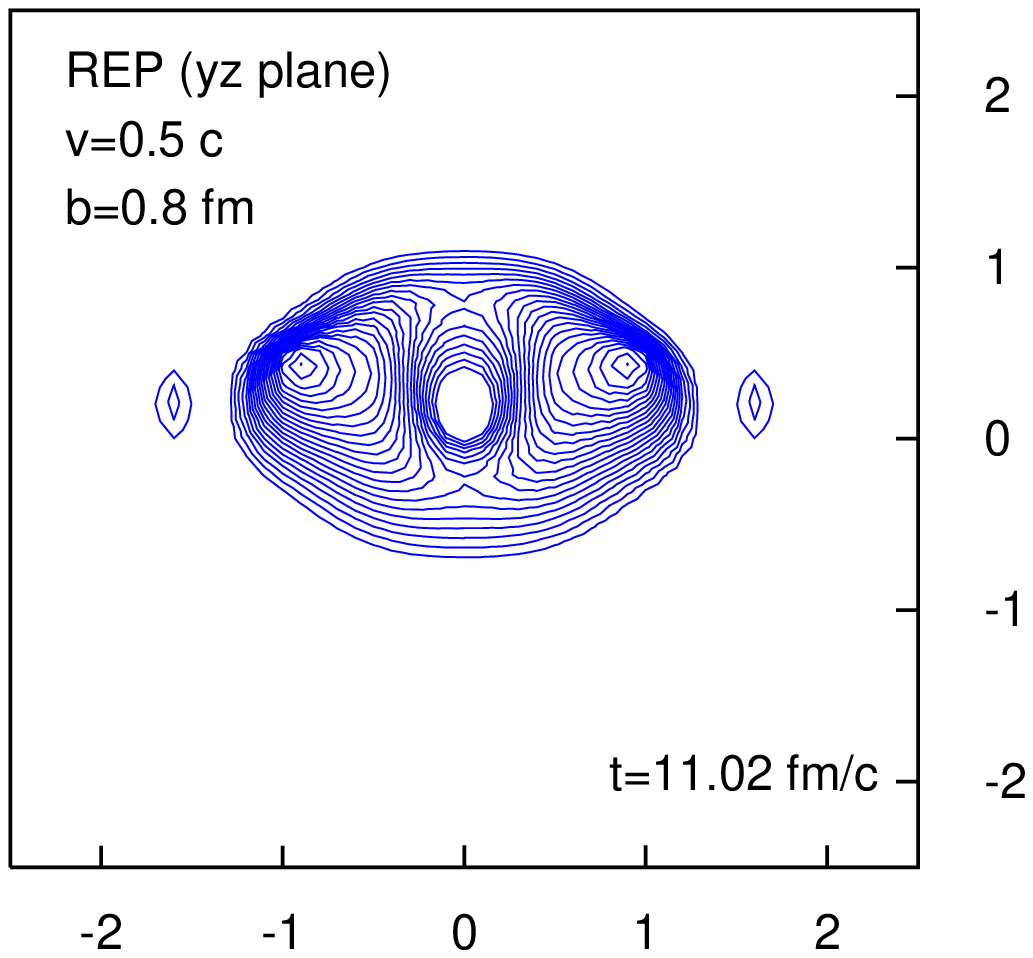,width=8cm,angle=0}
\hspace {-3.2cm}
\psfig{file=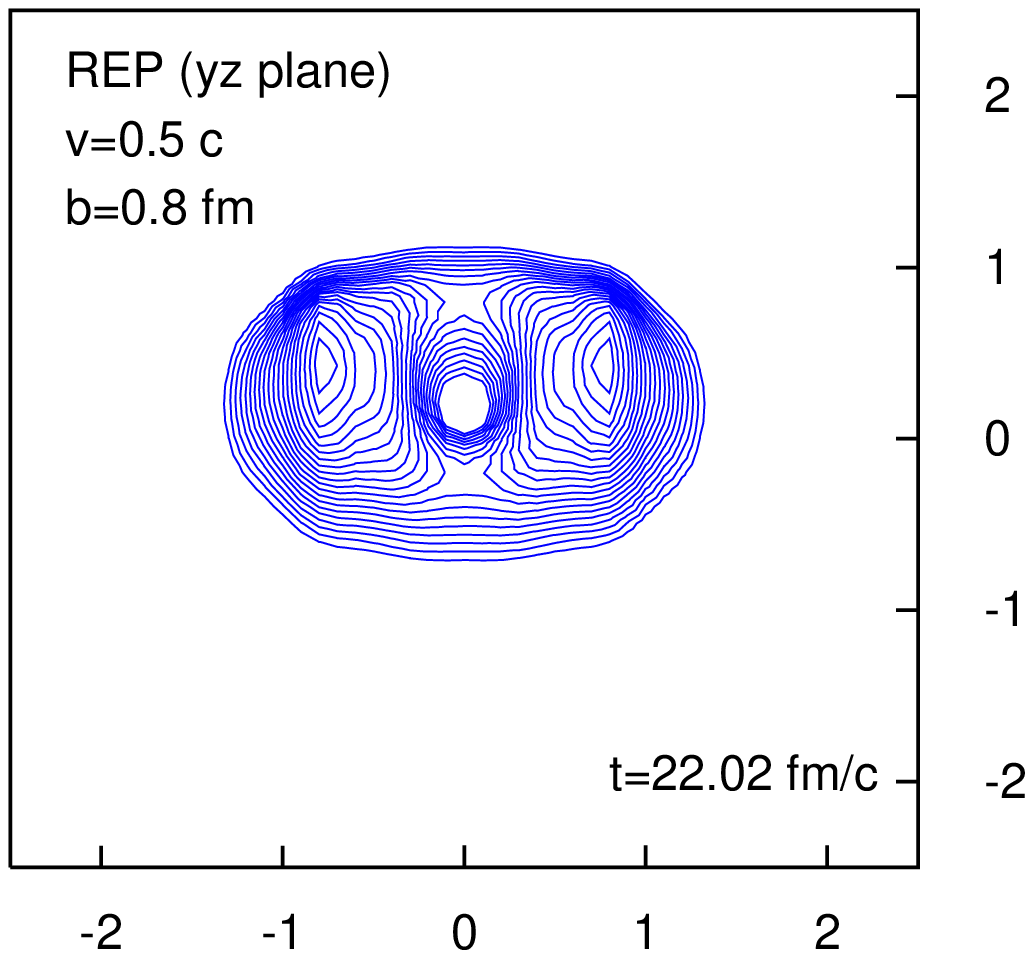,width=8cm,angle=0}
\hspace {-3.2cm}
\psfig{file=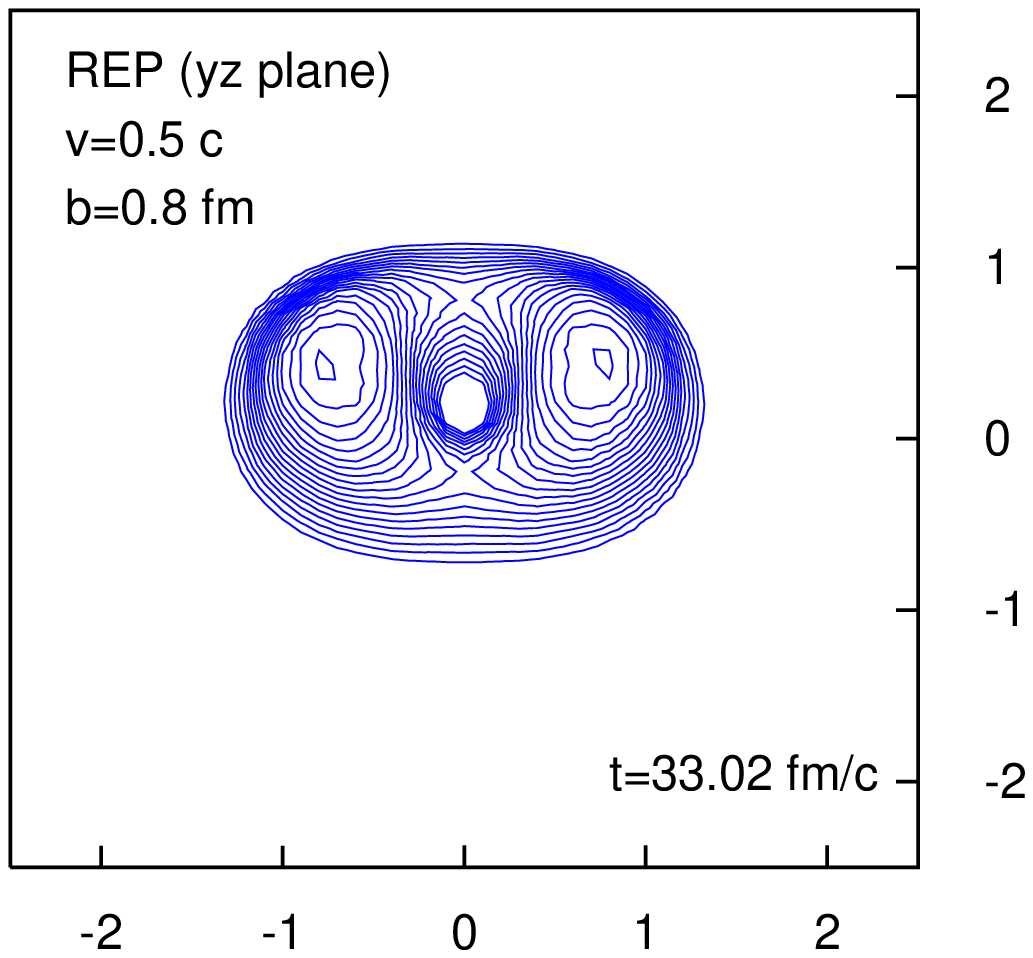,width=8cm,angle=0}
}
\vspace{-1.1cm}
}}
\caption{Contour plots of the energy density in the $ yz $ plane
for scattering in the repulsive channel with impact parameter $b=0.8~{\rm fm}$.
The spacing between the contours is $20~{\rm MeV/fm^3}$.
The first contour is at the $100~{\rm MeV/fm^3}$.
The length on both axes is measured in fermi.
Note that the plots are slightly tilted to indicate the height of the
contour lines.}
\label{REP_b0.4_yz}
\end{figure}

We now have a bound $B=2$ configuration, and therefore 
should expect the appearance of a torus.
The grooming in this case is about the $x$ axis.
The torus should now appear in the $yz$ plane, and
should rotate about the $z$ axis to carry the initial angular momentum.
Energy contours in the $yz$ plane corresponding to three frames from
Figure \ref{REP_b0.4_series} are shown in Figure \ref{REP_b0.4_yz}.
We recognize the familiar pattern of the doughnut-shaped
$B=2$ bound state. 
These snapshots correspond to situations when the doughnut is
aligned with the $yz$ plane.
The doughnut is strongly deformed,
with the two skyrmions clearly distinguishable. This deformation is only slightly
alleviated during the $20 ~{\rm fm/c}$ (and a full $\rm 360^o$ rotation)
between the first and the last frame in Figure \ref{REP_b0.4_yz}.

\begin{figure}
\centerline{
\vbox{
\vspace{-2.1cm}
\hbox{
\psfig{file=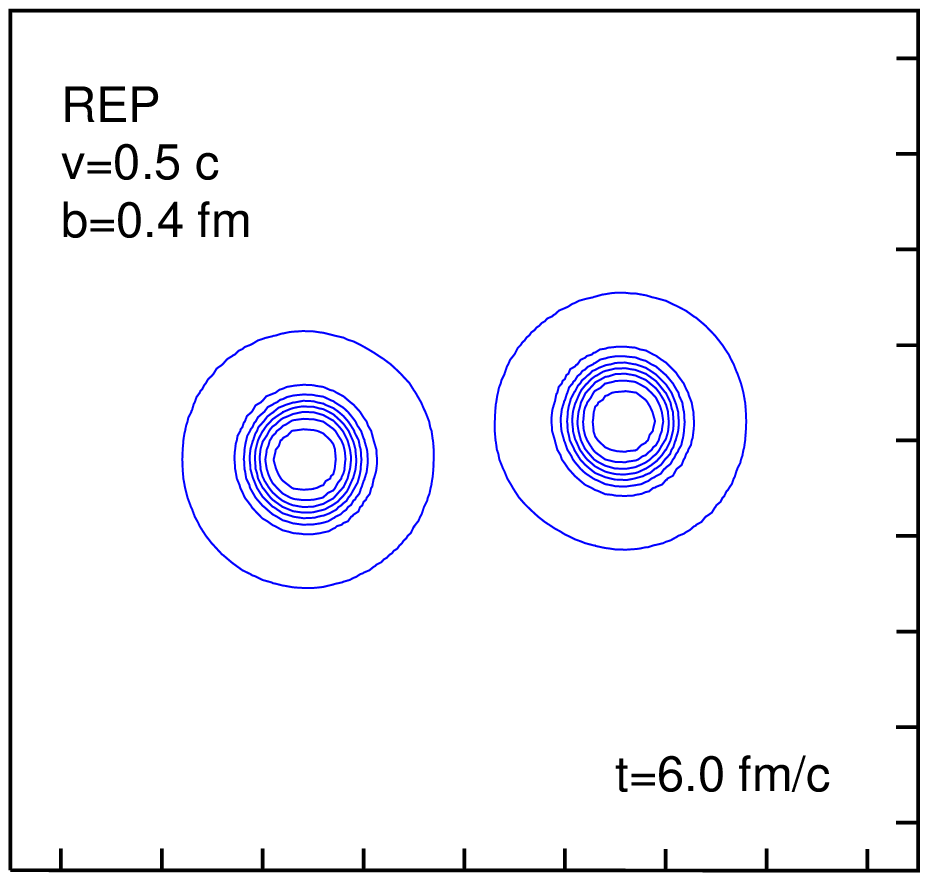,width=8.2cm,angle=0}
\hspace {-4.18cm}
\psfig{file=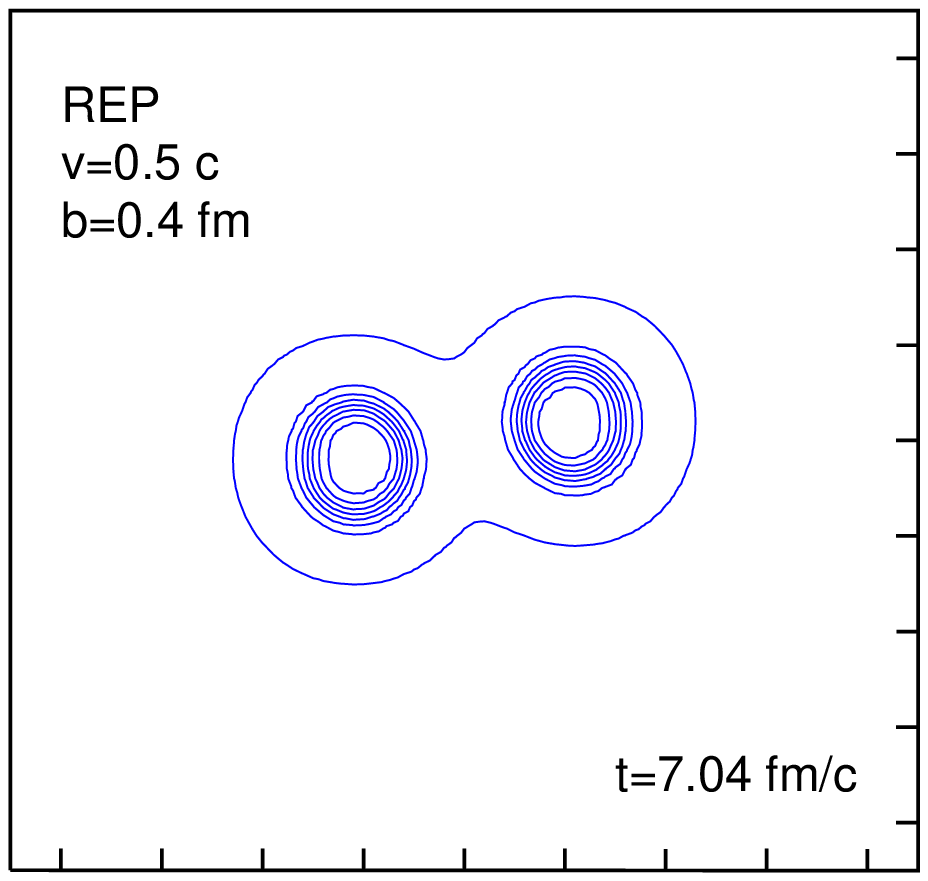,width=8.2cm,angle=0}
\hspace {-4.18cm}
\psfig{file=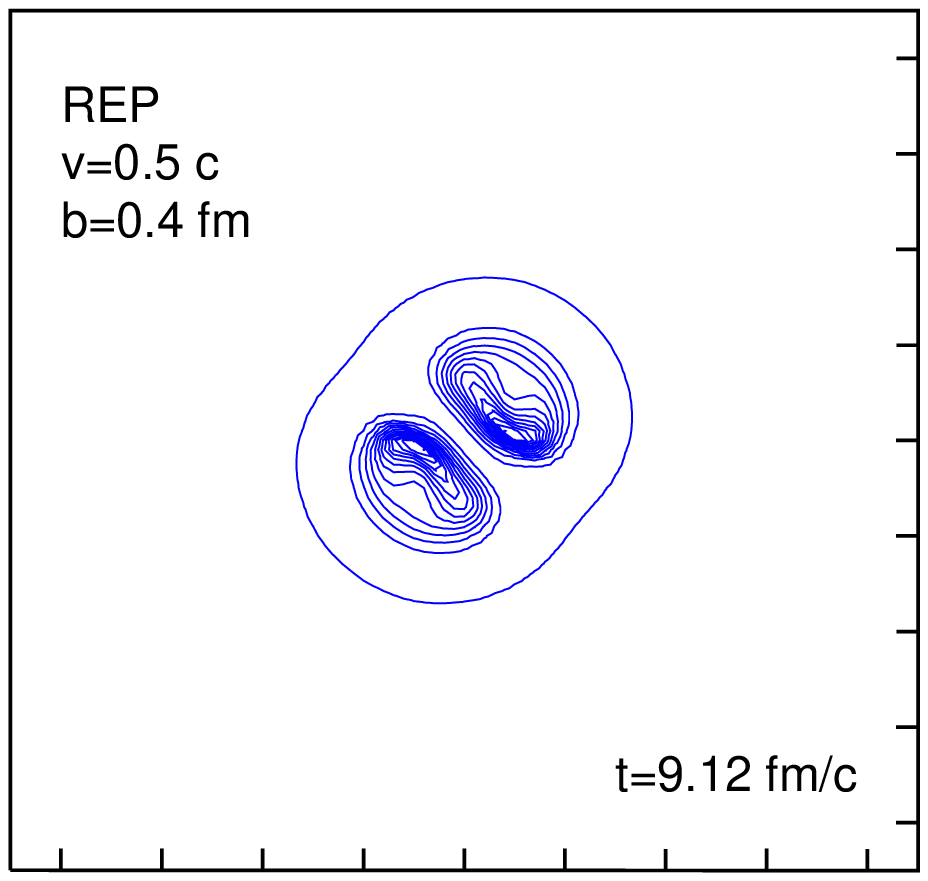,width=8.2cm,angle=0}
\hspace {-4.18cm}
\psfig{file=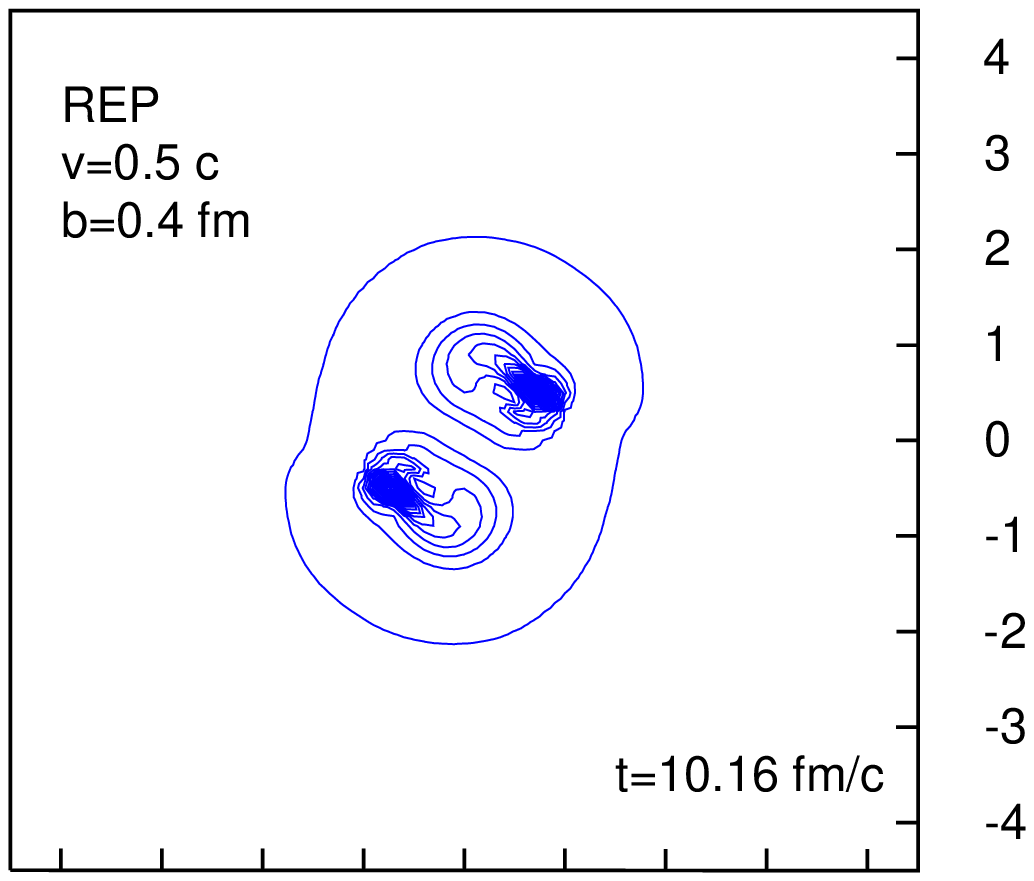,width=8.2cm,angle=0}
}
\vspace{-4.2cm}
\hbox{
\psfig{file=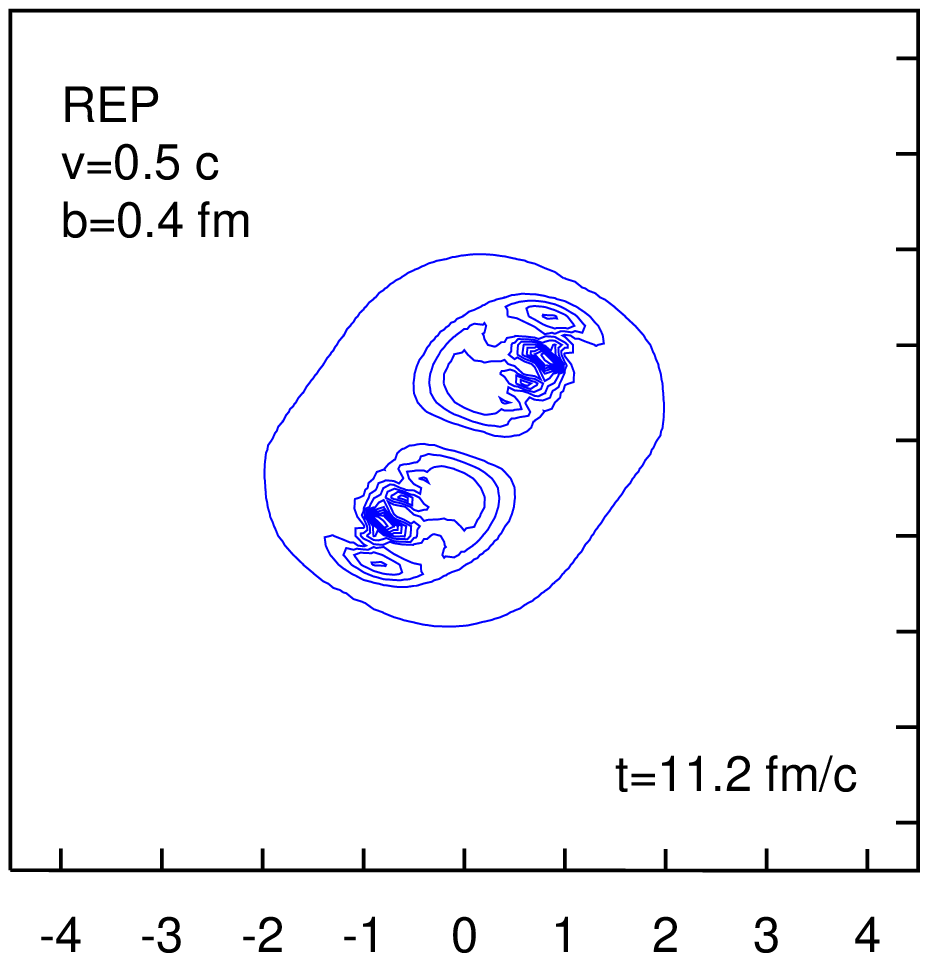,width=8.2cm,angle=0}
\hspace {-4.18cm}
\psfig{file=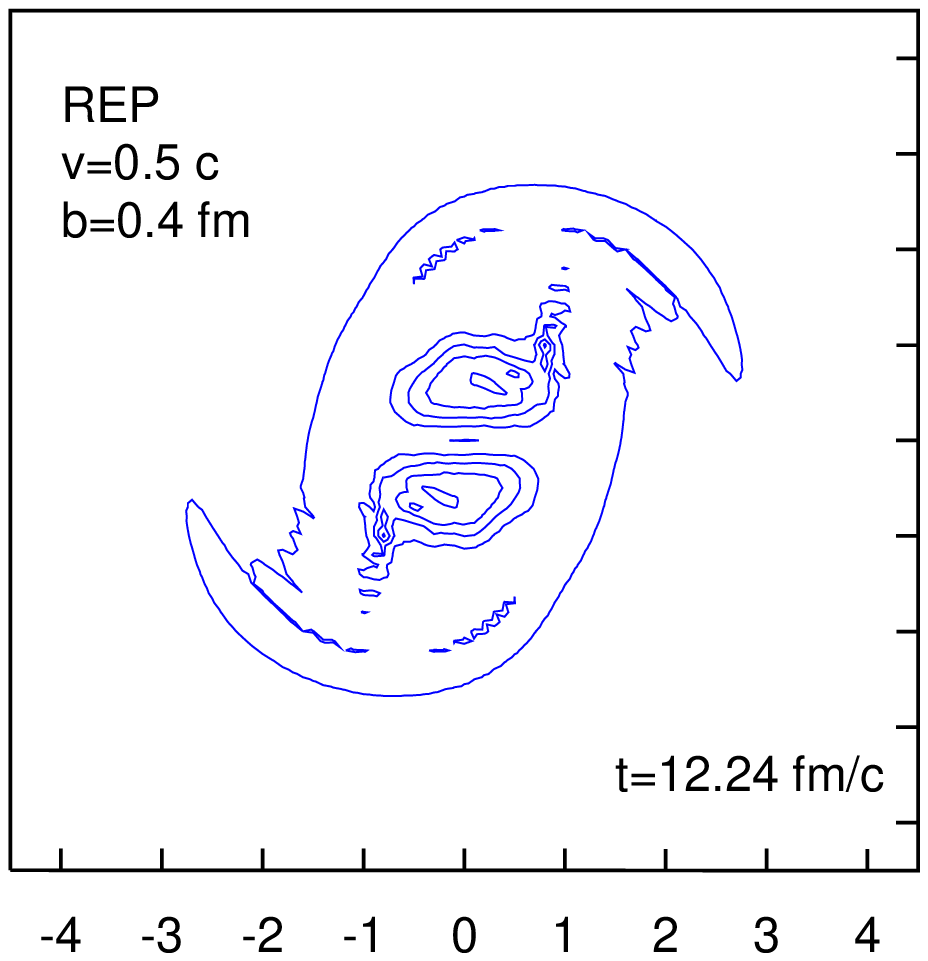,width=8.2cm,angle=0}
\hspace {-4.18cm}
\psfig{file=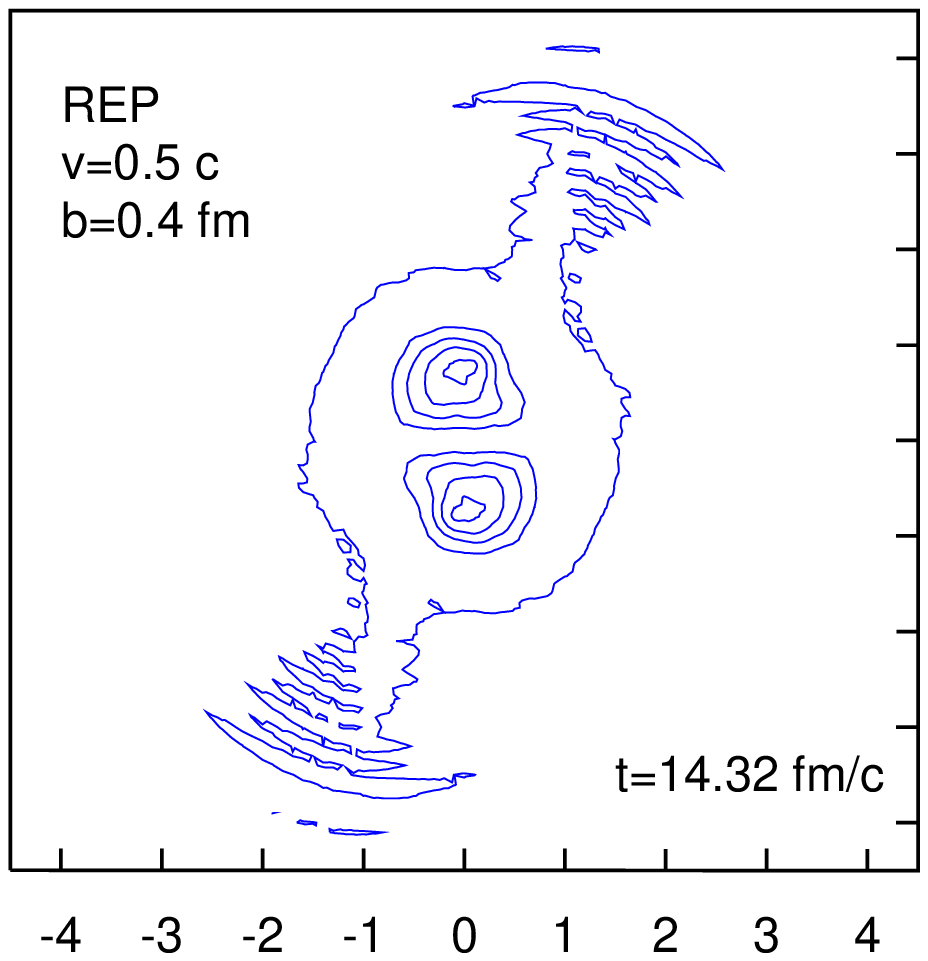,width=8.2cm,angle=0}
\hspace {-4.18cm}
\psfig{file=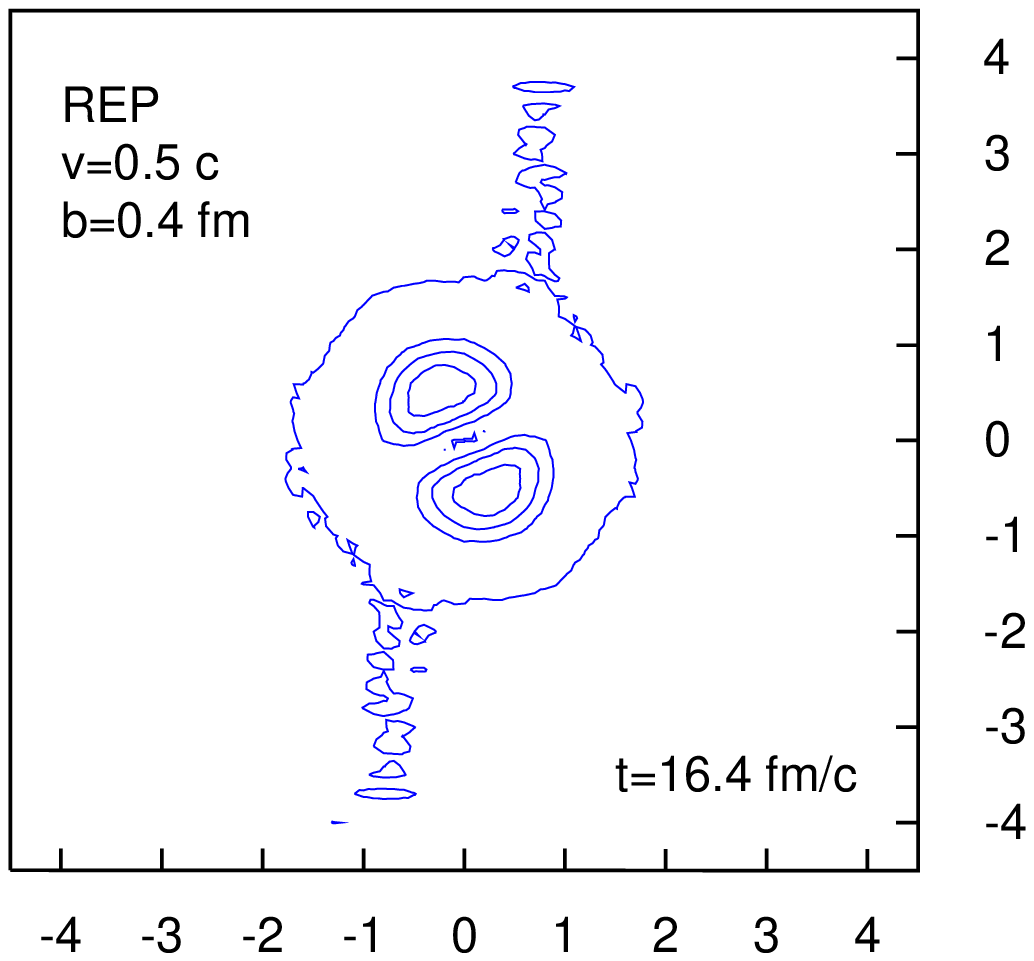,width=8.2cm,angle=0}
}
\vspace{-1.3cm}
}
}
\caption{Contour plots of the energy density in the $xy$ plane
for scattering in the repulsive channel with impact parameter $b=0.4~{\rm fm}$.
The spacing between the contours is $100~{\rm MeV/fm^3}$.
The first contour is at the $5~{\rm MeV/fm^3}$ level so that
the low-amplitude waves corresponding to outgoing radiation
can be observed.
The length on both axes is measured in fermi.
Note that the frames are not evenly spaced in time.}
\label{REP_b0.2_series}
\end{figure}

The appearance of the $B=2$ bound state is even clearer in
the $b=0.4~{\rm fm}$ case. 
The $xy$ plane energy contours 
are very similar to those in the $b=0.8~{\rm fm}$ process. There is again
quite some radiation early on, as illustrated in Figure \ref{REP_b0.2_series}.
As can be seen from the trajectories in Figure \ref{REPpath}, 
the attraction is strong 
in this case (once the skyrmions are past the initial repulsion), 
and the topological centers come very close.

\begin{figure}
\centerline{
\vbox{
\vspace{-2.1cm}
\hbox{
\psfig{file=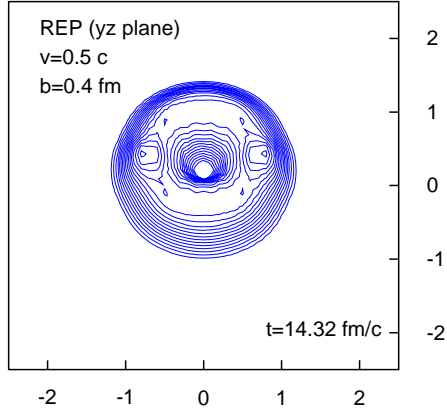,width=10cm,angle=0}
}
\vspace{-1.1cm}
}}
\caption{Contour plot of the energy density in the $ yz $ plane
corresponding to the seventh frame in Figure \ref{REP_b0.2_series}.
The spacing between the contours here is $20~{\rm MeV/fm^3}$.
The first contour is at the $100~{\rm MeV/fm^3}$.
The length on both axes is measured in fermi.
Notice that the plot is slightly tilted to emphasize the height of the
contours.
}
\label{REP_b0.2_yz}
\end{figure}

The $yz$ plane contour plot in Figure \ref{REP_b0.2_yz},
showing the doughnut just after its formation, exhibits 
more axial symmetry than the corresponding ones in the $b=0.8~{\rm fm}$ case.
We conclude that we have an example of two skyrmions merging into a $B=2$
axial symmetric configuration. The angular momentum 
remaining after the initial radiation burst 
is carried by the 
rotation of the doughnut around the $z$ axis. 
The individual skyrmions 
lose their identity early on and the movement of the topological centers
is always very close to the symmetry center. Periodically, they move out of the
$xy$ plane in the $z$ direction. We interpret this as oscillations of the torus.
We assume that eventually the kinetic energy and the angular momentum will be 
radiated away.

In the $b=1.6~{\rm fm}$ case the skyrmions first feel some repulsion, 
which turns to attraction as they pass.
This results in small persistent transverse oscillations. 
For $b=2.8~{\rm fm}$ there is only a very small
attractive interaction.

\subsection{Scattering out of the plane of motion: the attractive (2) channel}

\begin{figure}[t]
\centering
\epsfig{file=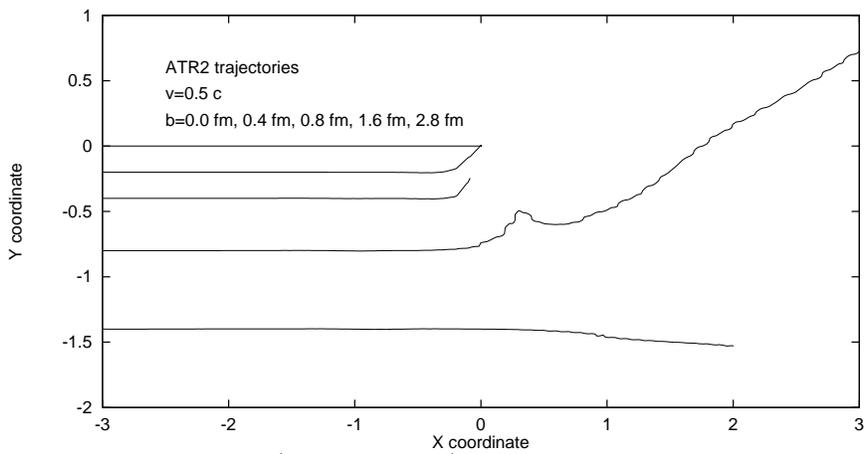,bbllx=80,bblly=50,bburx=750,bbury=410,width=11cm}
\caption{
Paths of the topological centers (in the $xy$ plane) of one of the 
colliding skyrmions
in the attractive case with grooming around the direction of the impact parameter.
For the $b=0.8~{\rm fm}$ case, the return of the trajectories to the $xy$ plane
is not shown.
The $x$ axis points to the right and the $y$ axis points upwards.
The length on both axes is measured in fermi.
}
\label{ATR2path}
\end{figure}

\begin{figure}
\centerline{
\vbox{
\hbox{
\psfig{file=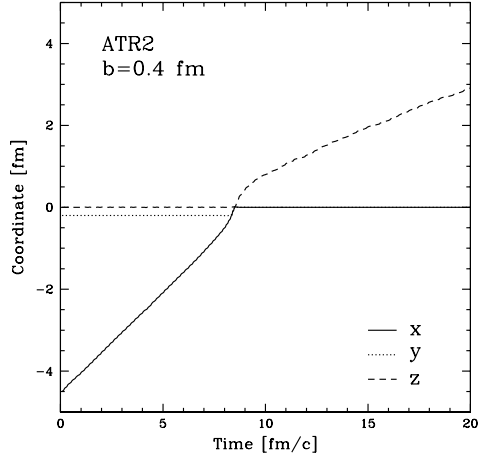,width=6.5cm,angle=0}
}
}}
\caption{
Time evolution of the three coordinates of the topological center of one of
the two skyrmions in the attractive (2) channel for  $b=0.4 ~{\rm fm}$.
}
\label{ATR2_b0.2_timepath}
\end{figure}

The scattering in the last remaining channel,
with grooming around the direction of the impact parameter,
 shows the most remarkable behavior of all. 
The corresponding trajectories are shown in Figure \ref{ATR2path}.
Recall that for zero impact parameter, this grooming leads to
right angle scattering in a direction normal to the plane formed
by the incident direction and the grooming axis. 
That direction is now \em normal \em to the scattering plane, which is 
the one that contains both the impact parameter and the incident momenta.
Usually one says that there cannot be scattering, 
for finite impact parameter, out of the  
scattering plane by angular momentum conservation. However, we have
meson radiation (mostly pion but also some $\omega$) 
that can carry off angular momentum, albeit 
inefficiently. 
For the impact parameter of $0.4~{\rm fm}$ this is just what happens. 
This remarkable trajectory can be better understood by studying
the cartesian components of the topological center's position
as a function of time. 
We take the $x$ direction as the incident
one, the impact parameter in the $y$ direction and the normal to the
scattering plane in the $z$ direction. 
The behavior of $x$, $y$, and $z$ as functions
of time for the topological center of each of the skyrmions in the case
of $b=0.4~{\rm fm}$ is shown in Figure \ref{ATR2_b0.2_timepath}. 
The skyrmions meet, interact, and then
go off normal to the scattering plane in the $z$ direction. 
Therefore they disappear from the plot of trajectories in
the $xy$ plane, Figure \ref{ATR2path}.
They disappear at the point $x=y=0$, with 
their final state trajectory along the $z$ axis
having \em no \em  impact parameter, 
as it must not, by angular momentum conservation.
Note that the slope of the $z$ trajectory in Figure  \ref{ATR2_b0.2_timepath}
is less than that of the $x$ trajectory, reflecting the energy carried off
by the radiation.

\begin{figure}
\centerline{
\vbox{
\vspace{-2.1cm}
\hbox{
\psfig{file=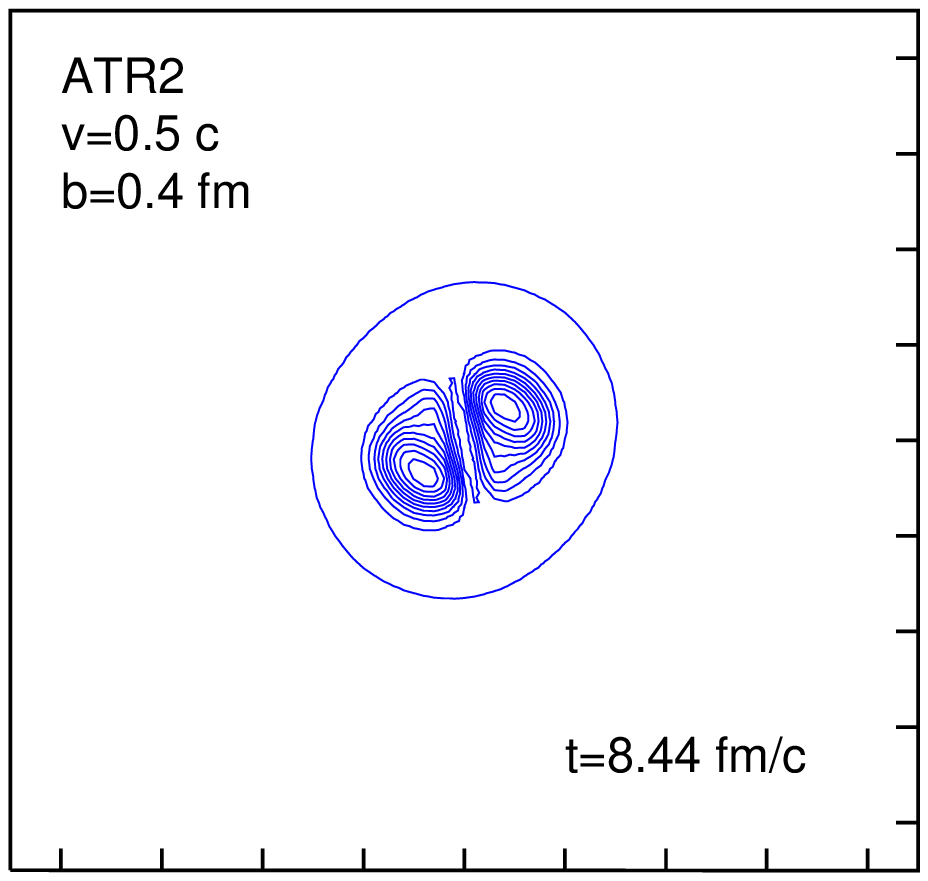,width=8.2cm,angle=0}
\hspace {-4.18cm}
\psfig{file=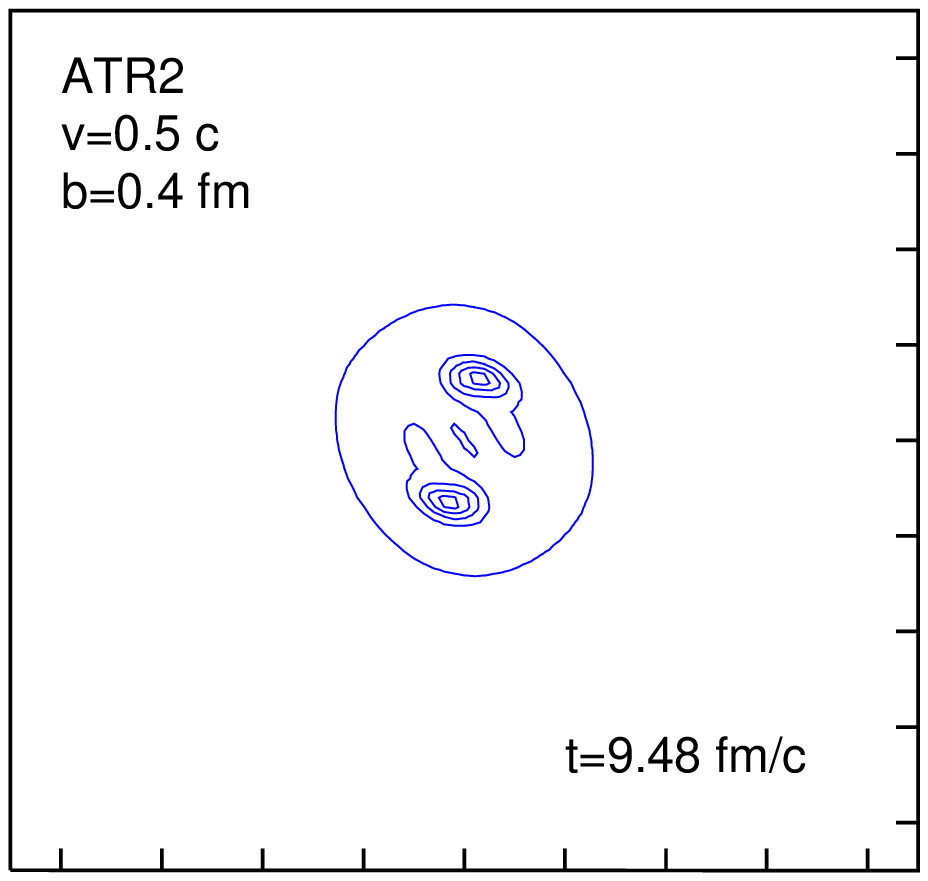,width=8.2cm,angle=0}
\hspace {-4.18cm}
\psfig{file=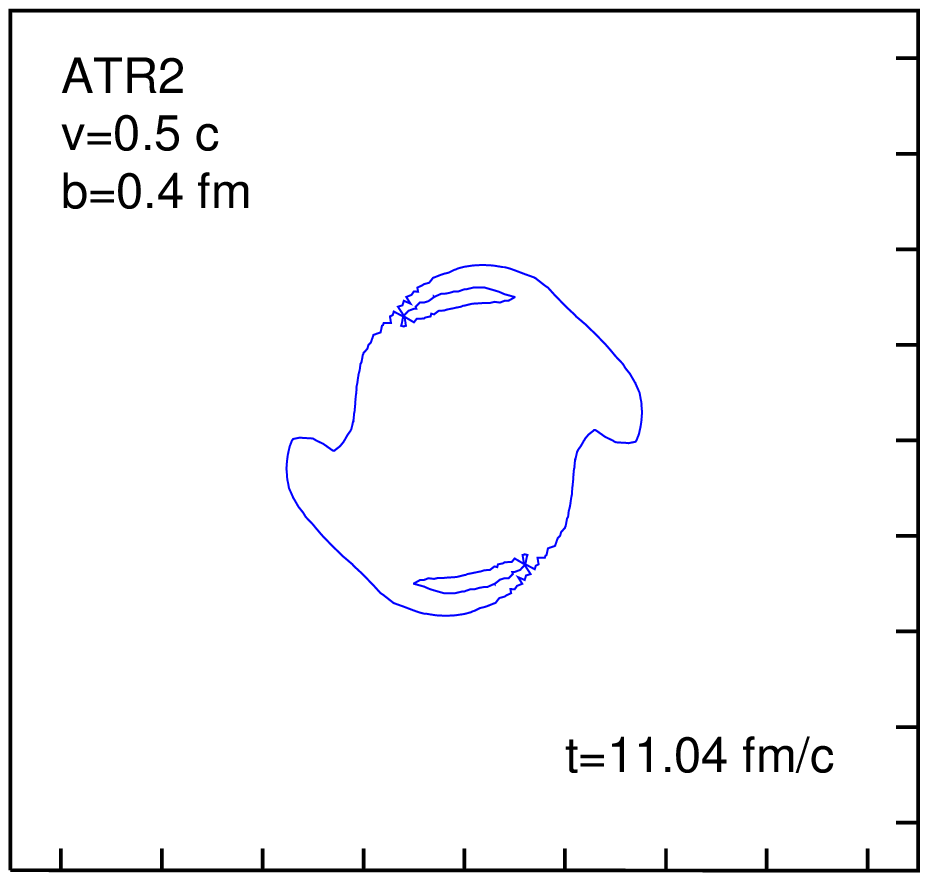,width=8.2cm,angle=0}
\hspace {-4.18cm}
\psfig{file=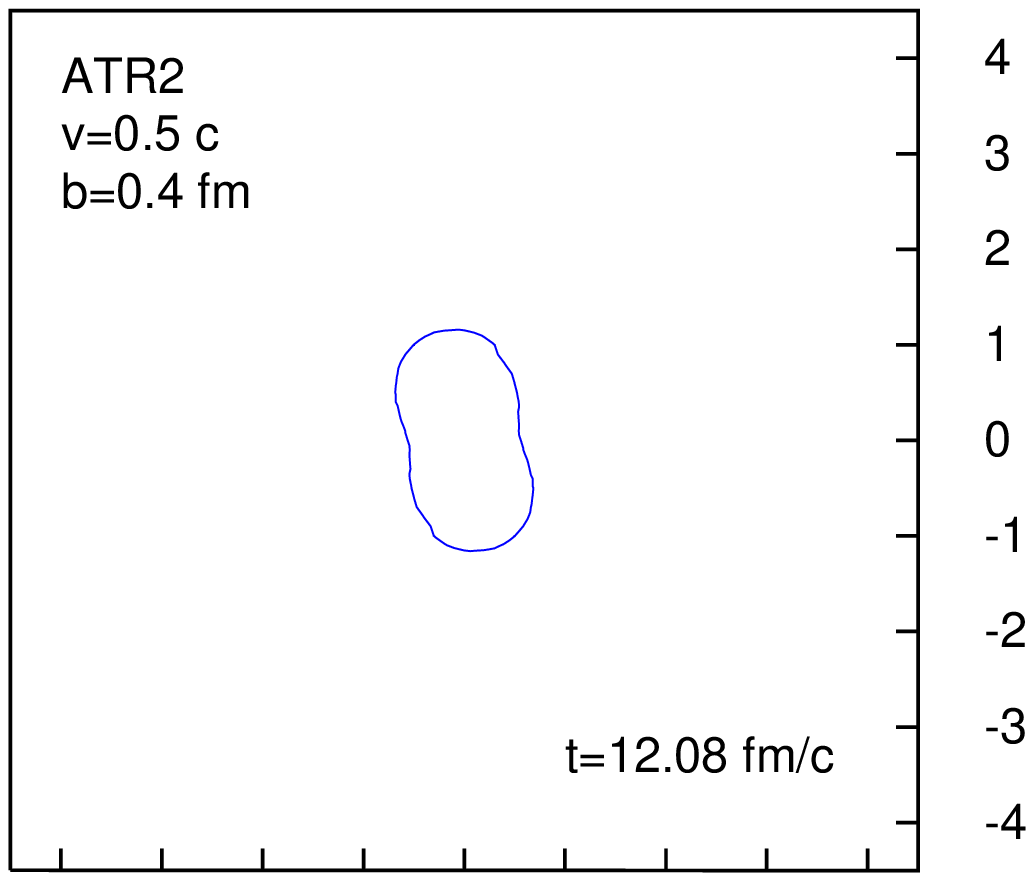,width=8.2cm,angle=0}
}
\vspace{-4.2cm}
\hbox{
\psfig{file=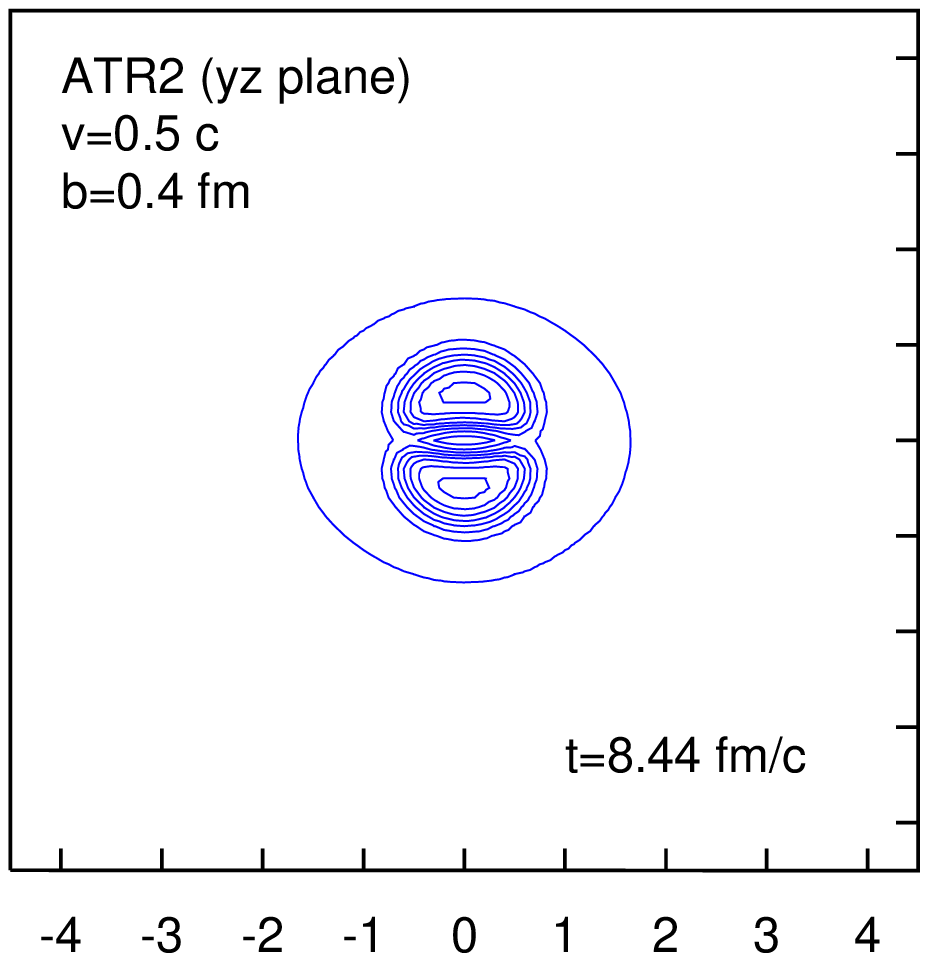,width=8.2cm,angle=0}
\hspace {-4.18cm}
\psfig{file=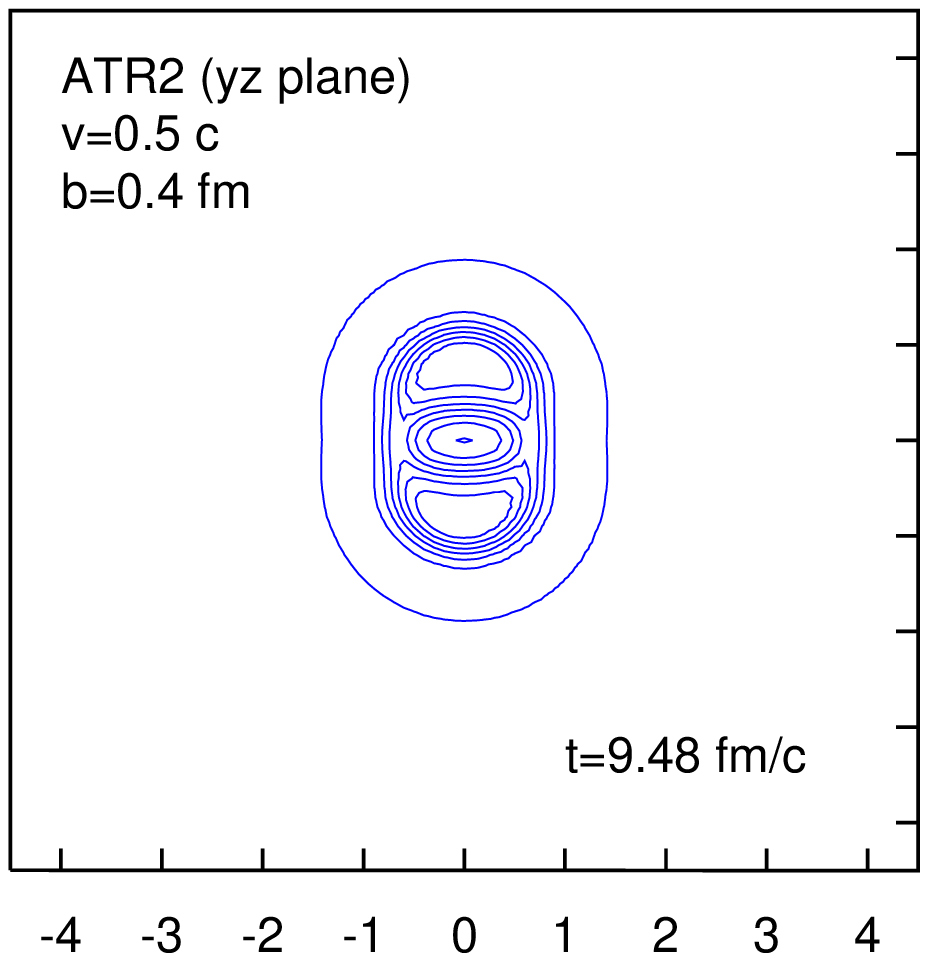,width=8.2cm,angle=0}
\hspace {-4.18cm}
\psfig{file=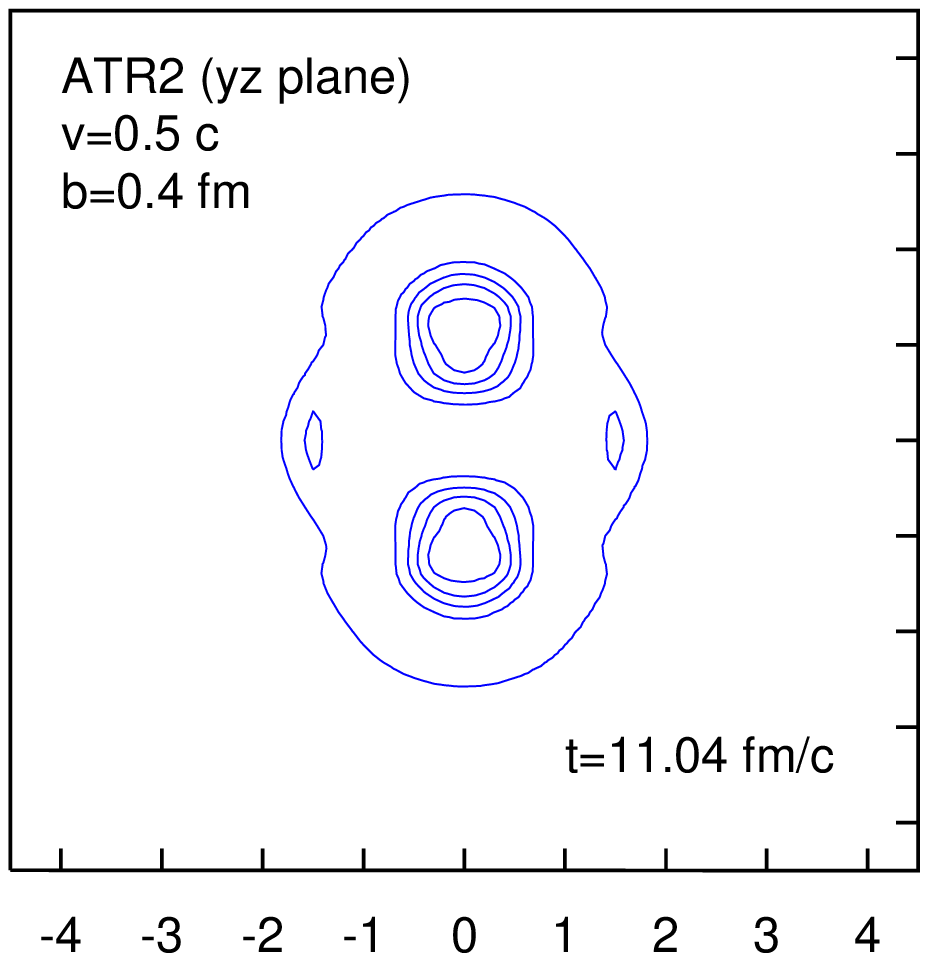,width=8.2cm,angle=0}
\hspace {-4.18cm}
\psfig{file=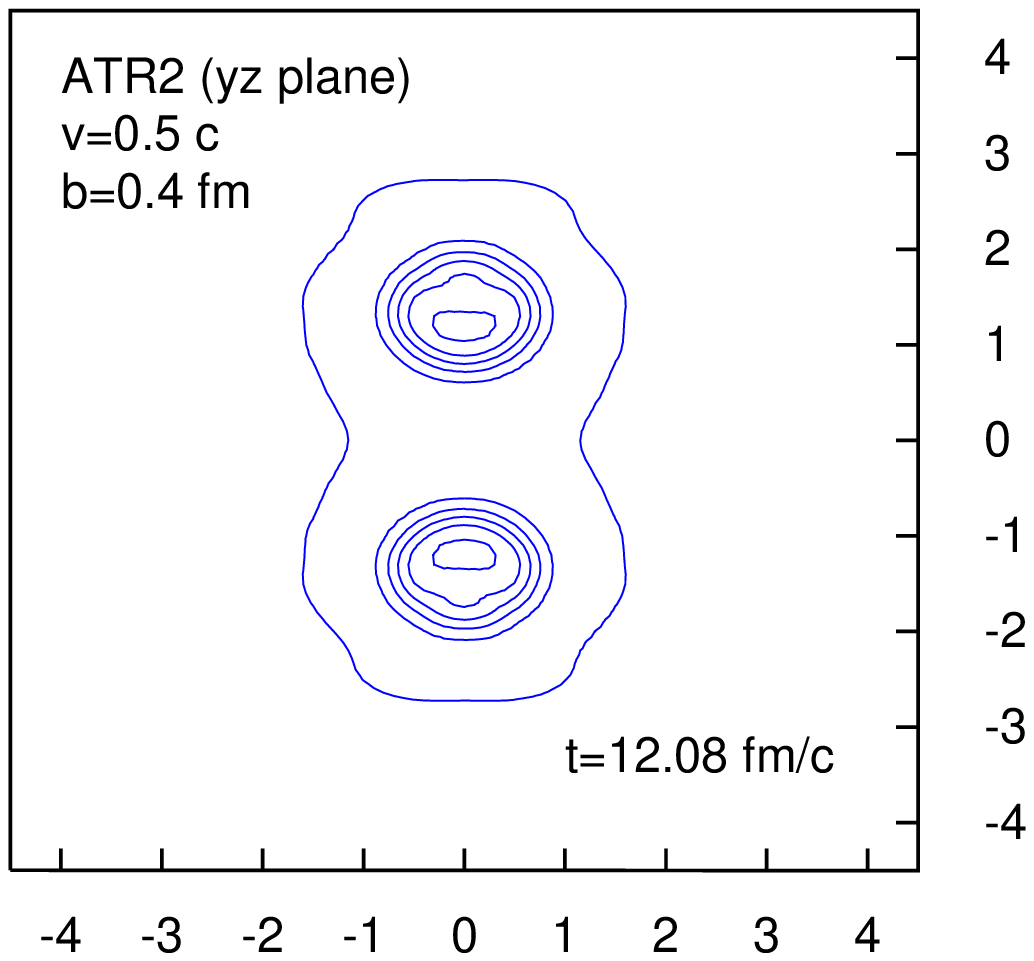,width=8.2cm,angle=0}
}
\vspace{-1.3cm}
}
}
\caption{Contour plots of the energy density in the $xy$ plane (top row)
and for the same times in the $yz$ plane (bottom row)
for the attractive (2) channel with impact parameter $b=0.4~{\rm fm}$.
The spacing between the contours is $100~{\rm MeV/fm^3}$.
The first contour is at the $5~{\rm MeV/fm^3}$ level.
The length on both axes is measured in fermi.
Note that the frames are not evenly spaced in time.}
\label{ATR2_b0.2_series}
\end{figure}

\begin{figure}
\centerline{
\vbox{
\vspace{-2.1cm}
\hbox{
\psfig{file=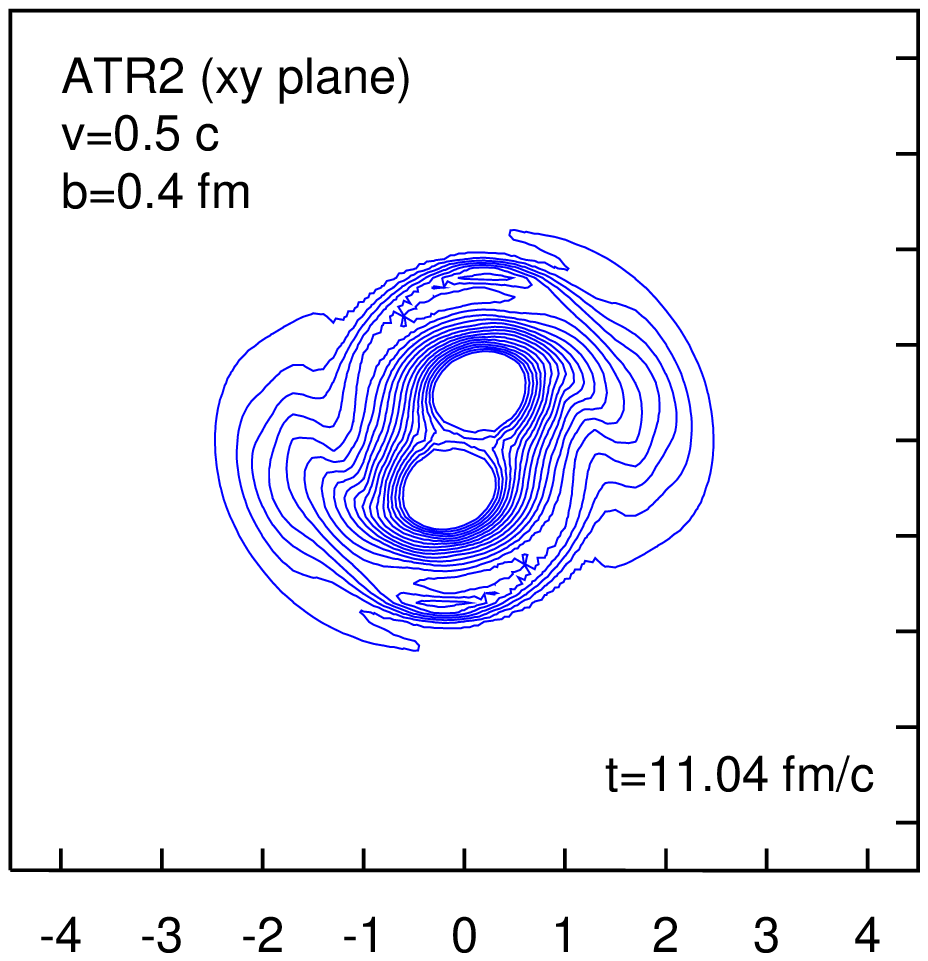,width=8.2cm,angle=0}
\hspace {-4.18cm}
\psfig{file=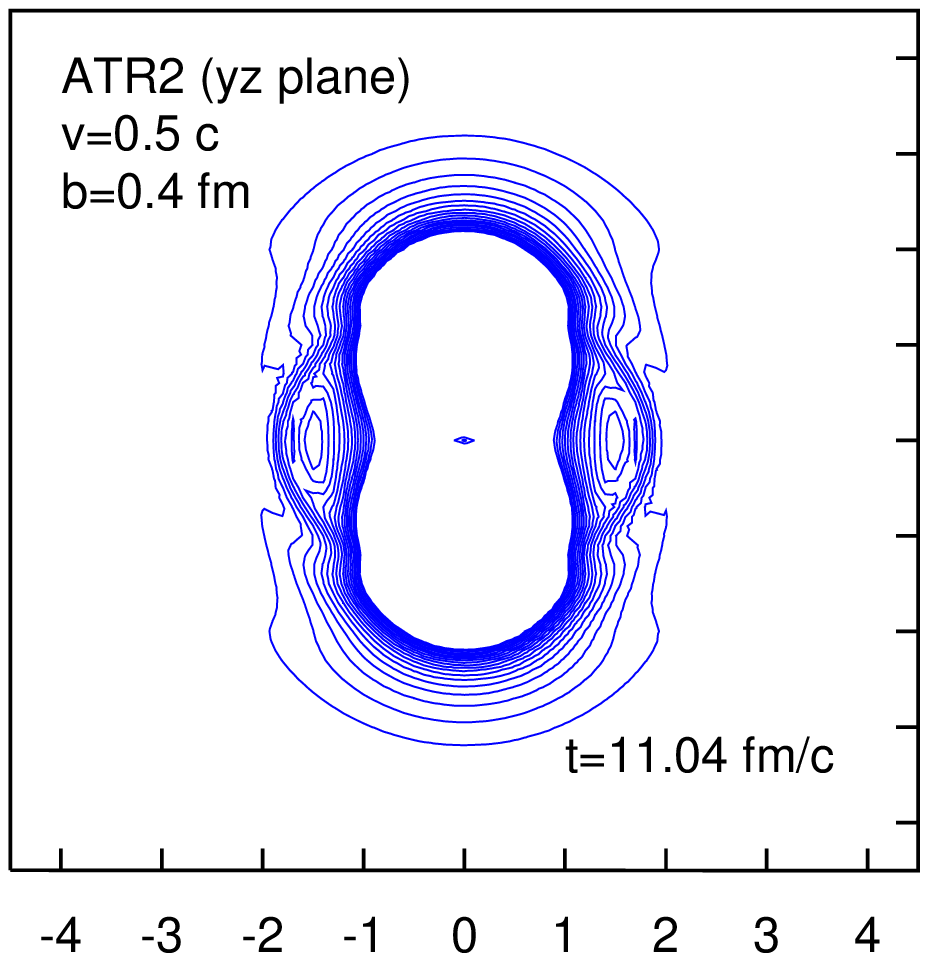,width=8.2cm,angle=0}
\hspace {-4.18cm}
\psfig{file=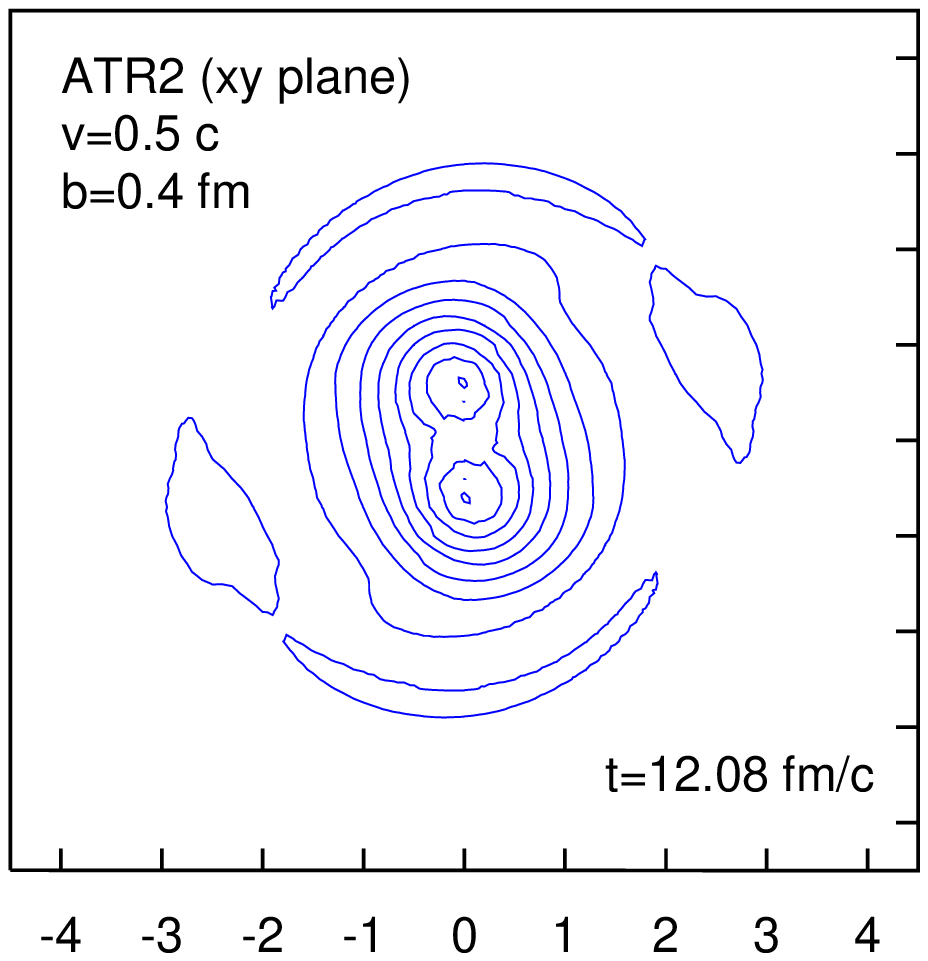,width=8.2cm,angle=0}
\hspace {-4.18cm}
\psfig{file=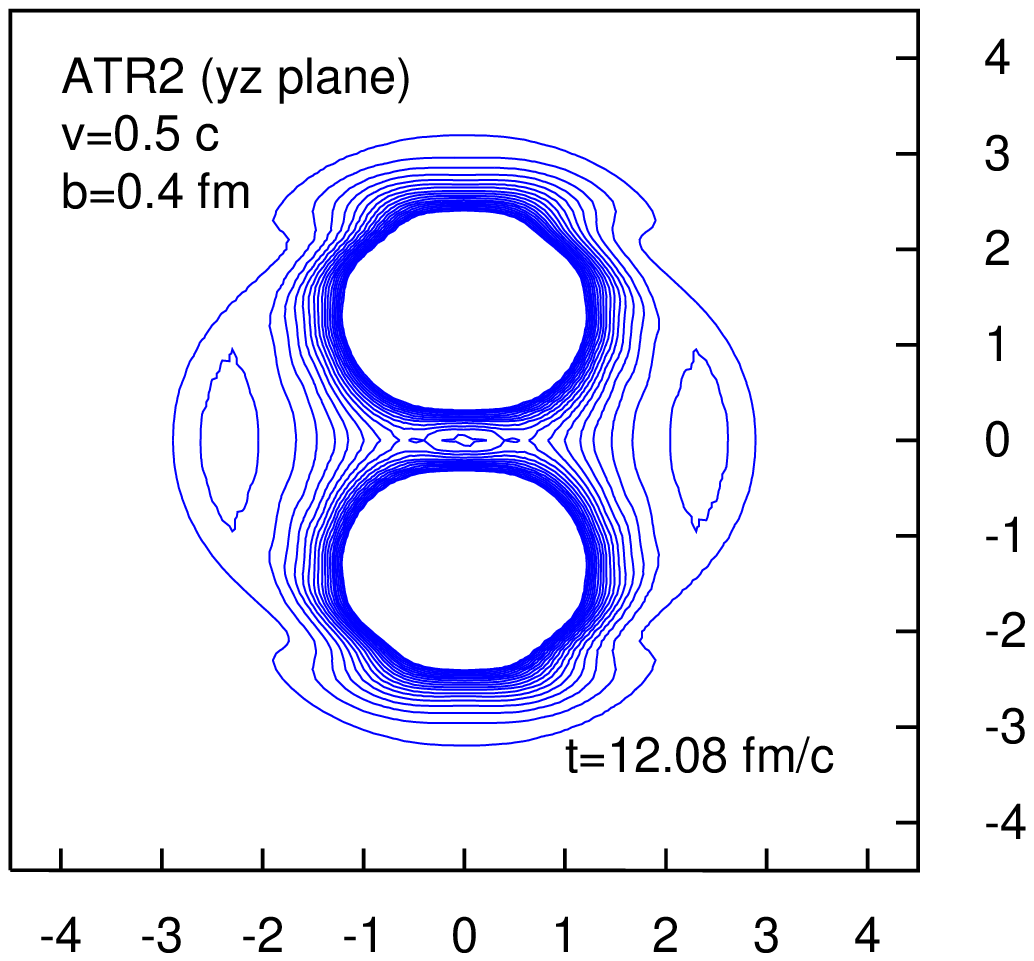,width=8.2cm,angle=0}
}
\vspace{-1.3cm}
}
}
\caption{
Contour plots of the energy density in the $xy$ plane
and for the same times in the $yz$ plane 
for the last two moments shown in Figure \ref{ATR2_b0.2_series}
(attractive (2) channel, $b=0.4 ~{\rm fm}$).
The  spacing between the contours is $1~{\rm MeV/fm^3}$,
from the $1~{\rm MeV/fm^3}$ level to the $20~{\rm MeV/fm^3}$ level. 
The higher energy surfaces are omitted.
The purpose of this fine energy scale is to show the radiation.
}
\label{ATR2_b0.2_radiation}
\end{figure}

The energy contour plots shown in Figure \ref{ATR2_b0.2_series} show
the skyrmions coming together and attempting to form a doughnut about a 
skewed axis in the $xy$ plane and then flying apart in the $z$ direction.
Figure \ref{ATR2_b0.2_radiation} shows the last two  configurations
at higher resolution in energy. This reveals the radiation that carries
off the angular momentum in the $xy$ plane.

\begin{figure}
\centerline{
\vbox{
\hbox{
\psfig{file=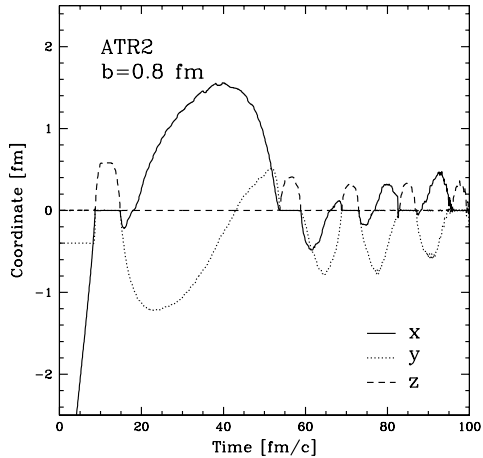,width=6.5cm,angle=0}
}
}}
\caption{
Time evolution of the three coordinates of the topological center of one of
the two skyrmions. 
}
\label{ATR2_b0.4_timepath}
\end{figure}

For the smallest non-zero impact parameter we investigated, $b=0.4 ~{\rm fm}$,
there is meson field radiation left behind in the scattering plane that carries off
the initial angular momentum.
When we go to an impact parameter of $0.8~{\rm fm}$, things change.
The cartesian coordinates for the  $b=0.8~{\rm fm}$ case 
are shown in Figure \ref{ATR2_b0.4_timepath}.
As the skyrmions meet there is some curvature and then
an attempt at uniform motion along $z$ at $x=y=0$. 
Now there is too much angular momentum  for the field to carry away. 
The skyrmions try to go off normal to the scattering
plane, but they have only a brief excursion in that direction while
the meson field is radiating. 
The skyrmions then return to the scattering
plane, but by now the field has taken off so much energy that they
are bound and they begin to orbit in the $xy$ plane, alternating with 
excursions in the $z$ direction, of slowly decreasing amplitude.
In Figure \ref{ATR2path} we have truncated the trajectory of the $b=0.8$ fm 
case at the point where the skyrmions first leave the $xy$ plane.
Presumably their final state would once
again be the static torus. Between $0.4~{\rm fm}$ and $0.8~{\rm fm}$ there must be a critical impact
parameter dividing the cases of skyrmions that escape normal to the 
scattering plane and those that are trapped in orbit in that 
plane. We are investigating this critical impact parameter and the 
nature of the solutions in its vicinity.

\begin{figure}
\centerline{
\vbox{
\vspace{-2.1cm}
\hbox{
\psfig{file=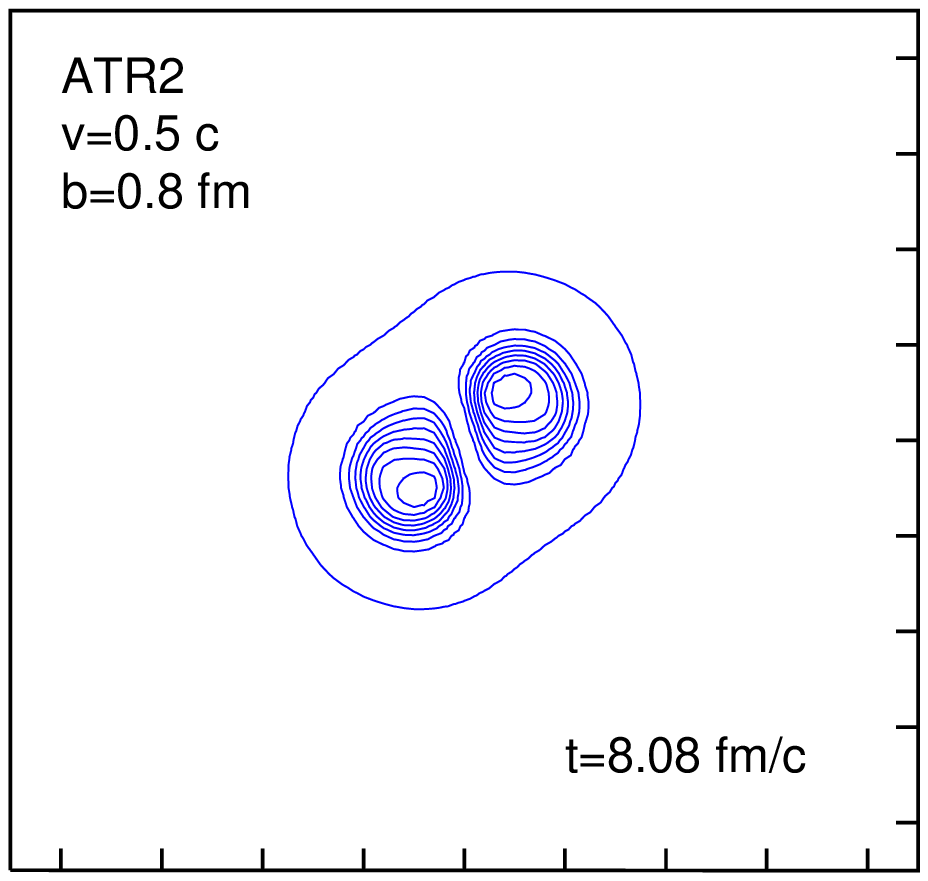,width=8.2cm,angle=0}
\hspace {-4.18cm}
\psfig{file=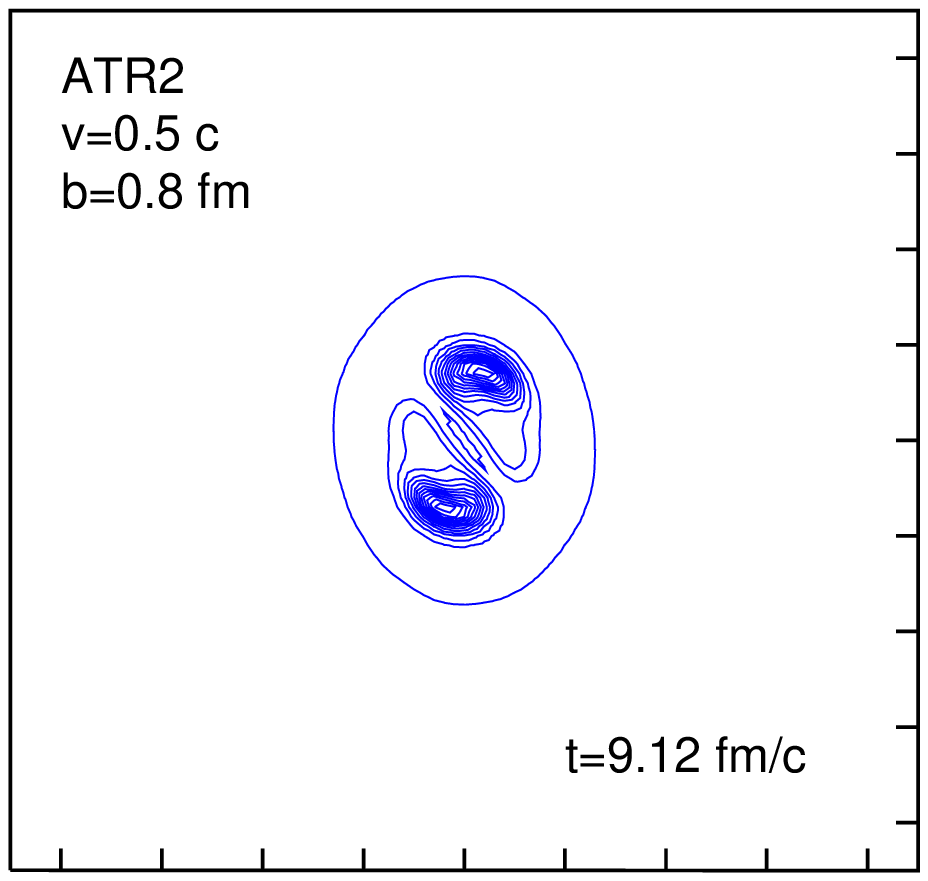,width=8.2cm,angle=0}
\hspace {-4.18cm}
\psfig{file=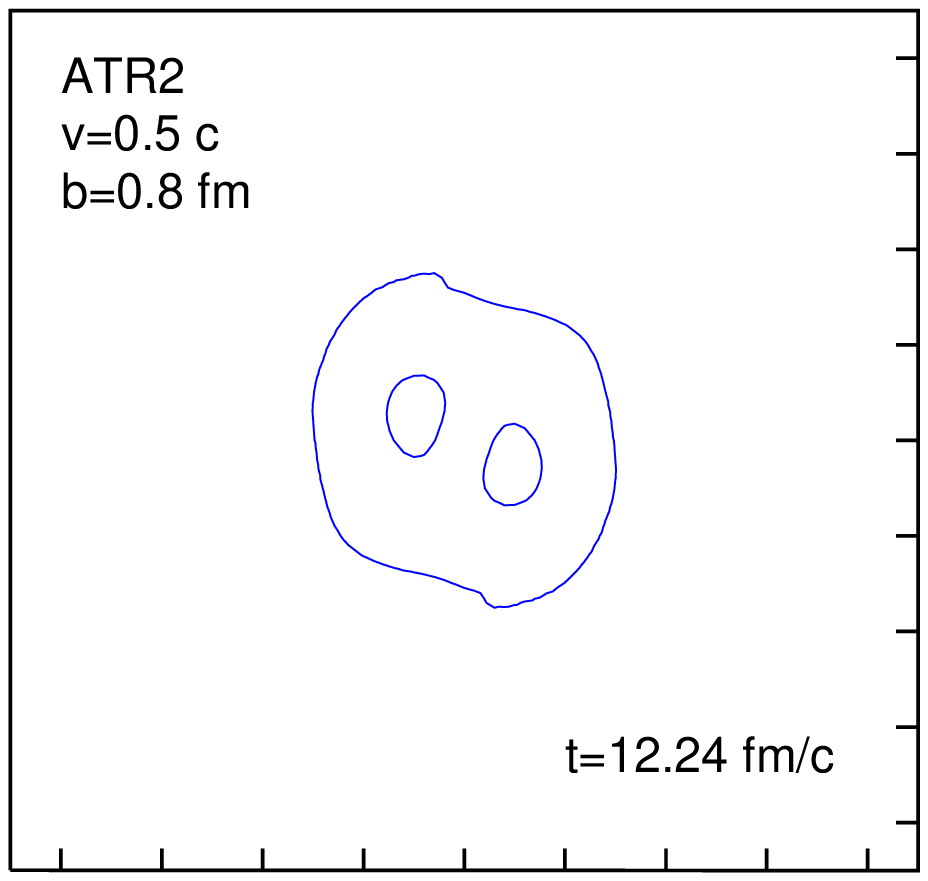,width=8.2cm,angle=0}
\hspace {-4.18cm}
\psfig{file=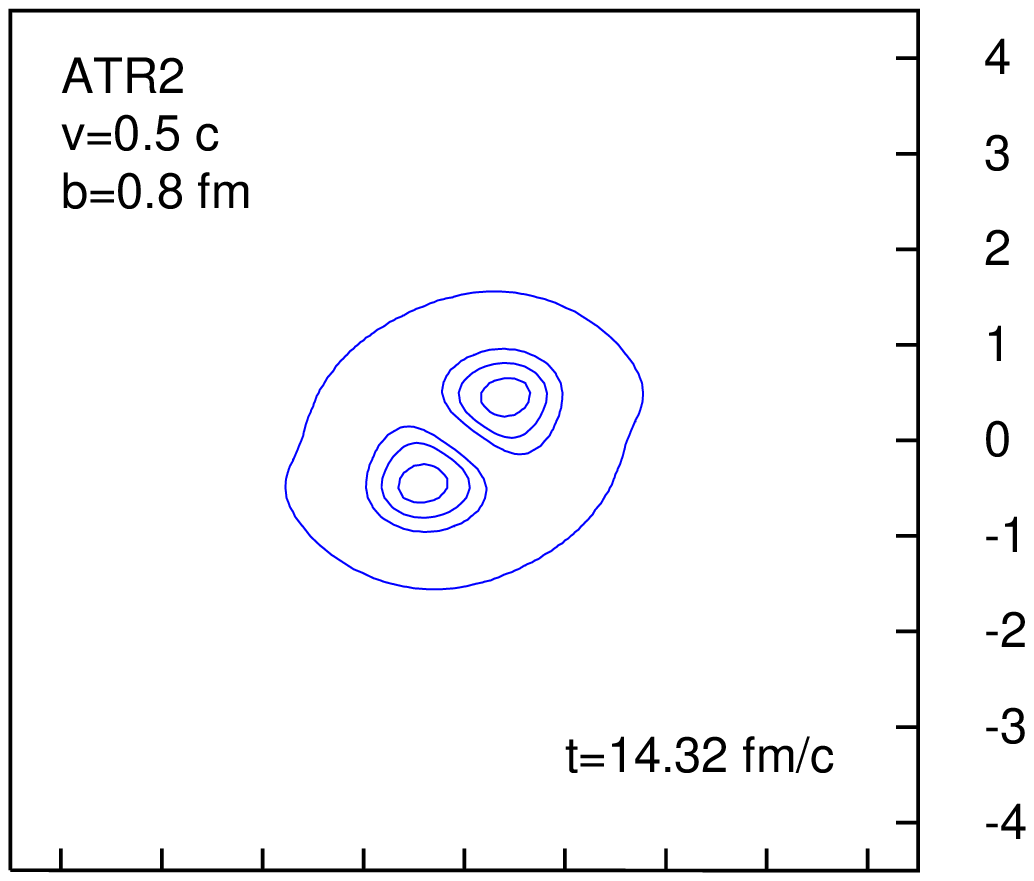,width=8.2cm,angle=0}
}
\vspace{-4.2cm}
\hbox{
\psfig{file=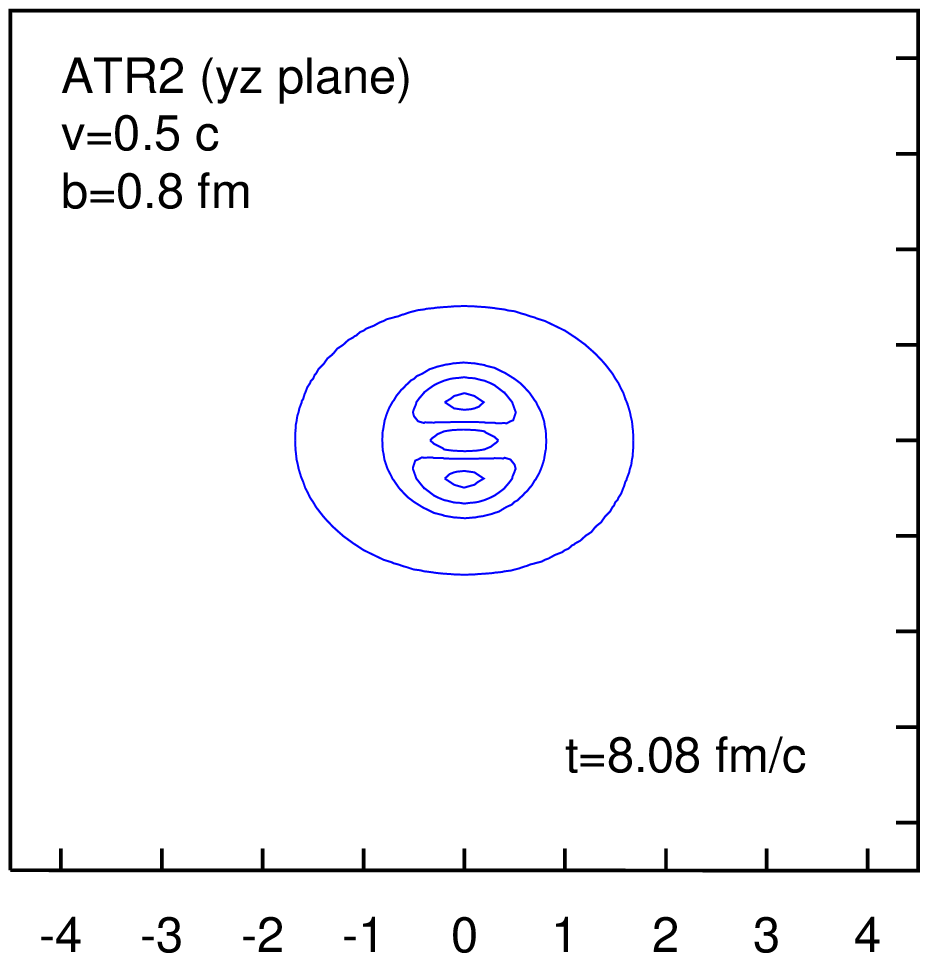,width=8.2cm,angle=0}
\hspace {-4.18cm}
\psfig{file=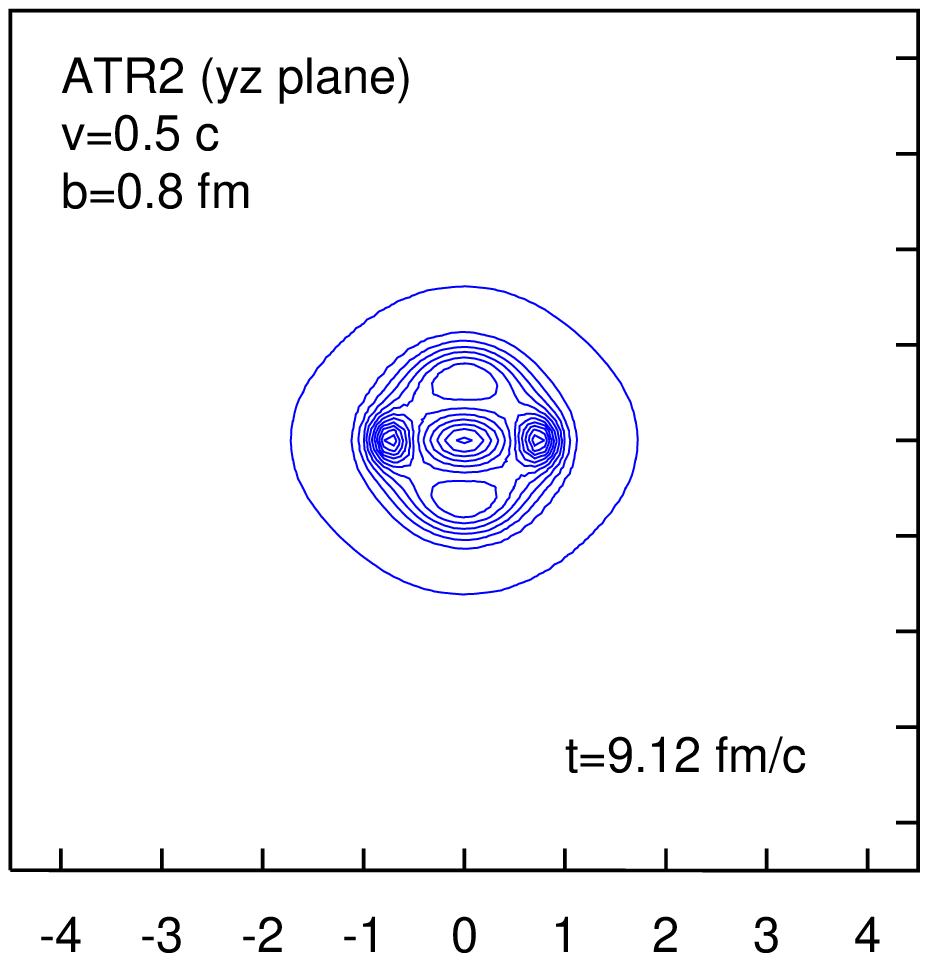,width=8.2cm,angle=0}
\hspace {-4.18cm}
\psfig{file=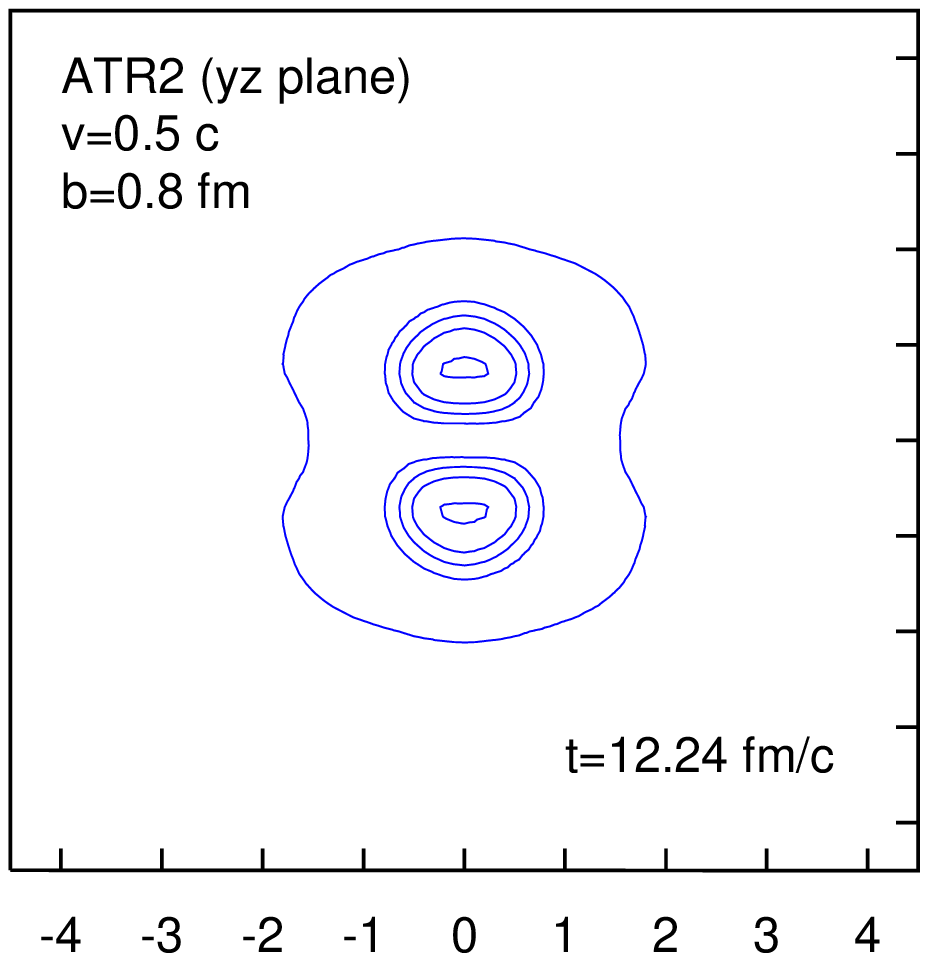,width=8.2cm,angle=0}
\hspace {-4.18cm}
\psfig{file=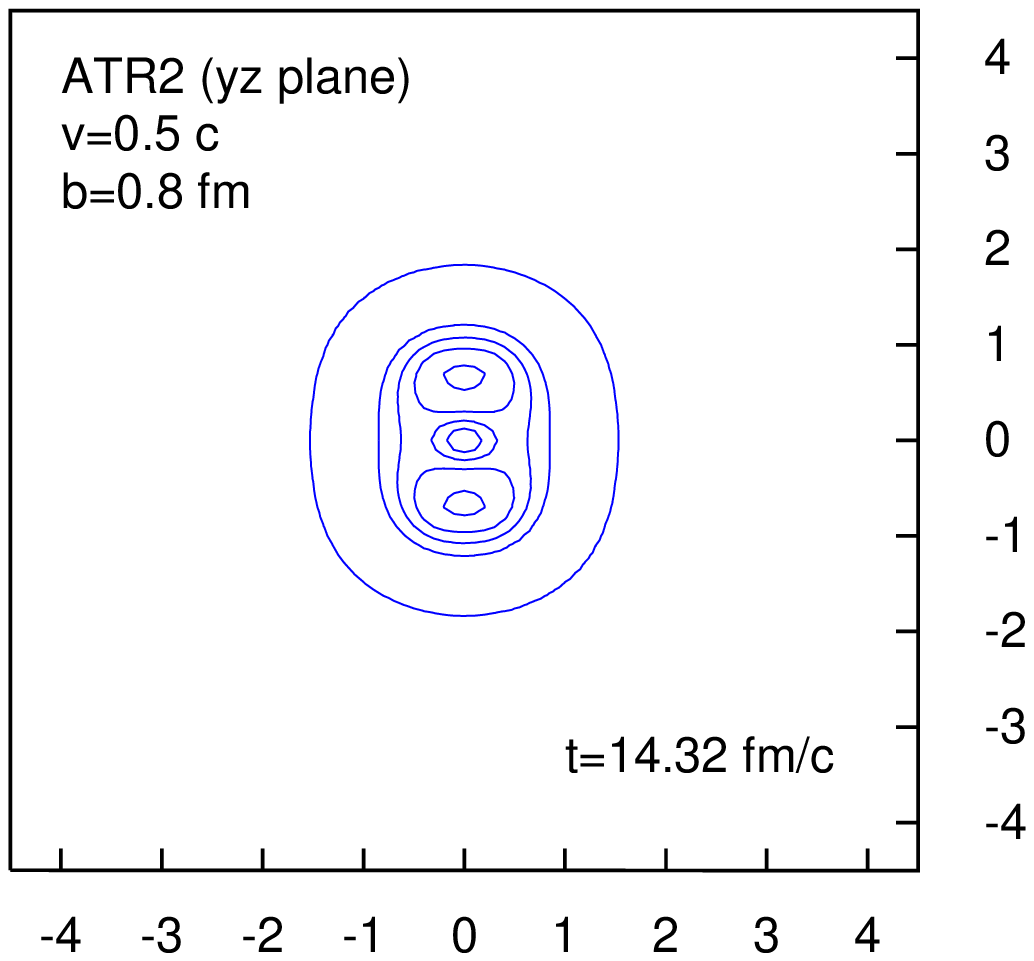,width=8.2cm,angle=0}
}
\vspace{-1.3cm}
}
}
\caption{Contour plots of the energy density in the $xy$ plane (top row)
and in the $yz$ plane (bottom row)
for scattering in the attractive (2) channel with impact parameter $b=0.8$ fm.
The spacing between the contours is $100~{\rm MeV/fm^3}$.
The first contour is at the $5~{\rm MeV/fm^3}$ level.
The length on both axes is measured in fermi.
Note that the frames are not evenly spaced in time.}
\label{ATR2_b0.4_series}
\end{figure}

\begin{figure}
\centerline{
\vbox{
\vspace{-2.1cm}
\hbox{
\psfig{file=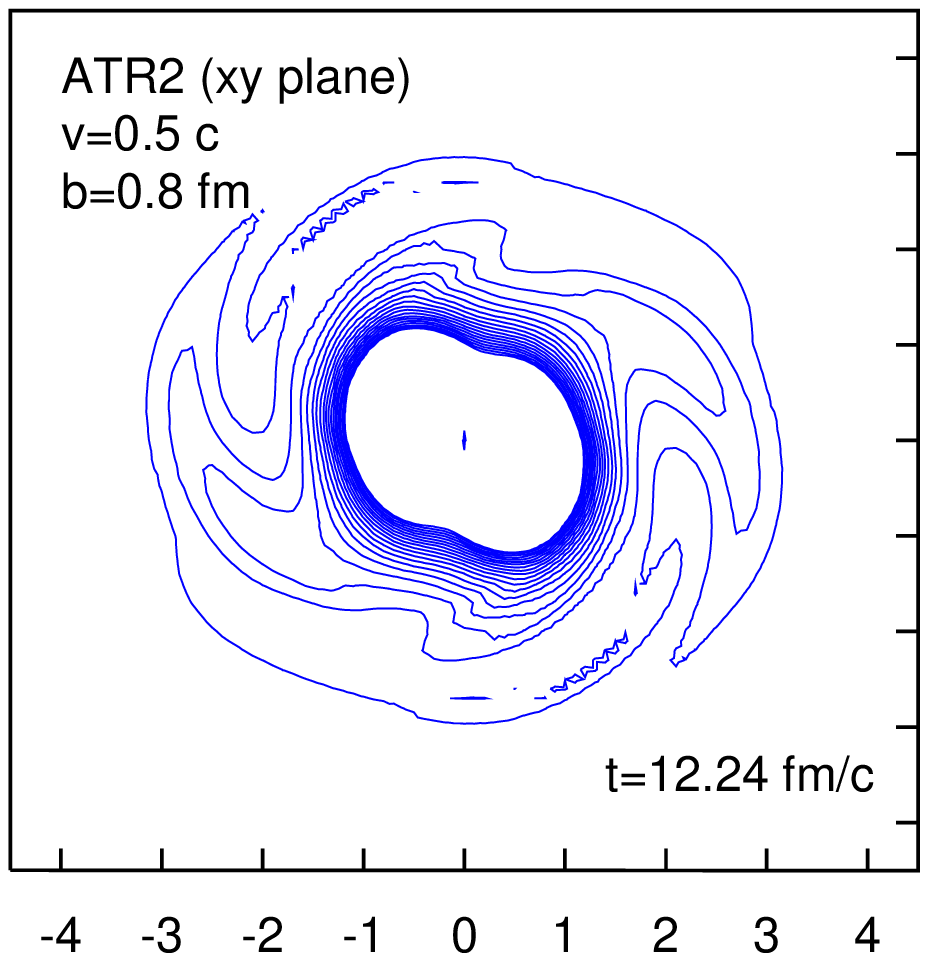,width=8.2cm,angle=0}
\hspace {-4.18cm}
\psfig{file=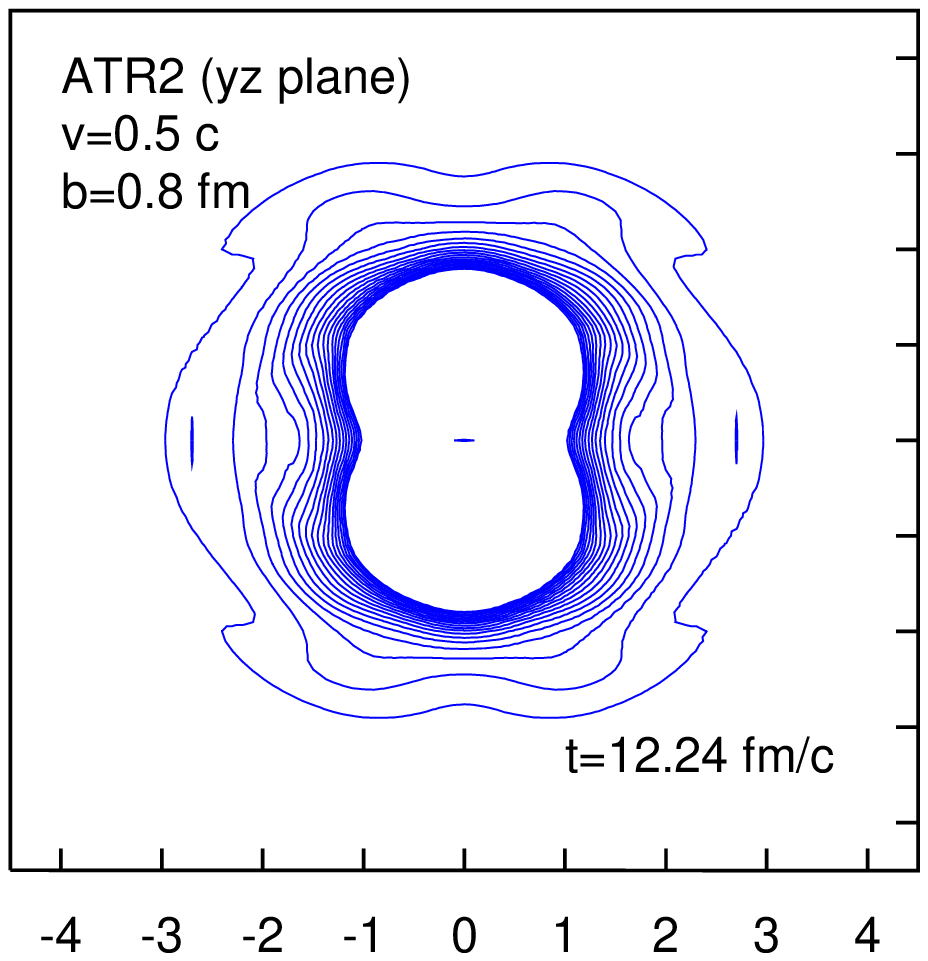,width=8.2cm,angle=0}
\hspace {-4.18cm}
\psfig{file=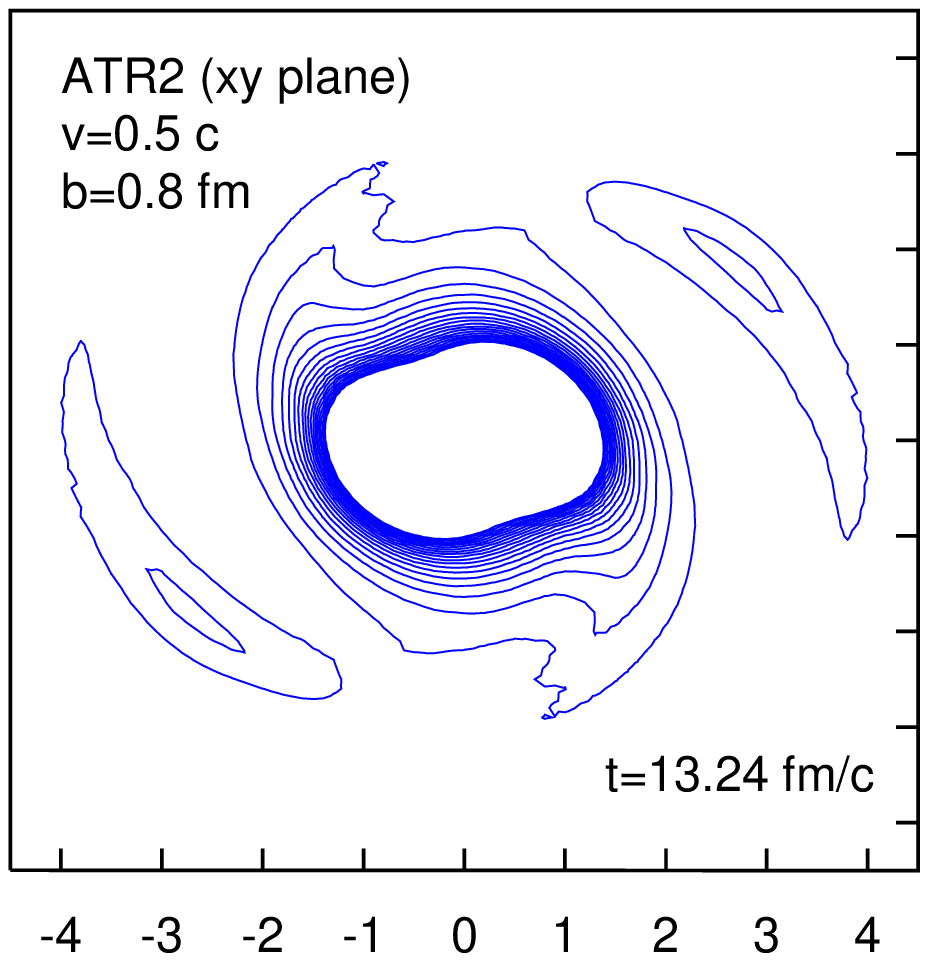,width=8.2cm,angle=0}
\hspace {-4.18cm}
\psfig{file=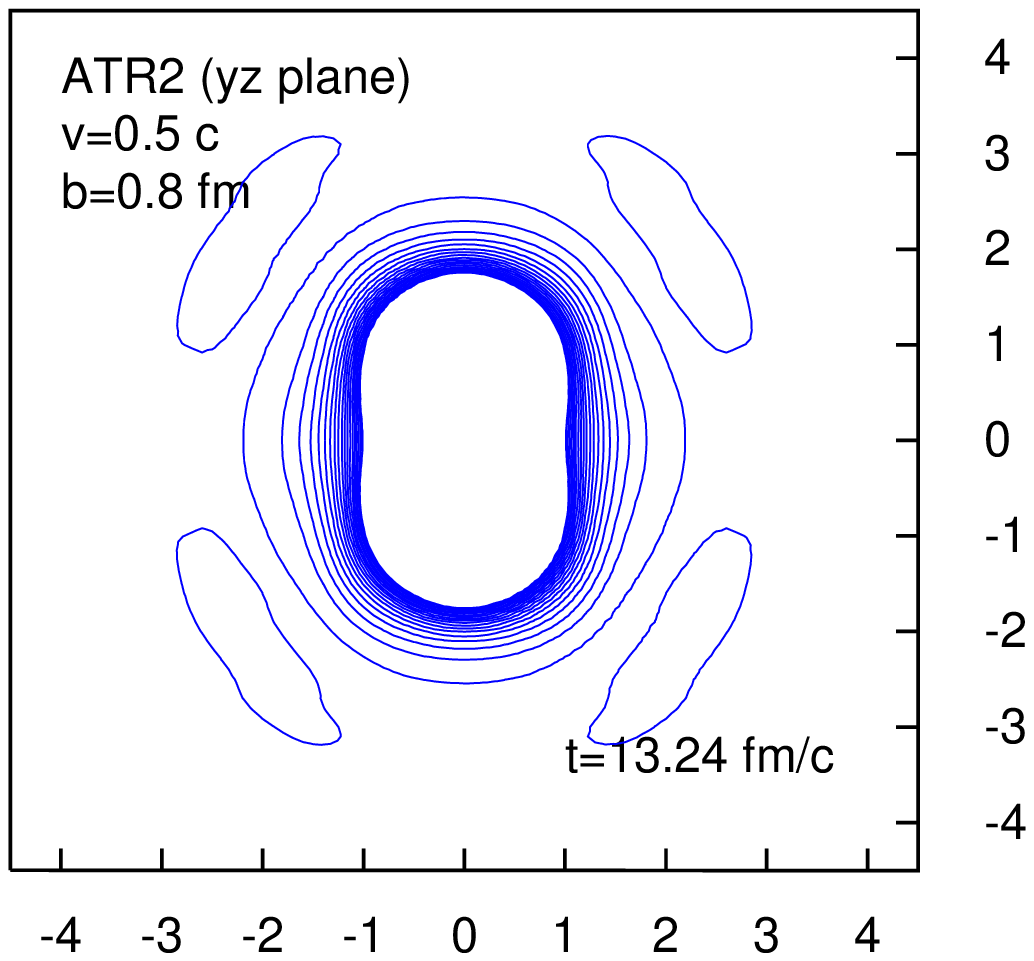,width=8.2cm,angle=0}
}
\vspace{-1.1cm}
}}
\caption{
Contour plots of the energy density in the $xy$ plane
and for the same times in the $yz$ plane 
for two moments of the process shown in Figure \ref{ATR2_b0.4_series}
(attractive (2) channel, $b=0.8 ~{\rm fm}$).
The  spacing between the contours is $1~{\rm MeV/fm^3}$,
from the $1~{\rm MeV/fm^3}$ level to the $20~{\rm MeV/fm^3}$ level. 
}
\label{ATR2_b0.4_radiation}
\end{figure} 

\begin{figure}
\centerline{
\vbox{
\vspace{-2.1cm}
\hbox{
\psfig{file=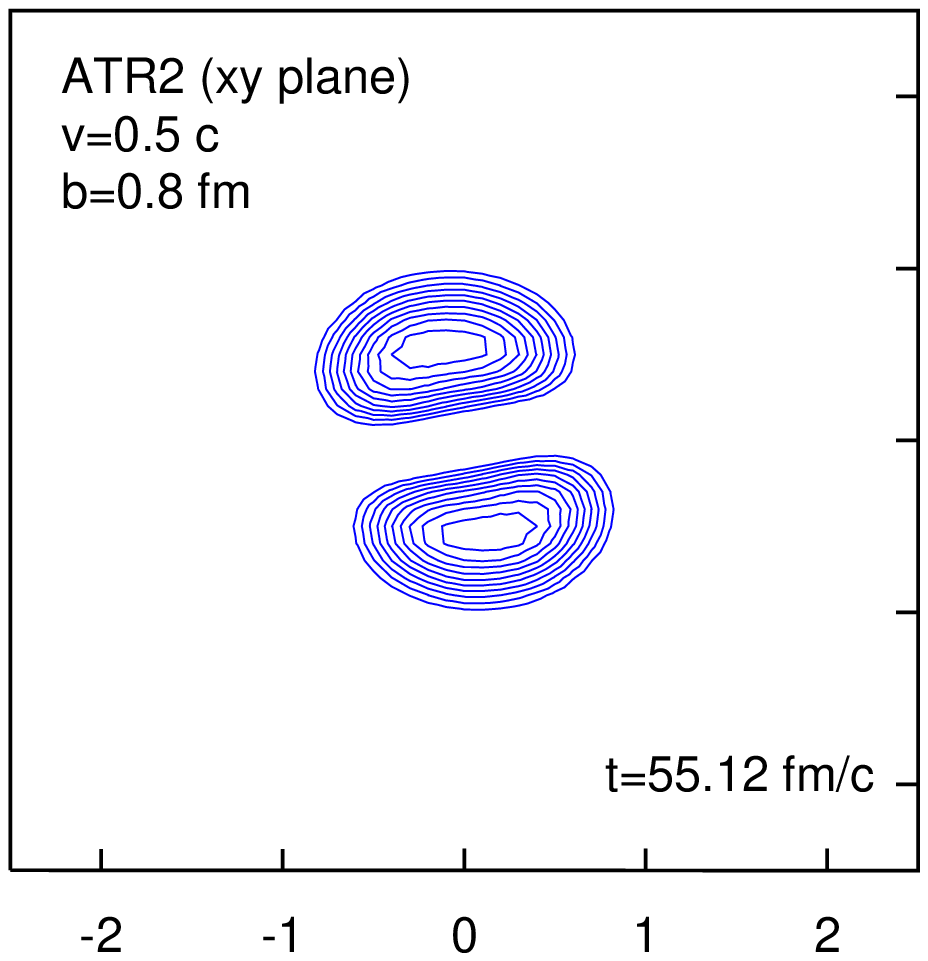,width=8.2cm,angle=0}
\hspace {-4.18cm}
\psfig{file=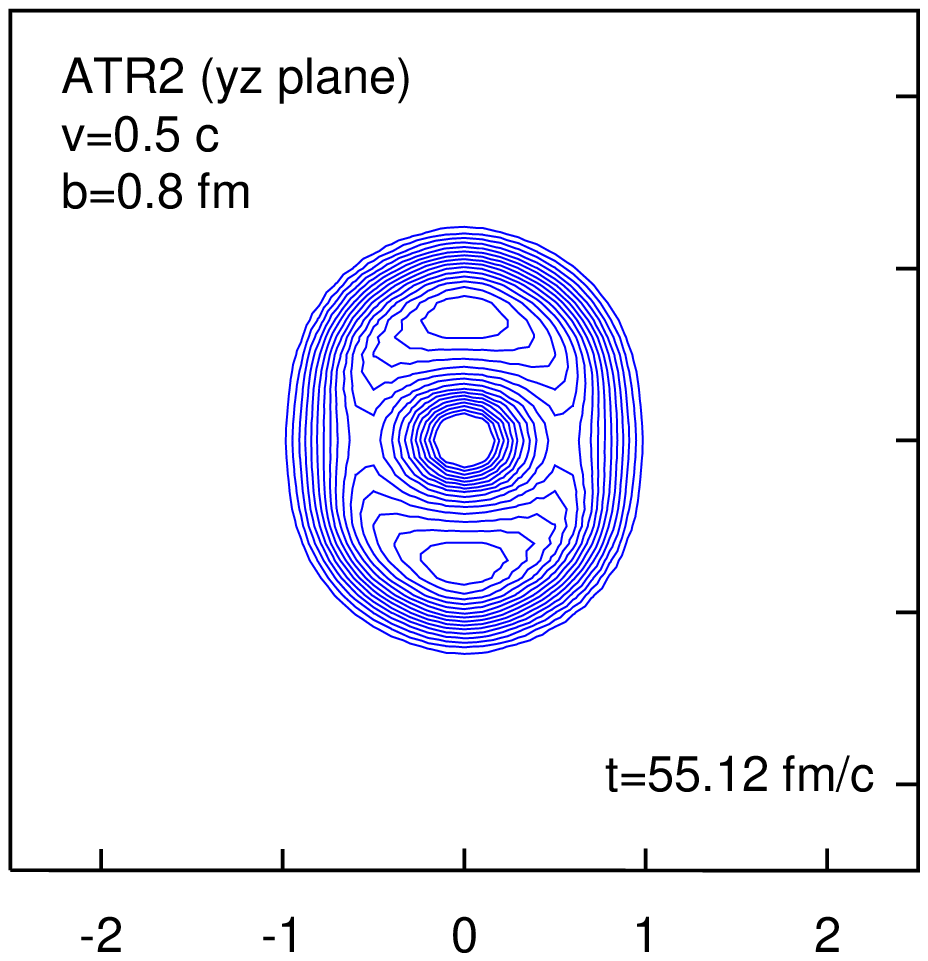,width=8.2cm,angle=0}
\hspace {-4.18cm}
\psfig{file=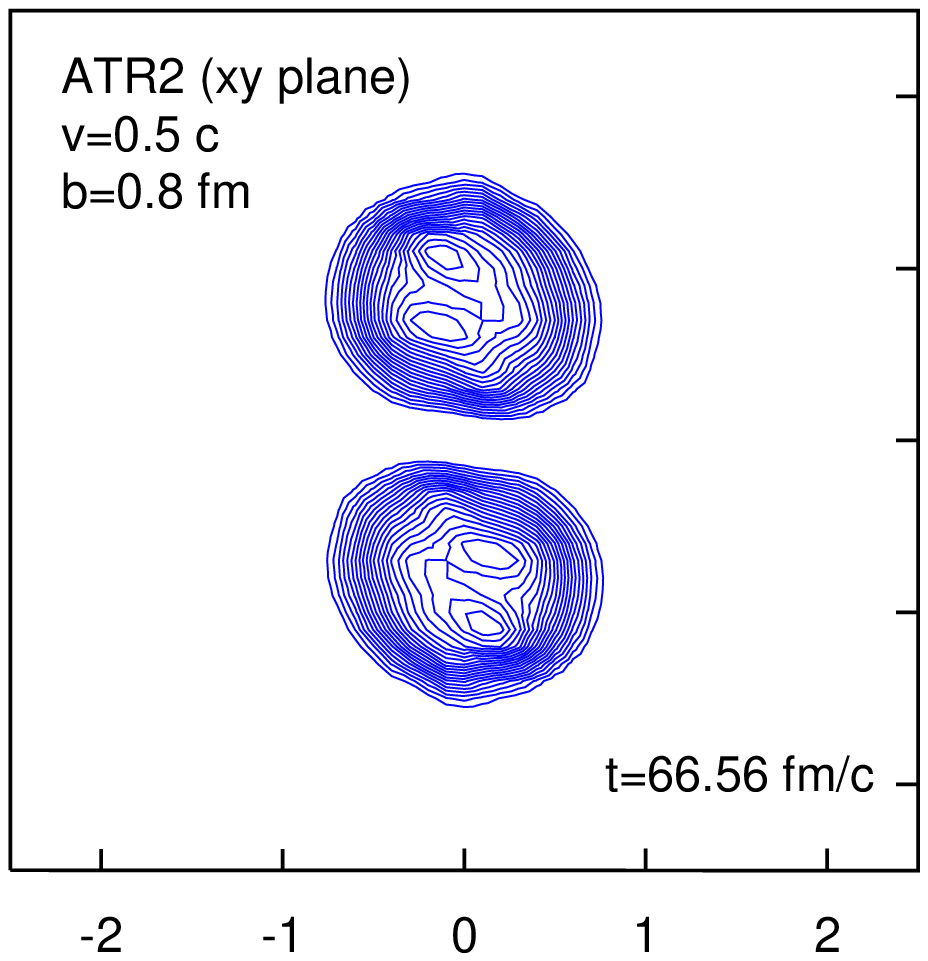,width=8.2cm,angle=0}
\hspace {-4.18cm}
\psfig{file=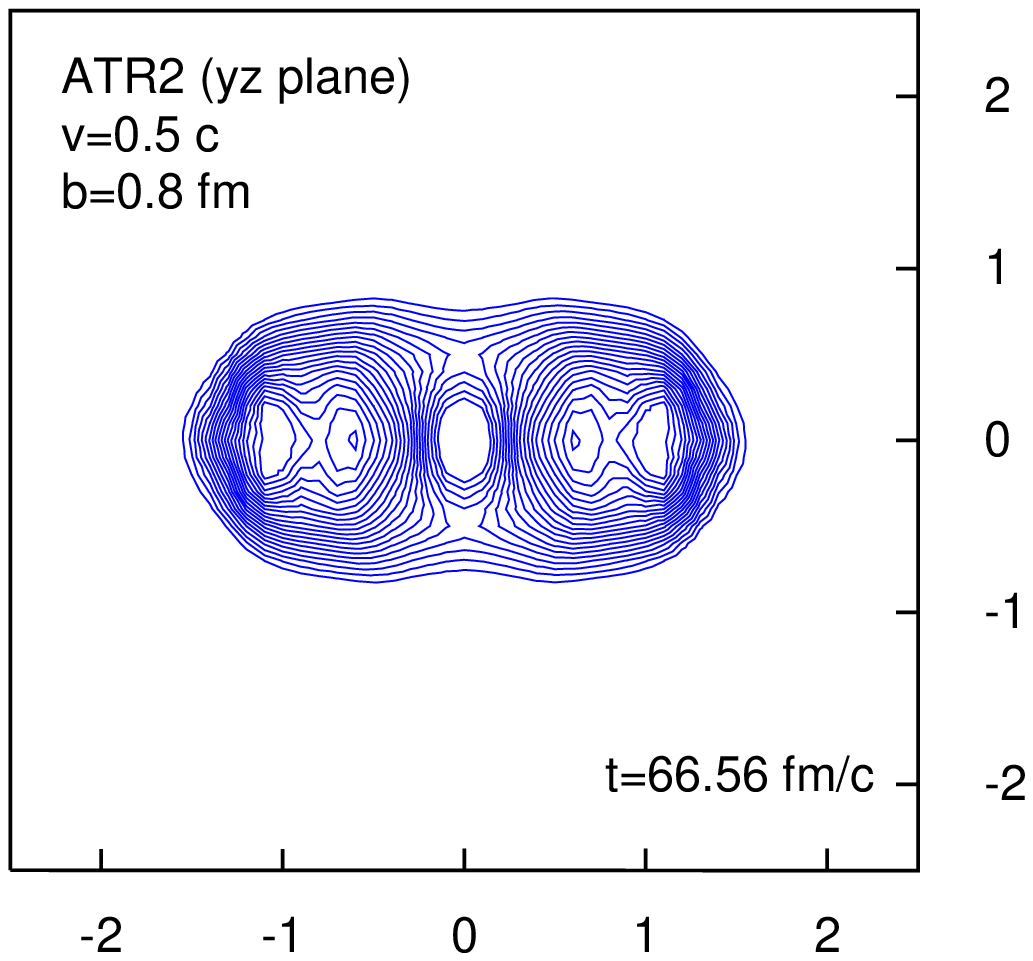,width=8.2cm,angle=0}
}
\vspace{-1.3cm}
}}
\caption{
Two snapshots from the late evolution of the two-skyrmion system 
in the attractive (2) case, for  $b=0.8~{\rm fm}$.
The $xy$ projections shows that the configuration is close to being aligned
with the $y$ axis. The corresponding $yz$ projections exhibit the ``doughnut''
structure. The fact that the deformation is once in the $z$ direction,
once in the $y$ direction indicates that besides spinning, the ``doughnut''
also oscillates.
The energy level contours here are at $20~{\rm MeV/fm^3}$, starting at $100~{\rm MeV/fm^3}$
and the view is slightly tilted.
}
\label{ATR2_b0.4_series2}
\end{figure}

Energy contour plots for the $b=0.8~{\rm fm}$ 
impact parameter are shown in Figure \ref{ATR2_b0.4_series}.
The first few frames are very similar to those in Figure \ref{ATR2_b0.2_series},
showing the skyrmions approaching, forming a doughnut practically in the $yz$ plane,
and attempting to escape in the $z$ direction.
As we see in Figure \ref{ATR2_b0.4_radiation}, 
there is considerable radiation at this time as the
field tries to carry off the angular momentum.
The field cannot, and the skyrmions come back first along $z$ and then through the doughnut
into the $xy$ plane, avoiding the center. 
Note that both the $z$ turn around and the subsequent $x$ turn around comes at no more than
$3.0$ fm separation, as it must. 
At larger separation, the skyrmions would
be out of effective force range and not able to come back. When 
the skyrmions  return to the $xy$ plane, they have lost so much energy
that they are bound. They begin to orbit, and then
make another unsuccessful attempt to escape along $z$. 
As can be seen from the time evolution plot, this sequence of
orbiting of the topological centers in the $xy$ plane alternating with
excursions out of the plane in the $z$ direction continues for a long time. 
We interpret this as follows.
The two skyrmions have essentially merged into a 
$B=2$ configuration. The residual angular momentum forces the torus
to rotate around the $z$ axis. 
The remainder of the momentum with which the
two skyrmions came into this configuration (more precisely, its component pointing
to the center), while not enough to push the skyrmions out in the perpendicular
direction, results in oscillations which deform the doughnut as illustrated in Figure
\ref{ATR2_b0.4_series2}. This central component of the momentum allows the skyrmions
to escape in the $b=0.4~{\rm fm}$ and the central case.

We are not able to follow this cycle
to completion, but believe that eventually the system will radiate enough
energy and angular momentum to settle into the static, axial-symmetric $B=2$ torus,
similarly to the small impact parameters for 
repulsive grooming discussed in the previous section.
For impact parameters of $1.6~{\rm fm}$ and $2.8~{\rm fm}$ in this channel, the scattering
is unremarkable by comparison, as is seen in Figure \ref{ATR2path}.
For the largest impact parameter we notice the medium-range \em repulsive \em
interaction, just the opposite of the corresponding case in the repulsive grooming
discussed above.
The two groomings are interchanged at the point where the centers cross the $x=0$ plane.

\section{Conclusion and Outlook}

The phenomenology of two-skyrmion scattering for various groomings
as it emerges from our qualitative study, may be summarized as follows.
In the \em absence of grooming \em , the scattering process is almost elastic
at the energy we considered. The kinematics is similar to that of point-particles
interacting via a central repulsive interaction. This is to be expected, since 
the interaction between two hedgehog configurations is central.
We know \cite{centralsc} that at higher velocities one expects to excite
vibrational modes of the individual skyrmions. We do see modest indications of that.

The head-on collision in the repulsive channel is quasi-elastic, similar to the
hedgehog-hedgehog processes.
The remaining collisions involving grooming of $180^o$ can be divided into two categories,
depending of whether or not the impact parameter, $b$, is small enough
for the formation of the $B=2$ (torus-shaped) bound state. 
If the impact parameter is large, the collisions are quasi-elastic with a weak 
attractive or repulsive character depending on the grooming.
If the impact parameter is small enough, 
the collision proceeds through a (sometimes deformed) doughnut configuration
in the plane normal to the grooming axis,
even if the final state is not bound. 

The cleanest example is the \em attractive (1) \em
channel, with the grooming axis normal to the plane of motion. 
For zero impact parameter it is well known \cite{rightangle} that
a torus is formed in the $xy$ plane, the skyrmions lose their identity, and fly out 
at right angles with respect to the incident direction and the grooming axis. 
For a nonzero impact
parameter, a deformed torus still appears in the $xy$ plane, and the scattering angle 
decreases continuously from $90^o$ as $b$ increases.

In the \em  repulsive \em case (grooming around the direction of motion, $x$), the torus
is formed close to the $yz$ plane. 
As the two incident skyrmions approach the
$y$ axis (that of the impact parameter), 
they find themselves in an attractive
configuration, since they are now groomed 
around an axis ($x$) perpendicular to the one ($y$)
connecting them. 
They form the doughnut initially in the plane perpendicular to the
grooming axis. The configuration carries some of the initial angular
momentum by rotating around the $z$ axis. The skyrmions do not tend to exit in the 
perpendicular direction because they came in with very little momentum \em in \em
the plane of the torus. 

In the attractive (1) and (2) cases, the initial momenta  
have a large component pointing to the center of the torus. This momentum is channeled 
into the perpendicular direction, always leading to scattering in the attractive (1) case.
In the \em attractive (2) \em case, 
with grooming around the impact parameter direction $y$,
the doughnut is formed close to the $xz$ plane. The $90^o$ scattering would therefore
happen in the $z$ direction, but this is strongly limited by the necessity of shedding 
the angular momentum around $z$. Radiation provides a mechanism for this, and for a
small enough impact parameter the skyrmions can escape in the $z$ direction. Otherwise
they go into a rotating doughnut configuration. 
In this case however, there is radiation and significant momentum in
the plane of the torus, which torus also exhibits quadrupole oscillations. 

While our investigation is by no means complete, we did identify a 
significant number of distinct patterns for the outcome of 
skyrmion-skyrmion collisions. One may define a number of critical 
configurations which separate the different outcomes, even for fixed velocity.
The picture might get even richer by sweeping a range of incident velocities.

\bigskip

The calculations of skyrmion-skyrmion scattering reported in the Section
above demonstrate two things. First they show that the $\omega$ meson 
stabilizes the Skyrme model and makes it numerically tractable out
to long times and through complex dynamical situations. Second the
calculation shows a rich mix of phenomena.
These include capture and orbiting, 
radiation, and excursions out of the scattering plane. Although
each of these arises naturally in the model, they have not been seen 
or demonstrated before, 
nor have most of them been suggested before. 
They therefore add to the  rich and often surprising mix of results
in the Skyrme model. 
Since this is a model of low energy
QCD at large $N_C$,
the new results give insight into aspects of QCD in the non-perturbative 
long wave length or low energy domain. 
This is a region in which our best hope for insight 
comes from effective theories like the Skyrme model. 

The success of these calculations also gives us confidence that the
method can be carried over, with only simple changes, to the annihilation
problem. In particular the stability of the calculations, thanks 
both to $\omega$ meson stabilization and a number of numerical
strategies, suggests that the annihilation calculation will also be stable.
For annihilation, meson field (pion and omega) radiation 
in the final state is all  
there is and the fact that we clearly see radiation in the skyrmion-skyrmion
case is reassuring. 

Our work here suggests a number of further avenues. We have already discussed
annihilation. It would also be interesting to explore the landscape of
skyrmion-skyrmion scattering as a function of energy as well as grooming
and impact parameter and also to examine more closely the behavior
of the model in the neighborhood of critical parameters where the 
scattering behavior changes abruptly. Also interesting would be to
try to extract information about nucleon-nucleon scattering from the
results for skyrmion-skyrmion scattering. We are investigating these
questions. 

\section{Acknowledgements}

We are grateful to Jac Verbaarschot and Yang Lu for many helpful discussions
on both physics and numerical issues. We are indebted to Folkert Tangerman
and Monica Holbooke for a number of suggestions regarding the calculation.
Jac Verbaarschot is also acknowledged for a critical reading of the manuscript.

This work was supported in part by the National Science Foundation. We are
very grateful to Prof. R. Hollebeek for making available to us the considerable
computing resources of the National Scalable Cluster Project Center at the
University of Pennsylvania, which center is also supported by the National
Science Foundation.

\end{document}